\documentclass[aps,showpacs,nofootinbib,superscriptaddress,preprintnumbers]{revtex4}

\usepackage{graphicx}
\usepackage{amssymb}
\usepackage{amsmath}
\usepackage{bm}

\newcommand{\g}{\gamma}
\newcommand{\uno}{1 \kern -0.3em {\rm l}}
\newcommand{\nn}{\nonumber}
\newcommand{\ds}{\displaystyle}
\newcommand{\sT}{\scriptscriptstyle T}
\newcommand{\half}{\textstyle {\frac{1}{2}}}
\newcommand{\intl}{\int\frac{d^4l}{(2\pi)^4}}
\newcommand{\pslash}{\kern 0.2 em p\kern -0.45em /}
\newcommand{\lslash}{\kern 0.2 em l\kern -0.45em /}
\newcommand{\Pslash}{\kern 0.2 em P\kern -0.5em /}
\newcommand{\Sslash}{\kern 0.2 em S\kern -0.5em /}

\begin{document}
\allowdisplaybreaks[2]

\title{Transverse-momentum distributions in a diquark spectator 
model}

\author{Alessandro Bacchetta}
\email{alessandro.bacchetta@jlab.org}
\affiliation{Theory Center, Jefferson Lab, 12000 Jefferson Av., Newport News, VA 23606, USA}

\author{Francesco Conti}
\email{francesco.conti@pv.infn.it}
\affiliation{Dipartimento di Fisica Nucleare e Teorica, Universit\`{a} di 
Pavia, I-27100 Pavia, Italy}
\affiliation{Istituto Nazionale di Fisica Nucleare, Sezione di Pavia, I-27100 Pavia, Italy}

\author{Marco Radici}
\email{marco.radici@pv.infn.it}
\affiliation{Istituto Nazionale di Fisica Nucleare, Sezione di Pavia, I-27100 
Pavia, Italy}

\begin{abstract}
All the leading-twist parton distribution functions are calculated in a 
spectator model of the nucleon, using scalar and axial-vector diquarks. Single 
gluon rescattering is used to generate T-odd distribution functions.
Different choices for the diquark polarization states are considered, as well
as a few options for the form factor at the nucleon-quark-diquark vertex. 
The results are listed in analytic form and interpreted in terms of light-cone 
wave functions. The model parameters are fixed by reproducing the 
phenomenological parametrization of unpolarized and helicity parton 
distributions at the lowest available scale. Predictions for the other parton 
densities are given and, whenever possible, compared with available 
phenomenological parametrizations. 
\end{abstract}



\preprint{JLAB-THY-08-841}

\maketitle

\section{Introduction}
\label{sec:intro}

Partonic transverse-momentum distributions (TMDs) --- also 
called unintegrated PDFs --- describe the probability to find in a hadron a 
parton with longitudinal momentum fraction $x$ and transverse momentum 
${p}_{\sT}$ with respect to the direction of the parent hadron
momentum~\cite{Collins:1982uw}. They give a three-dimensional view of the 
parton distribution in momentum space, complementary to 
what can be obtained through generalized parton
distributions~\cite{Burkardt:2000za,Diehl:2002he,Ji:2003ak,Belitsky:2003nz,Boffi:2007yc}. 

In 
the last years a lot of theoretical and experimental activity related to TMDs 
has taken place. Crucial steps were made in the understanding of 
factorization theorems involving TMDs ($k_{\sT}$
factorization)\cite{Ji:2004wu,Ji:2004xq}.  
Some of the properties of TMDs 
have been investigated from the theoretical standpoint. 
For instance, positivity bounds were presented in
Ref.~\cite{Bacchetta:1999kz}.  Relations among these 
functions in the large-$N_c$ limit of QCD were put forward in 
Ref.~\cite{Pobylitsa:2003ty}. Their behavior at large $x$ was studied in 
Ref.~\cite{Brodsky:2006hj}, and at high transverse momentum 
in Ref.~\cite{Bacchetta:2008xw}. Last but not least, 
it was also demonstrated~\cite{Brodsky:2002cx}
that TMDs that  are odd under na\"ive 
time-reversal transformations (for brevity, T-odd) 
can be nonzero and must be included in the  complete list of 
leading-twist TMDs (see, e.g., 
Ref.~\cite{Boer:1997nt,Bacchetta:2006tn}). 
Their universality properties are different from the standard
PDFs~\cite{Collins:2002kn}. 

In the meanwhile, 
several azimuthal asymmetries were measured in semi-inclusive deep inelastic
scattering (SIDIS) and elsewhere (see Ref.~\cite{D'Alesio:2007jt} and 
references therein), and  more experimental measurements are planned. 
However, 
not much phenomenological information concerning TMDs is available as yet
(see, e.g., Ref.~\cite{D'Alesio:2004up} and references therein). 
The analysis of azimuthal spin
asymmetries both in hadron-hadron collisions and in SIDIS led to the extraction
of the Sivers function~\cite{Sivers:1990cc}, denoted 
as $f_{1T}^\perp$, a T-odd TMD that describes how
the parton distribution is distorted by the transverse polarization of the
parent hadron  
(see Ref.~\cite{Anselmino:2005an} for a comparison of
various parametrizations). A recent attempt to extract the T-odd Boer-Mulders
function, $h_1^{\perp}$~\cite{Boer:1998nt}, a T-odd TMD describing the
distribution of transversely polarized partons in an unpolarized hadron, 
was presented in 
Ref.~\cite{Zhang:2008nu}. 
All of
the above studies assume a flavor-independent 
Gaussian distribution of the transverse momentum,
although there is no compelling reason for this choice. 

In this context, building a relatively simple model to compute TMDs and to 
allow for numerical estimates is of great importance. From the theoretical 
side, this can help understanding some of the essential features of TMDs,
for instance their relation to the orbital angular momentum of partons
(see, e.g., Refs.~\cite{Brodsky:2002cx,Brodsky:2002rv,Burkardt:2003je,Goeke:2006ef,Lu:2006kt,Meissner:2007rx,Qiu:2007ey,Burkardt:2007xm}).
From the experimental side, a model could be useful to estimate the size of 
observables in different processes and kinematical 
regimes~\cite{Gamberg:2003ey,Gamberg:2003eg,Lu:2004au,Gamberg:2005ip,Barone:2006ws,Lu:2007ev,Gamberg:2007wm} 
and to set up Monte Carlo
simulations~\cite{Bianconi:2004wu,Bianconi:2005bd,Bianconi:2005yj,Bianconi:2006hc,Radici:2007vc}.

Although many model calculations of integrated PDFs are available, there are 
not so many for TMDs. In Ref.~\cite{Jakob:1997wg} all the leading-twist 
T-even functions were calculated in a spectator model with scalar and 
axial-vector diquarks. Recently, an analogous calculation has been
  performed in a light-cone quark model~\cite{Pasquini:2008ax}. 
T-odd functions were calculated in the spectator model 
with scalar 
diquarks~\cite{Brodsky:2002cx,Boer:2002ju,Goldstein:2002vv,Gamberg:2003ey}, 
with scalar and vector diquarks~\cite{Bacchetta:2003rz,Gamberg:2007wm}, in the 
MIT bag model~\cite{Yuan:2003wk,Cherednikov:2006zn}, in a constituent quark
model~\cite{Courtoy:2008vi}  and in the spectator model for the 
pion~\cite{Lu:2004hu}. A complete calculation of all the leading-twist TMDs 
in a spectator model with scalar diquarks was presented in 
Ref.~\cite{Meissner:2007rx}. 

In this work, we choose a more phenomenological approach. We consider also 
axial-vector diquarks (in the following often called simply vector diquarks), 
necessary for a realistic flavor analysis, and we 
further distinguish between isoscalar ($ud$-like) and isovector ($uu$-like) 
spectators. We generate the relative phase necessary to produce T-odd 
structures by approximating the gauge link operator with a one gluon-exchange 
interaction. We consider several choices of form factors at the 
nucleon-quark-diquark vertex and several choices for the polarization states of 
the diquark. All results are presented in analytic form and interpreted also in 
terms of overlaps of light-cone wave functions, leading to a detailed analysis 
of the quantum numbers of the quark-diquark system. The free parameters of the 
model are fixed by reproducing the phenomenological parametrization of 
unpolarized and longitudinally polarized parton distributions at the lowest 
available scale.

The paper is organized as follows. In Sec.~\ref{sec:phi}, the analytic form for 
all the leading-twist TMDs is discussed for the dipolar 
nucleon-diquark-quark form factor and for the light-cone choice of the diquark 
propagator, postponing the results for the other explored combinations to the 
Appendices~\ref{sec:A} (T-even TMDs) and~\ref{sec:B} (T-odd TMDs). 
In Sec.~\ref{sec:fit}, numerical results are shown and 
compared with phenomenological parametrizations, whenever available 
in the literature. In Sec.~\ref{sec:end}, some conclusions are drawn.


\section{Analytical results for transverse-momentum-dependent parton densities}
\label{sec:phi}

In this section we present the fundamentals of the model and we give in
analytical form the results for the light-cone wave functions (LCWFs) 
and the TMDs
obtained in the model.

\subsection{General framework}

In the following we will make use of light-cone coordinates.
We introduce the light-like vectors $n_\pm$ satisfying $n_\pm^2 = 0, \, n_+\cdot
n_-=1,$ and we describe a generic 4-vector $a$ as
\begin{equation}
a = [a^-,a^+,\bm{a}_{\sT}] 
\end{equation} 
where $a^\pm = a\cdot n_\mp$. We will make use of the transverse tensor
$\epsilon_{\sT}^{ij}= \epsilon^{\mu \nu i j} n_{+\mu} n_{-\nu}$, whose only nonzero
components are $\epsilon_{\sT}^{12}=-\epsilon_{\sT}^{21} = 1$. We choose a frame where the
hadron momentum $P$ has no transverse components, i.e.,
\begin{equation}
P=\biggl[\frac{M^2}{2 P^+},P^+,\bm{0}\biggr]. 
\end{equation} 
The quark momentum can be written as
\begin{equation}
p=\biggl[\frac{p^2+\bm{p}_{\sT}^2}{2 x P^+},x P^+,\bm{p}_{\sT}\biggr]. 
\end{equation} 

In a hadronic state $|P,S \rangle$ with momentum $P$ and spin $S$, the density 
of quarks  can be defined starting from the quark-quark correlator (see, 
e.g., Ref.~\cite{Bacchetta:2006tn}) 
\begin{equation} 
\Phi (x,{\bm p}_{\sT};S)= \int \frac{d\xi^- d\bm{\xi}_{\sT}}{(2\pi)^3}\; 
       e^{i p \cdot \xi}\,\langle P,S|\bar{\psi}(0)\,{\cal U}_{[0,\xi]}\,
       \psi(\xi)|P,S \rangle \Big|_{\xi^+=0} \; ,
\label{eq:Phi-tree}
\end{equation}
where 
\begin{equation} 
U_{[0,\xi]} = {\cal P}\, e^{-ig \int_0^\xi dw \cdot A(w)}
\label{eq:link}
\end{equation}
is the so-called gauge link operator, or Wilson line, connecting the two 
different space-time points $0,\xi$, by all possible ordered paths followed by 
the gluon field $A$, which couples to the quark field $\psi$ through the 
coupling $g$. The gauge link ensures that the matrix element of
Eq.~(\ref{eq:Phi-tree}) is color-gauge invariant and arises from the
interaction of the outgoing quark field with the spectators inside the hadron. 
The 
leading contributions of the path $[0,\xi]$ in space-time are selected by
the hard process in which the parton distributions appear, thus breaking
standard universality of the parton densities. For instance, in 
SIDIS the gauge link path in light-cone coordinates 
runs along
\begin{equation} 
[0,\xi] \equiv (0,0,{\bm 0}_{\sT})\to (0,\infty ,{\bm 0}_{\sT})\to (0,\infty ,
\infty_{\sT})\to (0,\infty ,{\bm \xi}_{\sT})\to (0,\xi^-,{\bm \xi}_{\sT}) \; , 
\label{eq:SIDISpath}
\end{equation}
while in the Drell--Yan case it runs in the opposite direction through
$-\infty$. This fact leads to a sign difference in T-odd parton densities, 
as mentioned for the first time in Ref.~\cite{Collins:2002kn}.

Similarly to Ref.~\cite{Jakob:1997wg}, we evaluate the correlator of 
Eq.~(\ref{eq:Phi-tree}) in the spectator approximation, i.e.\ we insert a 
completeness relation and at tree-level we truncate the sum over final states
to a single
on-shell spectator state with mass $M_X$, thus getting the analytic form
\begin{equation} 
\Phi(x,{\bm p}_{\sT},S) \sim \frac{1}{(2\pi)^3}\,\frac{1}{2(1-x)P^+}\, 
\overline{\mathcal{M}}^{(0)}(S)\, \mathcal{M}^{(0)}(S) 
\Big\vert_{p^2=\tau (x,{\bm p}_{_T})}\; , 
\label{eq:Phi-tree-spect}
\end{equation}
where $p$ is the momentum of the active quark, $m$ its mass, and the on-shell
condition $(P-p)^2=M_X^2$ for the spectator implies for the quark the off-shell
condition
\begin{align} 
p^2 \equiv \tau (x,\bm{p}_{\sT}) &=-\frac{\bm{p}_{\sT}^2+L_X^2(m^2)}{1-x}+m^2 \; , 
&L_X^2(m^2)&=x M_X^2 + (1-x) m^2 - x (1-x) M^2 \; ,
\label{eq:offshell} 
\end{align} 
with $M$ the hadron mass.

\begin{figure}[h]
\begin{center}
\includegraphics[width=6cm]{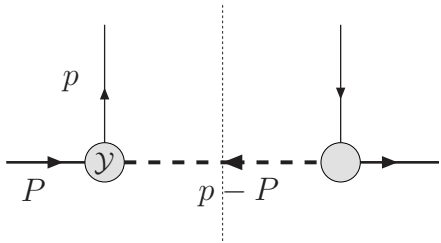} 
\end{center}
\caption{Tree-level cut diagram for the calculation of T-even 
leading-twist parton densities. The dashed line indicates both 
scalar and axial-vector diquarks.}
\label{fig:sidis}
\end{figure}

We assume the spectator to be point-like, with the quantum numbers of a
diquark. Hence, the proton can couple to a quark and to a spectator diquark 
with spin 0 (scalar $X=s$) or spin 1 (axial-vector $X=a$), as well as with 
isospin 0 (isoscalar $ud$-like system) or isospin 1 (isovector $uu$-like 
system). Therefore, the tree-level ``scattering amplitude'' $\mathcal{M}^{(0)}$ 
is given by (see Fig.~\ref{fig:sidis}) 
\begin{equation} 
\mathcal{M}^{(0)}(S) = \langle P-p |\psi(0) |P,S\rangle = \
  \begin{cases}
  \displaystyle{\frac{i}{\pslash-m}}\, {\cal Y}_s \, U(P,S)& 
        \text{scalar diquark,} \\ 
  \displaystyle{\frac{i}{\pslash-m}}\, \varepsilon^*_{\mu}(P-p,\lambda_a)\,
        {\cal Y}^{\mu}_a \, U(P,S)& \text{axial-vector diquark,} 
  \end{cases}
\label{eq:m-tree}
\end{equation}
and is actually a Dirac spinor because of the understood spinorial indices of 
the quark field $\psi$. The $\varepsilon_\mu(P-p,\lambda_a)$ is the 4-vector 
polarization of the spin-1 vector diquark with momentum $P-p$ and 
helicity states $\lambda_a$. When summing over all polarizations 
states, several choices have been used for 
$d^{\mu \nu}= \sum_{\lambda_a} \varepsilon^{\ast \mu}_{(\lambda_a)} \, 
\varepsilon^\nu_{(\lambda_a)}$: 
\begin{equation} 
d^{\mu \nu}(P-p)
= \left\{ 
  \begin{array}{cc}
  -g^{\mu \nu} + 
  \ds{\frac{(P-p)^\mu n_-^\nu+(P-p)^\nu n_-^\mu}{(P-p)\cdot n_-}}  
   - \ds{\frac{M_a^2}{[(P-p)\cdot n_-]^2}}\,n_-^\mu\,n_-^\nu 
  & \quad \mbox{(see Ref.~\cite{Brodsky:2000ii})}, \\[0.4cm]
  -g^{\mu \nu} + \ds{\frac{(P-p)^\mu\,(P-p)^\nu}{M_a^2}} 
  & \quad \mbox{(see Ref.~\cite{Gamberg:2007wm}),} \\[0.4cm]
  -g^{\mu \nu} + \ds{\frac{P^\mu\,P^\nu}{M_a^2}} 
  & \quad \mbox{(see Ref.~\cite{Jakob:1997wg}),}   \\[0.4cm]
  -g^{\mu \nu} 
  & \quad \mbox{(see Ref.~\cite{Bacchetta:2003rz}).}
  \end{array}  \right. 
\label{eq:lc} 
\end{equation}

The different forms for the diquark propagator correspond to different
physical theories and lead to different results for the 
parton distribution functions. We have analyzed all of them except 
for the third 
one, which was extensively studied already in
Ref.~\cite{Jakob:1997wg}. However, 
we think that the first one is preferable to the others. 
The motivation is that in the 
spectator model we have to take into account that 
the diquarks have an electric 
charge and can couple to the virtual photon in DIS.  
In other words, in this model 
the quarks are not the only charged partons in the proton: the 
diquarks are also charged partons and they have spin different from $\half$. 
The scalar diquark couples only 
to longitudinally polarized photons and gives contribution to the structure 
function $F_L$. This leads to a violation of the Callan--Gross relation, but 
leaves unchanged the (leading-order) interpretation of the structure function 
$F_{T}$ as a charge-weighted sum of quark distribution functions. 
This seems the best way to reduce the phenomenological impact of the problem
represented by the presence of the diquarks. 
For the vector diquark, 
we checked that the same situation occurs when only the light-cone transverse
polarization states of the diquark are propagated, i.e., when the first  
choice of Eq.~(\ref{eq:lc}) for the polarization sum is used. 
In the other cases, the diquark would give a contribution also to the 
structure function $F_{T}$. 
On top of this, we remark that the last choice of Eq.~\eqref{eq:lc} for the
polarization sum introduces unphysical polarization states of the vector
diquark (see discussion in next section).
In conclusion, in the following we shall consider only light-cone transverse
polarizations of the diquark, make only a few comments on the other choices,
 and 
leave the complete list of results in the 
Appendices.

\begin{figure}[h]
\begin{center}
\includegraphics[width=6cm]{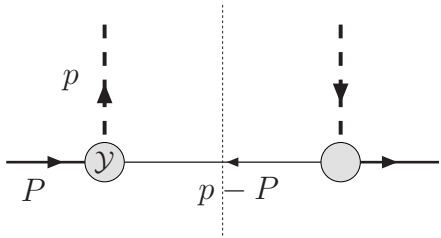}
\end{center}
\caption{Tree-level cut diagram for the calculation of T-even 
leading-twist parton densities for an active scalar or vector diquark
(dashed line), with a spectator quark (solid line).}
\label{fig:DISdiquark}
\end{figure}

Equation~(\ref{eq:m-tree}) can be further elaborated by choosing the 
nucleon-quark-diquark vertex ${\cal Y}$. We choose the scalar and vector 
vertices to be 
\begin{align} 
{\cal Y}_s &= i g_s(p^2)\, \uno \; , &{\cal Y}^{\mu}_a &= 
i \frac{g_a(p^2)}{\sqrt{2}}\,\g^\mu \, \g_5\, ,
\label{eq:tree-vertices}
\end{align} 
where $g_X(p^2)$ is a suitable form factor. Other choices are possible (see,
e.g., Refs~\cite{Jakob:1997wg,Gamberg:2007wm}), but we
limit ourselves to these ones, which are the simplest. 
For the form factor, we explored three possible choices: 
\begin{equation} 
g_X(p^2) = \left\{ 
   \begin{array}{ccl}
   g_X^{p.l.} & \quad &\text{point-like,} \\[0.2cm]
   g_X^{dip}\, \ds{\frac{p^2-m^2}{{|p^2-\Lambda_X^2|}^2}} &\quad 
     &\text{dipolar,} \\[0.4cm]
   g_X^{exp}\, e^{\,(p^2-m^2)/\Lambda_X^2} &\quad &\text{exponential,} 
   \end{array}  \right. 
\label{eq:NqDff}
\end{equation}
where $g_X$ and $\Lambda_X$ are appropriate coupling constants and cutoffs, 
respectively, to be considered as free parameters of the model together with 
the mass of the diquark $M_X$. All these parameters can in principle be 
different for each type of diquark. Only the point-like coupling can be derived 
from a specific Lagrangian with protons, quarks and diquarks as fundamental 
degrees of freedom, and meant to effectively describe QCD in the 
nonperturbative regime. Since our interest here is mainly phenomenological, we 
prefer to introduce form factors.  
They smoothly suppress the influence of high ${\bm p}_{\sT}$ ---
where our theory cannot be trusted --- and eliminate the logarithmic
divergences arising after  
${\bm p}_{\sT}$ integration when using a point-like coupling.
For later use, we note that 
the dipolar form factor can be usefully rewritten, using 
Eq.~(\ref{eq:offshell}), as
\begin{equation} 
g_X(p^2) = g_X^{dip}\,\frac{p^2-m^2}{{|p^2-\Lambda_X^2|}^2} = g_X^{dip}\,
\frac{(p^2-m^2)\,(1-x)^2}{\left( {\bm p}_{\sT}^2+L_X^2(\Lambda_X^2)\right)^2} 
\; . 
\label{eq:ffdip}
\end{equation}

In summary, we have analyzed in total nine combinations of 
nucleon-quark-diquark form factors and forms for the diquark propagator. 
As mentioned above, we will discuss analytical and 
numerical results involving the dipolar form factor and the first
choice of Eq.~(\ref{eq:lc}) (transverse diquark polarizations only), 
listing the formulae for the other cases in the 
Appendices~\ref{sec:A} and \ref{sec:B}. To keep the notation lighter, we will
denote the coupling $g_X^{dip}$ simply as $g_X$ from now on.


\subsection{Light-cone wave functions}
\label{sec:LCwf}

A convenient way to compute parton distribution functions is by making use of
light-cone wave functions (LCWFs), as done for instance in 
Ref.~\cite{Brodsky:2000ii}. For the scalar diquark, LCWFs can be defined as
\begin{equation} 
\psi^{\lambda_N}_{\lambda_q}  (x,\bm{p}_{\sT}) = 
   \sqrt{\frac{p^+}{(P-p)^+}}\,\frac{\bar{u}(p,\lambda_q)}{p^2 -m^2}\, 
   {\cal Y}_s \, U(P,\lambda_N) \; ,
\label{eq:lcwf-s}
\end{equation}
where the indices $\lambda_N, \, \lambda_q,$ refer to the helicity of the 
nucleon and of the quark, respectively, and are constrained by angular momentum 
conservation to the ``spin sum rule'' $\lambda_N = \lambda_q + L_z$, where 
$L_z$ is the projection of the relative orbital angular momentum 
between the quark and the diquark. We use the conventions of 
Ref.~\cite{Lepage:1980fj} (see also Ref.~\cite{Meissner:2007rx}). In standard 
representation, the spinors can be written as
\begin{align}
u(p,+)&=\frac{1}{\sqrt{2^{3/2}\,p^+}}\,
  \begin{pmatrix}
   \sqrt{2}\,p^++m\\
   \quad\ p_x+i p_y\quad\ \\
   \sqrt{2}\,p^+-m\\
   p_x+i p_y
  \end{pmatrix} \; , 
&u(p,-) &=\frac{1}{\sqrt{2^{3/2}\,p^+}}\,
  \begin{pmatrix}
   -p_x+i p_y\\
   \sqrt{2}\,p^++m\\
   \quad\ p_x-i p_y\quad\ \\
   -\sqrt{2}\,p^++m 
  \end{pmatrix} \; ,
\label{eq:qspinors}
\end{align}
and similarly for the nucleon spinors (changing $p,m$, to $P,M$, respectively). 
We obtain
\begin{align} 
\psi^{+}_{+} (x,\bm{p}_{\sT}) &= (m+ x M)\,\phi/x & &(L_z=0), \\
\psi^{+}_{-} (x,\bm{p}_{\sT}) &= -(p_x+ i p_y)\,\phi/x & &(L_z=+1), \\
\psi^{-}_{+} (x,\bm{p}_{\sT}) &= -\left[\psi^{+}_{-} (x,\bm{p}_{\sT})\right]^* 
& &(L_z=-1), \\
\psi^{-}_{-} (x,\bm{p}_{\sT}) &= \psi^{+}_{+} (x,\bm{p}_{\sT}) & &(L_z=0), \\
\phi(x,\bm{p}_{\sT}^2) &= -\frac{g_s}{\sqrt{1-x}}\, 
    \frac{x (1-x)}{\bm{p}_{\sT}^2+L_s^2(m^2)} \; ,
\label{eq:lcwf-s2}
\end{align}
which correspond to Eqs.~(44,46) of Ref.~\cite{Brodsky:2000ii}. 

For the vector diquark, LCWFs can be defined as
\begin{equation} 
\psi^{\lambda_N}_{\lambda_q \lambda_a}  (x,\bm{p}_{\sT}) =  
  \sqrt{\frac{p^+}{(P-p)^+}}\,\frac{\bar{u}(p,\lambda_q)}{p^2-m^2}\,
  \varepsilon^*_{\mu}(P-p,\lambda_a)\,{\cal Y}^{\mu}_a \, U(P,\lambda_N) \; ,
\label{eq:lcwf-a}
\end{equation}
where the index $\lambda_a$ refers to the helicity of the vector 
diquark and is constrained by $\lambda_N = \lambda_q + \lambda_a + L_z$. 
The light-cone transverse 
polarization vectors are given by~\cite{Brodsky:2000ii}
\begin{align}
\varepsilon (P-p,+) & = \biggl[-\frac{(P-p)_x + i (P-p)_y}{\sqrt{2}\,(P-p)^+}, 
        0,-\frac{1}{\sqrt{2}},-\frac{i}{\sqrt{2}}\biggr] = 
        \biggl[\frac{p_x + i p_y}{\sqrt{2}\,(1-x) \,P^+} , 0 ,
            -\frac{1}{\sqrt{2}},-\frac{i}{\sqrt{2}}  \biggr] \; ,  \\
\varepsilon (P-p,-) & = \biggl[-\frac{p_x - i p_y}{\sqrt{2}\,(1-x)\,P^+} , 
        0,\frac{1}{\sqrt{2}},-\frac{i}{\sqrt{2}}\biggr] \; . 
\label{eq:polvect}
\end{align} 
They satisfy the usual properties\footnote{Note 
that $(\bm{P}-\bm{p}) \cdot \bm{\varepsilon}(\pm) \neq 0$,
since $\varepsilon(\pm)$ do not describe transverse polarization
with respect to the diquark momentum.}  
$\varepsilon(\pm) \cdot \varepsilon^*(\pm)=-1$, 
$\varepsilon(\pm) \cdot \varepsilon^*(\mp)=0$, and 
$(P-p) \cdot \varepsilon(\pm) =0$.
They are consistent with the 
polarization sum being expressed by the first option in 
Eq.~(\ref{eq:lc}). The LCWFs become
\begin{align} 
\psi^{+}_{++} (x,\bm{p}_{\sT}) &= \frac{p_x- i p_y}{1-x}\,\phi/x  & &(L_z=-1),
\label{eq:lz-1} \\
\psi^{+}_{+-} (x,\bm{p}_{\sT}) &= -x \, \frac{p_x+ i p_y}{1-x}\,\phi/x  & &(L_z=+1), 
\label{eq:lz1} \\
\psi^{+}_{-+} (x,\bm{p}_{\sT}) &= (m+ x M)\,\phi/x & &(L_z=0), \\
\psi^{+}_{--} (x,\bm{p}_{\sT}) &= 0 & &(L_z=+2), \\
\psi^{-}_{++} (x,\bm{p}_{\sT}) &= 0 & &(L_z=-2), \\
\psi^{-}_{+-} (x,\bm{p}_{\sT}) &= -\psi^{+}_{-+} (x,\bm{p}_{\sT}) & &(L_z=0), \\
\psi^{-}_{-+} (x,\bm{p}_{\sT}) &= \left[\psi^{+}_{+-} (x,\bm{p}_{\sT})\right]^*
& &(L_z=-1), \\
\psi^{-}_{--} (x,\bm{p}_{\sT}) &= \left[\psi^{+}_{++} (x,\bm{p}_{\sT})\right]^* 
& &(L_z=+1), \\
\phi(x,\bm{p}_{\sT}^2) &= -\frac{g_a}{\sqrt{1-x}}\, 
     \frac{x (1-x)}{\bm{p}_{\sT}^2+L_a^2(m^2)} \; ,
\label{eq:lcwf-a2}
\end{align}
and are analogous to Eqs.~(21,24) in Ref.~\cite{Brodsky:2000ii}, the
differences being due to the fact that here the diquark is an axial-vector 
particle rather than a vector one.
Note that in our model we can only have wavefunctions with at most one
  unit of orbital angular momentum ($p$ wave). The LCWFs with two units of
  orbital angular momentum 
 ($d$-wave), $\psi^+_{--}$ and
  $\psi^-_{++}$, vanish.

If we add to $\varepsilon(P-p,\pm)$ also the third longitudinal 
polarization vector 
\begin{equation} 
\varepsilon(P-p,0) = \frac{1}{M_a} \biggl[
   \frac{\bm{p}_{\sT}^2-M_a^2}{2\,(1-x)\,P^+},(1-x)\,P^+,-p_x,-p_y \biggr] \; ,
\label{eq:polvect0}
\end{equation}
satisfying\footnote{Note that $\bm{\varepsilon}(0)$ is not 
parallel to $(\bm{P}-\bm{p})$ because it describes longitudinal polarization 
states in the light-cone.} 
$\varepsilon(0) \cdot \varepsilon^{*}(0)=-1$, 
$\varepsilon(0) \cdot \varepsilon^{*}(\pm)=0$, and 
$(P-p) \cdot \varepsilon(0) =0$, 
the corresponding additional LCWFs are
\begin{align} 
\psi^{+}_{+0} (x,\bm{p}_{\sT}) &= 
   \frac{\bm{p}_{\sT}^2 - x M_a^2 - mM\,(1-x)^2}{\sqrt{2}\,(1-x)\,M_a}\,\phi/x 
   & &(L_z=0), \\
\psi^{+}_{-0} (x,\bm{p}_{\sT}) &= \frac{(m+M)\,(p_x+ip_y)}{\sqrt{2}\,M_a}\,
   \phi/x & &(L_z=+1), \\
\psi^{-}_{+0} (x,\bm{p}_{\sT}) &= \left[\psi^{+}_{-0} (x,\bm{p}_{\sT})\right]^*
   \,\phi/x & &(L_z=-1), \\
\psi^{-}_{-0} (x,\bm{p}_{\sT}) &= -\psi^{+}_{+0} (x,\bm{p}_{\sT})\,\phi/x \;  & &(L_z=0).
\label{eq:lcwf-a3}
\end{align} 

From the above combinations we deduce, for 
example, that the proton with positive helicity $+\half$ can be in a state with 
probability density proportional to $|\psi^{+}_{-}|^2$, where the quark has 
opposite helicity and $L_z=+1$ orbital angular momentum with respect to a 
scalar diquark. This configuration is relativistically enhanced with respect to 
$|\psi^{+}_{+}|^2$ with $L_z=0$, where proton and quark helicities are aligned; 
thus, it suggests a possible explanation of the proton ``spin puzzle'' in terms 
of the relativistic aspects of the motion of quarks inside 
hadrons~\cite{Brodsky:2000ii}. 

For the purpose of this work, it is also important to note that a nonvanishing
relative orbital angular momentum between the quark and the diquark implies that
the partons do not necessarily occupy the lowest-energy available orbital (with 
quantum numbers $J^P={\half}^+$ and $L_z=0$). 
Hence, in this version of the
spectator diquark model the nucleon wave function does not show a 
SU(4)=SU(2)$\otimes$SU(2) spin-isospin symmetry, contrary to what is usually 
assumed~\cite{Jakob:1997wg}.

Finally, we mention that the completeness relation for the last
choice of the polarization sum in Eq.~(\ref{eq:lc}) should be written
\begin{equation}
\sum_{\lambda_a=\pm,0} \varepsilon^{\ast \mu}(P-p,\lambda_a) \, 
\varepsilon^\nu (P-p,\lambda_a) - \varepsilon^{\ast \mu}(P-p,t) \, 
\varepsilon^\nu(P-p,t) = - g^{\mu \nu} ,
\end{equation} 
where the unphysical time-like polarization state
$\varepsilon^{\mu}(P-p,t) = (P-p)^{\mu}/M_a$ appears. The associated LCWFs read
\begin{align} 
\psi^{+}_{+t} (x,\bm{p}_{\sT}) &= 
   \frac{\bm{p}_{\sT}^2 + x M_a^2 - mM\,(1-x)^2}{\sqrt{2}\,(1+x)\,M_a}\,\phi/x 
   & &(L_z=0), \\
\psi^{+}_{-t} (x,\bm{p}_{\sT}) &= \frac{(m+M)\,(p_x+ip_y)}{\sqrt{2}\,M_a}\,
   \phi/x & &(L_z=+1), \\
\psi^{-}_{+t} (x,\bm{p}_{\sT}) &= \left[\psi^{+}_{-t} (x,\bm{p}_{\sT})\right]^*
   \,\phi/x & &(L_z=-1), \\
\psi^{-}_{-t} (x,\bm{p}_{\sT}) &= -\psi^{+}_{+t} (x,\bm{p}_{\sT})\,\phi/x  & &(L_z=0).
\label{eq:lcwf-at}
\end{align}


\subsection{T-even functions}
\label{sec:teven}

The simplest example of T-even parton density 
is the unpolarized quark distribution
$f_1(x,{\bm p}_{\sT})$, defined as
\begin{equation} \begin{split}  
f_1(x,{\bm p}_{\sT}) &= \frac{1}{4}\,\mathrm{Tr}\left[ \left( 
\Phi(x,{\bm p}_{\sT},S) + \Phi(x,{\bm p}_{\sT},-S) \right) \, \g^+ \right] 
+ \mathrm{h.c.}  \\
&= \frac{1}{4}\,\frac{1}{(2\pi)^3}\,\frac{1}{2(1-x)P^+}\, \mathrm{Tr}\left[ 
\left( \overline{\mathcal{M}}^{(0)}(S)\, \mathcal{M}^{(0)}(S) + 
\overline{\mathcal{M}}^{(0)}(-S)\, \mathcal{M}^{(0)}(-S) \right) \, \g^+ 
\right] + \mathrm{h.c.} \; .
\label{eq:f1}
\end{split} \end{equation} 
By inserting in $\mathcal{M}^{(0)}$ of Eq.~(\ref{eq:m-tree}) the 
rules~(\ref{eq:tree-vertices}) for the nucleon-quark-diquark vertex, the 
dipolar form factor of Eq.~(\ref{eq:ffdip}), and the first choice in 
Eq.~(\ref{eq:lc}) for the sum of the polarization states of the diquark
(transverse polarizations only), we get 
\begin{align} 
f_1^{q(s)}(x,{\bm p}_{\sT}) &= \frac{g_s^2}{(2\pi)^3}\, 
\frac{[(m+xM)^2+{\bm p}_{\sT}^2]\,(1-x)^3}
     {2\,[{\bm p}_{\sT}^2+L_s^2(\Lambda_s^2)]^4}  
\label{eq:f1sspect}
\\
f_1^{q(a)} (x,{\bm p}_{\sT}) &= \frac{g_a^2}{(2\pi)^3}\, 
\frac{[{\bm p}_{\sT}^2\,(1+x^2)+(m+xM)^2\,(1-x)^2]\,(1-x)}
     {2\,[{\bm p}_{\sT}^2 + L_a^2(\Lambda_a^2)]^4} \; . 
\label{eq:f1aspect}
\end{align} 

The same result can be recovered through the alternative definition
\begin{align}  
f_1^{q(s)}(x,\bm{p}_{\sT}^2) &= \frac{1}{16 \pi^3}\,\frac{1}{2} 
   \sum_{\lambda_N=\pm}\,\sum_{\lambda_q=\pm} \, 
   |\psi^{\lambda_N}_{\lambda_q}|^2 
  = \frac{1}{16 \pi^3} \Bigl(|\psi^{+}_{+}|^2 + |\psi^{+}_{-}|^2 \Bigr)  \\
f_1^{q(a)}(x,\bm{p}_{\sT}^2) &= \frac{1}{16 \pi^3}\,\frac{1}{2} 
   \sum_{\lambda_N=\pm} \, \sum_{\lambda_q=\pm}\,\sum_{\lambda_a=\pm} \, 
   |\psi^{\lambda_N}_{\lambda_q \lambda_a}|^2 
  = \frac{1}{16 \pi^3} \Bigl(|\psi^{+}_{++}|^2 + |\psi^{+}_{+-}|^2 + 
   |\psi^{+}_{-+}|^2 + |\psi^{+}_{--}|^2 \Bigr)  \; , 
\label{eq:f1lcwf}
\end{align} 
and replacing the results for the LCWFs using Eqs.~(\ref{eq:lcwf-s2}) and 
(\ref{eq:lcwf-a2}) for the scalar and vector diquark, respectively. 

If we use, instead, the second option of Eq.~(\ref{eq:lc}) 
for the sum over polarizations of
the vector diquark (transverse and longitudinal polarizations), we obtain
\begin{equation} 
f_1^{q(a)}(x,\bm{p}_{\sT})+\frac{1}{16 \pi^3} \Bigl(|\psi^{+}_{+0}|^2 + |\psi^{+}_{-0}|^2 \Bigr)  \; . 
\label{eq:f1lcwf2}
\end{equation}
The complete expression is given in Eq.~(\ref{eq:pTTeven-a-pl-covar}) and
corresponds to Eq.~(10) of Ref.~\cite{Gamberg:2007wm} with $R_g=0$.

Finally, the results with the last choice of Eq.~(\ref{eq:lc}) (transverse, 
longitudinal, and time-like polarizations) can be written as
\begin{equation}
f_1^{q(a)}(x,\bm{p}_{\sT})+\frac{1}{16 \pi^3} \Bigl(|\psi^{+}_{+0}|^2 +
|\psi^{+}_{-0}|^2 \Bigr)  
-\frac{1}{16 \pi^3} \Bigl(|\psi^{+}_{+t}|^2 + |\psi^{+}_{-t}|^2 \Bigr)  \; . 
\end{equation} 
Note that the contribution of the diquark 
time-like polarization states enters with an overall
negative sign. The complete expression is given in
Eq.~(\ref{eq:pTTeven-a-pl-feyn})  and
corresponds to Eq.~(8) of Ref.~\cite{Bacchetta:2003rz}.

Turning back to our preferred choice, 
i.e.\  the first option of Eq.~(\ref{eq:lc}) (light-cone transverse
polarizations only), we now compute all
other T-even, leading-twist TMDs. Their definition in terms of traces of the
quark-quark correlator can be derived from, e.g., Eqs.~(3.19) and ff.\ in
Ref.~\cite{Bacchetta:2006tn}. To write them in terms of LCWFs, we need to
introduce the polarization state in a generic direction
 $\hat{\bm{S}}_{\sT} = (\cos\phi_S, \sin\phi_S)$ in the transverse plane  
\begin{align} 
U(P,\uparrow) &= \frac{1}{\sqrt{2}}\, \left(U(P,+) + e^{i \phi_S}U(P,-)\right) \; ,
\label{eq:Tup}
\\
U(P,\downarrow) &= \frac{1}{\sqrt{2}}\, \left(U(P,+) +
  e^{i(\phi_S+\pi)}U(P,-)\right) \; . 
\label{eq:Tdown}
\end{align} 
For $\phi_S=0, \pi/2,$ we recover the (positive) polarizations along the 
$\hat{x}$ and $\hat{y}$ axis, respectively~\cite{Barone:2003fy}. 
For the quark, we will use similar decompositions and use the notation
$\hat{\bm{S}}_{q\sT}$ and $\phi_{S_q}$, i.e.,
\begin{align} 
\bar{u}(p,\uparrow) &= \frac{1}{\sqrt{2}}\, \left(\bar{u}(p,+) + e^{-i \phi_{S_q}}\bar{u}(p,-)\right) \; ,
\label{eq:Tupq}
\\
\bar{u}(p,\downarrow) &= \frac{1}{\sqrt{2}}\, 
    \left(\bar{u}(p,+) + e^{-i(\phi_{S_q}+\pi)}\bar{u}(p,-)\right) \; . 
\label{eq:Tdownq}
\end{align} 

With this conventions and 
keeping in mind that $\lambda_X$ is absent 
for the scalar
diquark and $\lambda_X=\pm $ for the vector diquark, 
we can write the TMDs in the following way
\begin{align}
g_{1L}(x,{\bm p}_{\sT}) &= \frac{1}{16 \pi^3} \, 
   \sum_{\lambda_X}\, \Bigl(|\psi^{+}_{+\lambda_X}|^2 - 
                                |\psi^{+}_{-\lambda_X}|^2 \Bigr) \; ,
\\
\frac{{\bm p}_{\sT}\cdot \hat{\bm{S}}_{\sT}}{M}\,g_{1T} (x,{\bm p}_{\sT}) &=
  \frac{1}{16\pi^3}\, \sum_{\lambda_X}\, 
  \Bigl(|\psi^{\uparrow}_{+\lambda_X}|^2 - 
        |\psi^{\uparrow}_{-\lambda_X}|^2 \Bigr) 
\\
\frac{{\bm p}_{\sT}\cdot \hat{\bm{S}}_{q\sT}}{M}\,h_{1L}^{\perp}(x,{\bm p}_{\sT}) &= 
\frac{1}{16\pi^3}\, 
   \sum_{\lambda_X}\, \Bigl(|\psi^{+}_{\uparrow\lambda_X}|^2 - 
        |\psi^{+}_{\downarrow\lambda_X}|^2 \Bigr) 
\; ,  
\\
\hat{\bm{S}}_{\sT}\cdot\hat{\bm{S}}_{q\sT}\;
h_{1T}(x,{\bm p}_{\sT}) &+ 
\frac{{\bm p}_{\sT}\cdot \hat{\bm{S}}_{\sT}}{M}\, 
\frac{\bm{p}_{\sT}\cdot\hat{\bm{S}}_{q\sT}}{M}\,
h_{1T}^{\perp}(x,{\bm p}_{\sT}) = \frac{1}{16\pi^3}\, 
   \sum_{\lambda_X}\, \Bigl(
         |\psi^{\uparrow}_{\uparrow \lambda_X}|^2 - 
         |\psi^{\uparrow}_{\downarrow \lambda_X}|^2 \Bigr) \; .
\end{align} 
The above results automatically fulfill positivity
bounds~\cite{Bacchetta:1999kz}.

The explicit expressions are
\begin{align} 
g_{1L}^{q(s)}(x,{\bm p}_{\sT}^2) &= 
\frac{g_s^2}{(2\pi)^3}\, 
\frac{[(m+xM)^2-{\bm p}_{\sT}^2]\,(1-x)^3}{2\,[{\bm p}_{\sT}^2+L_s^2(\Lambda_s^2)]^4} \; , 
\label{eq:g1Ls} 
\\
g_{1L}^{q(a)}(x,{\bm p}_{\sT}^2) &= \frac{g_a^2}{(2\pi)^3}\, 
\frac{[{\bm p}_{\sT}^2\,(1+x^2)-(m+xM)^2\,(1-x)^2]\,(1-x)}{2\,[{\bm p}_{\sT}^2+
L_a^2(\Lambda_a^2)]^4} \; , \label{eq:g1La} 
\\ \nn \\
g_{1T}^{q(s)} (x,{\bm p}_{\sT}^2) &= 
\frac{g_s^2}{(2\pi)^3}
\, \frac{M\,(m+xM)\,(1-x)^3}{[{\bm p}_{\sT}^2 + L_s^2(\Lambda_s^2)]^4} \; ,  
\\
g_{1T}^{q(a)} (x,{\bm p}_{\sT}^2) &= 
\frac{g_a^2}{(2\pi)^3}
\, \frac{xM\,(m+xM)\,(1-x)^2}{[{\bm p}_{\sT}^2+L_a^2(\Lambda_a^2)]^4} \; , 
\label{eq:g1Tspect} 
\\ \nn \\
h_{1L}^{\perp\,q(s)}(x,{\bm p}_{\sT}^2) &= 
-\frac{g_s^2}{(2\pi)^3}\, 
\frac{M\,(m+xM)\,(1-x)^3}{[{\bm p}_{\sT}^2 + L_s^2(\Lambda_s^2)]^4} \; , 
\\
h_{1L}^{\perp\,q(a)}(x,{\bm p}_{\sT}^2) &= 
\frac{g_a^2}{(2\pi)^3}\, 
\frac{M\,(m+xM)\,(1-x)^2}{[{\bm p}_{\sT}^2+L_a^2(\Lambda_a^2)]^4} \; , 
\label{eq:h1Lperpspect} 
\\ \nn \\
h_{1T}^{q(s)}(x,{\bm p}_{\sT}^2) &=
\frac{g_s^2}{(2\pi)^3}\, 
\frac{[{\bm p}_{\sT}^2 + (m+xM)^2]\,(1-x)^3}{2\,[{\bm p}_{\sT}^2+L_s^2(\Lambda_s^2)]^4} \:, 
\\
h_{1T}^{q(a)}(x,{\bm p}_{\sT}^2) &=
-\frac{g_a^2}{(2\pi)^3}\, 
\frac{{\bm p}_{\sT}^2\,x(1-x)}{[{\bm p}_{\sT}^2+L_a^2(\Lambda_a^2)]^4} \; ,
\label{eq:h1Tspect} 
\\ \nn \\
h_{1T}^{\perp\,q(s)}(x,{\bm p}_{\sT}^2)  &= 
- \frac{g_s^2}{(2\pi)^3}\, 
\frac{M^2\,(1-x)^3}{[{\bm p}_{\sT}^2+L_s^2(\Lambda_s^2)]^4} \; , 
\\
h_{1T}^{\perp\,q(a)}(x,{\bm p}_{\sT}^2) &= 
0 \;.
\label{eq:h1Tperpspect}
\end{align} 
From the last two formulae we deduce also the expressions for
the transversity distribution:
\begin{align} 
h_1^{q(s)}(x,{\bm p}_{\sT}^2) &= h_{1T}^{q(s)}(x,{\bm p}_{\sT}^2) + 
\frac{{\bm p}_{\sT}^2}{2M^2}\, h_{1T}^{\perp\,q(s)}(x,{\bm p}_{\sT}^2) = 
\frac{g_s^2}{(2\pi)^3}\,
\frac{(m+xM)^2\,(1-x)^3}{2\,[{\bm p}_{\sT}^2+L_s^2(\Lambda_s^2)]^4} 
\label{eq:h1spects} 
\\
h_1^{q(a)}(x,{\bm p}_{\sT}^2) &= -\frac{g_a^2}{(2\pi)^3}\,
\frac{{\bm p}_{\sT}^2\,x(1-x)}{[{\bm p}_{\sT}^2+L_a^2(\Lambda_a^2)]^4} \; .
\label{eq:h1spect}
\end{align} 
Note that the functions $g_{1T}$ and $h_{1L}^{\perp}$ arise from
  the interference of LCWFs with $|L_z|=1$ and $L_z=0$. The function
  $h_{1T}^{\perp}$ requires the interference of two LCWFs that differ by two
  units of $L_z$. This
  condition is necessary but not sufficient to have $h_{1T}^{\perp}\neq
  0$. In fact, the vector diquark spectator gives $h_{1T}^{\perp}=0$ even if LCWFs 
with $L_z=\pm 1$ are present. 

Some interesting relations can be evinced from the above expressions. For
example, the transversity with scalar diquark 
saturates the Soffer bound, while for axial-vector diquarks the relation is 
more involved:
\begin{align} 
h_1^{q(s)}(x,{\bm p}_{\sT}^2) &=\frac{1}{2} \, \left( f_1^{q(s)}(x,{\bm p}_{\sT}^2) + g_1^{q(s)}(x,{\bm p}_{\sT}^2) \right) \; ,  \\
h_1^{q(a)}(x,{\bm p}_{\sT}^2) &=-\frac{x}{1+x^2}\, \frac{1}{2} \, \left( f_1^{q(a)}(x,{\bm p}_{\sT}^2) + g_1^{q(a)}(x,{\bm p}_{\sT}^2) 
\right) \; . 
\label{eq:Tevenrelations}
\end{align} 
When restricting to the results with scalar diquark, the $g_{1T}$ 
distribution is connected to two other partners by the relations
\begin{align} 
g_{1T}^{q(s)}(x,{\bm p}_{\sT}^2 ) &= - h_{1L}^{\perp\, q(s)}(x,{\bm p}_{\sT}^2 ) \; , 
&
g_{1T}^{q(s)}(x,{\bm p}_{\sT}^2 ) &= \frac{2 M}{m+xM}\, h_1^{q(s)}(x,{\bm p}_{\sT}^2 ) \; , 
\label{eq:Tevenrelations2}
\end{align} 
while for axial-vector diquarks we have
\begin{equation} 
g_{1T}^{q(a)}(x,{\bm p}_{\sT}^2 ) = x\, h_{1L}^{\perp\, q(a)}(x,{\bm p}_{\sT}^2 ) \; .
\label{eq:Tevenrelations3}
\end{equation}
This relation is however different when considering spectator diquarks with
more degrees of freedom (see App.~\ref{sec:A}). Our results seem to indicate 
that no general
relation exists between $g_{1T}$ and $h_{1L}^{\perp}$, contrary to what is
proposed in Ref.~\cite{Pasquini:2008ax}. The reason is connected to the
difference between LCWFs with $L_z=1$ and $L_z=-1$, as in Eqs.~\eqref{eq:lz-1} and
\eqref{eq:lz1}. 
We also observe that in the vector-diquark case
$g_{1L} - h_1$ and $h_{1T}^{\perp}$ are not simply related through the relation 
suggested in
Ref.~\cite{Avakian:2008dz}.  We are
led to conclude that such a relation is not general.

The ${\bm p}_{\sT}$-integrated results are
\begin{align} 
f_1^{q(s)} (x) &=\frac{g_s^2}{(2\pi)^2}\,
\frac{[2\,(m+xM)^2 + L_s^2(\Lambda_s^2) ]\,(1-x)^3}{24\,L_s^6(\Lambda_s^2)}  
\label{eq:f1sdip_int}
\\
f_1^{q(a)} (x) &=\frac{g_a^2}{(2\pi)^2}\,
\frac{[2\,(m+xM)^2\,(1-x)^2+(1+x^2)\,L_a^2(\Lambda_a^2) ]\,(1-x)}
     {24\,L_a^6(\Lambda_a^2)}  
\label{eq:f1adip_int}
\\ \nn \\
g_1^{q(s)}(x) &=\frac{g_s^2}{(2\pi)^2}\,
\frac{[2\,(m+xM)^2-L_s^2(\Lambda_s^2)]\,(1-x)^3}{24\,L_s^6(\Lambda_s^2)}  
\label{eq:g1sdip_int}
\\
g_1^{q(a)}(x) &=-\frac{g_a^2}{(2\pi)^2}\, 
\frac{[2\,(m+xM)^2\,(1-x)^2-(1+x^2)\,L_a^2(\Lambda_a^2)]\,(1-x)}
     {24\,L_a^6(\Lambda_a^2)}  
\label{eq:g1adip_int}
\\ \nn \\
h_1^{q(s)}(x) &=\frac{g_s^2}{(2\pi)^2}\, 
\frac{(m+xM)^2\,(1-x)^3}{12\,L_s^6(\Lambda_s^2)}  
\\
h_1^{q(a)}(x) &=-\frac{g_a^2}{(2\pi)^2}\, \frac{x(1-x)}{12\,L_a^4(\Lambda_a^2)} 
\; . \label{eq:h1xspecta}
\end{align}


\subsection{T-odd functions}
\label{sec:todd}

\begin{figure}[h]
\begin{center}
\includegraphics[width=6cm]{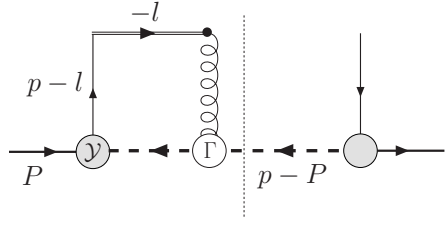}
\end{center}
\caption{Interference between the one-gluon exchange diagram in eikonal
  approximation and the tree level diagram in the spectator model. The
  Hermitean conjugate diagram is not shown.}
\label{fig:eikonal}
\end{figure}

The two leading-twist T-odd structures are the Sivers and Boer-Mulders
distributions. They are defined as
\begin{align} 
\frac{\varepsilon_{\sT}^{ij}p_{\sT i}S_{\sT j}}{M}\,
f_{1T}^{\perp}(x,{\bm p}_{\sT}^2) &= -\frac{1}{4}\, \mathrm{Tr} \left[ \left( 
\Phi (x,{\bm p}_{\sT},S) -\Phi (x,{\bm p}_{\sT},-S)\right) \, \g^+ \right] + 
\mathrm{h.c.} \; , \label{eq:Sivers} \\
\frac{\varepsilon_{\sT}^{ij}p_{\sT j}}{M}\,
h_1^{\perp}(x,{\bm p}_{\sT}^2) &= \frac{1}{4}\, \mathrm{Tr} \left[ \left( 
\Phi (x,{\bm p}_{\sT},S) +\Phi (x,{\bm p}_{\sT},-S)\right) \,
i\sigma^{i+}\,\g_5 \right] + \mathrm{h.c.} \; . \label{eq:Boer-Mulders} 
\end{align}
At tree level, these expressions vanish because 
there is no residual interaction between the active quark and the spectators; 
equivalently, there is no interference between two competing channels producing 
the complex amplitude whose imaginary part gives the T-odd contribution. We can 
generate such structures by considering the interference between the tree-level 
scattering amplitude and the single-gluon-exchange scattering amplitude in 
eikonal approximation, as shown in Fig.~\ref{fig:eikonal} (the Hermitean 
conjugate partner must also be considered). This corresponds just to the 
leading-twist one-gluon-exchange approximation of the gauge link operator of 
Eq.~(\ref{eq:link})~\cite{Boer:2003cm}.

For the moment, we use Abelian gluons.  The QCD color 
structure will be recovered at the end. 
The Feynman rules to be used for the eikonal vertex and propagator 
are~\cite{Collins:1981uk,Collins:1982uw}
\begin{align}
\raisebox{-6mm}{\includegraphics[height=1.2cm]{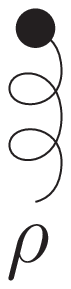}}
  &=  - i e_c \,n_-^{\rho} \,,
&
\includegraphics[width=1.6cm]{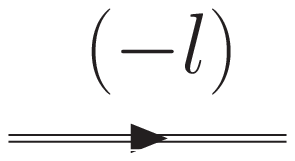}
  &= \frac{i}{-l^+ +i \epsilon} \,,
\end{align} 
where $e_c$ is the color charge of the quark and
the sign of $i \epsilon$ for the eikonal line corresponds to the 
gauge link of SIDIS.  In cut diagrams one must take
the complex conjugate of these expressions for vertices and
propagators on the right of the final-state cut.

The explicit form of the contribution $\Phi^{(1)}$ to the correlation function
corresponding to Fig.~\ref{fig:eikonal} is 
\begin{equation} 
\Phi^{(1)}(x,{\bm p}_{\sT},S) \sim \frac{1}{(2\pi)^3}\,\frac{1}{2(1-x)P^+}\, 
\Bigl(\overline{\mathcal{M}}^{(0)}(S)\, \mathcal{M}^{(1)}(S)+  
\overline{\mathcal{M}}^{(1)}(S)\, \mathcal{M}^{(0)}(S) \Bigr)
\Big\vert_{p^2=\tau (x,{\bm p}_{_T})}\; , 
\label{eq:Phi+-spect}
\end{equation}
where $\tau (x,{\bm p}_{_{\sT}})$ is defined in Eq.~(\ref{eq:offshell}) and 
\begin{equation} 
\mathcal{M}^{(1)}(S) = 
\begin{cases}
  \displaystyle{-\intl \,
                \frac{i e_c\, \Gamma_{s\, \rho} \, n_-^{\rho} (\pslash-\lslash+m)\, 
		      {\cal Y}_s \, U(P,S)}
                        {(D_1+i \varepsilon)\,(D_2-i \varepsilon)\,
                                (D_3+i \varepsilon)\,(D_4+i \varepsilon)}}
& 
       \text{scalar diquark,} \\[0.5cm]
  \displaystyle{-\intl \,
                \frac{i e_c\,\varepsilon^*_{\sigma}(P-p,\lambda_a)\, 
                      \Gamma_{a\, \rho}^{\nu \sigma} \, n_-^{\rho} 
		      (\pslash-\lslash+m)\, d_{\mu\nu}(p-l-P) \,{\cal Y}^{\mu}_a 
		      \, U(P,S)}
                        {(D_1+i \varepsilon)\,(D_2-i \varepsilon)\,
                                (D_3+i \varepsilon)\,(D_4+i \varepsilon)}}
& 
       \text{axial-vector diquark,} 
\end{cases}
\label{eq:m-1loop}
\end{equation}
where for convenience we have introduced the notation
\begin{equation} \begin{split} 
D_1 &= l^2-m_g^2,\\
D_2 &= l^+, \\
D_3 &= (p-l)^2-m^2, \\
D_4 &= (P-p+l)^2-M_X^2\; . 
\label{eq:denoms}
\end{split} \end{equation} 

In order to explicitly calculate ${\cal M}^{(1)}$, we need to model the
gluon vertex with the scalar ($\Gamma_s$) and axial vector ($\Gamma_a$) diquark 
in Fig.~\ref{fig:eikonal}:
\begin{align} 
\Gamma_s^\rho &= i e_c\, (2P-2p+l)^\rho \nn \\
\Gamma_{a\,\rho}^{\nu \sigma} &= -i e_c\, \left[ (2P-2p+l)_\rho\,
g^{\nu \sigma} -(P-p+(1+\kappa_a)l)^\sigma g_\rho^\nu-(P-p-\kappa_a l)^\nu \, 
g_\rho^\sigma \right] \; , 
\label{eq:Dgvertex}
\end{align} 
where $e_c$ is the diquark color charge, which is
the same for scalar and vector ones and identical to that of the quark;
$\kappa_a$ is the diquark anomalous chromomagnetic moment. The structure of the
vector diquark-gluon vertex resembles the one for the coupling between the
photon and a spin-1 particle (see, e.g., Ref.~\cite{Robinson:1986ui}); 
for $\kappa_a=1$ the standard point-like
photon-$W$ coupling is recovered 
(see, e.g., Ref.~\cite{Peskin:1995ev}). 

The Sivers and Boer-Mulders functions can then be computed as
\begin{align} 
\frac{\varepsilon_{\sT}^{ij}p_{\sT i}S_{\sT j}}{M}\,
f_{1T}^{\perp}(x,{\bm p}_{\sT}^2) &= -\frac{1}{4}\,\frac{1}{(2\pi)^3}\,
\frac{1}{2(1-x)P^+}\,\mathrm{Tr}\, \left[ \left(
\mathcal{M}^{(1)}(S) \overline{\mathcal{M}}^{(0)}(S) - \mathcal{M}^{(1)}(-S) 
\overline{\mathcal{M}}^{(0)}(-S) \right)\, \g^+ \right] + \mathrm{h.c.} \; , 
\label{eq:Siversspect} \\
\frac{\varepsilon_{\sT}^{ij}p_{\sT j}}{M}\,
h_1^{\perp}(x,{\bm p}_{\sT}^2) &= \frac{1}{4}\,\frac{1}{(2\pi)^3}\,
\frac{1}{2(1-x)P^+}\,\mathrm{Tr}\, \left[ \left(
\mathcal{M}^{(1)}(S) \overline{\mathcal{M}}^{(0)}(S) + \mathcal{M}^{(1)}(-S) 
\overline{\mathcal{M}}^{(0)}(-S) \right)\, i\sigma^{i+}\,\g_5 \right] + 
\mathrm{h.c.} \; . 
\label{eq:BoerMuldersspect}
\end{align} 

Again, results have been produced for the three 
different choices of both Eq.~(\ref{eq:NqDff}) for the form factors at the 
nucleon-quark-diquark vertex, as well as of the axial-vector diquark 
propagator on each side of the diquark-gluon vertex in Fig.~\ref{fig:eikonal}. 
Consistently with the case of T-even parton densities, here we show the results
for the dipolar form factor of Eq.~(\ref{eq:ffdip}) and for the light-cone
transverse polarizations of the vector diquark, i.e.\ the
first choice in Eq.~(\ref{eq:lc}), the other combinations being listed in
App.~\ref{sec:B}. Combining the rules~(\ref{eq:tree-vertices}) with the
(\ref{eq:Dgvertex}) ones, we can rewrite Eq.~(\ref{eq:Siversspect}) and
(\ref{eq:BoerMuldersspect}) as
\\ 
\begin{align} 
f_{1T}^{\perp\,q(s)}(x,{\bm p}_{\sT}^2) &= - \frac{g_s}{4}\,
\frac{1}{(2\pi)^3}\, \frac{M\,e_c^2}{2(1-x)P^+}\,
\frac{(1-x)^2}{[{\bm p}_{\sT}^2+L_s^2(\Lambda_s^2)]^2}\,2\,\mathrm{Im}\,J_1^s 
 \label{eq:Siversspect2s} 
\\
f_{1T}^{\perp\,q(a)}(x,{\bm p}_{\sT}^2) &= \frac{g_a}{4}\,
\frac{1}{(2\pi)^3}\, \frac{M\,e_c^2}{4(1-x)P^+}\,
\frac{(1-x)^2}{[{\bm p}_{\sT}^2+L_a^2(\Lambda_a^2)]^2}\,2\,\mathrm{Im}\,J_1^a 
\; , \label{eq:Siversspect2a} 
\\ \nn \\
h_1^{\perp\,q(s)}(x,{\bm p}_{\sT}^2) &= f_{1T}^{\perp\,q(s)}(x,{\bm p}_{\sT}^2) 
\label{eq:BoerMuldersspect2s}
\\
h_1^{\perp\,q(a)}(x,{\bm p}_{\sT}^2) &=
-\frac{1}{x}\,f_{1T}^{\perp\,q(a)}(x,{\bm p}_{\sT}^2).
\label{eq:BoerMuldersspect2a}
\end{align} 
Note that for scalar diquarks the spectator model gives the same result for the
Sivers and the Boer-Mulders functions, independent of the choice of
the nucleon-quark-diquark form factor (see App.~\ref{sec:B}). 

In Eqs.~(\ref{eq:Siversspect2s}) and ff., the expressions $J_1$
contain the integral over the
loop momentum, the denominators $D_{1,2,3,4}$ defined in
Eq.~\eqref{eq:denoms}, and
the evaluation of the trace of 
the projected amplitude. For instance (see App.~\ref{sec:B_scalar}) 
\begin{equation}
J_1^{s}=\intl\,\frac{g_s\bigl((p-l)^2\bigr)}{(D_1+i \varepsilon)\,(D_2-i \varepsilon)\,(D_3+i
  \varepsilon)\,(D_4+i \varepsilon)}\;4 i\Big(l^++2(1-x)P^+\Big)\Big(l^+M
-P^+(m+xM)\,\frac{{\bm l}_{\sT}\cdot {\bm p}_{\sT}}{{\bm
p}_{\sT}^2}\Big).
\end{equation} 
To calculate its imaginary part, it is sufficient to make the replacements
\begin{align}
\frac{1}{D_2- i \varepsilon} &\to 2 \pi i \delta(D_2),
& 
\frac{1}{D_4+ i \varepsilon} &\to -2 \pi i \delta(D_4) 
\end{align} 
which corresponds to applying the 
Cutkosky rules~\cite{Cutkosky:1960sp}, 
cutting the diquark propagator ($D_4$) and the 
eikonalized quark one ($D_2$). 
We then get 
\begin{equation} \begin{split} 
2\,\mathrm{Im}\,J_1^{s} &=\intl\,\frac{g_s\bigl((p-l)^2\bigr)}{D_1\,D_3}\,4\,\left( l^+ +
2(1-x)P^+ \right) \, \left(l^+M-P^+(m+xM)\,
\frac{{\bm l}_{\sT}\cdot {\bm p}_{\sT}}{{\bm p}_{\sT}^2} \right)\,
(2\pi i)\, \delta(D_2)\,(-2\pi i)\, \delta(D_4)  \\
&=-4P^+\,(m+xM)\,(1-x)\,g_s\,\mathcal{I}_1\,.
\label{eq:J1s}
\end{split} \end{equation}
The calculation of $\mathcal{I}_1$ depends on the form factor used. Their
calculation can be found in App.~\ref{sec:C}.  For the case of the dipolar
form factor we obtain
\begin{equation} 
- 4P^+\,(m+xM)\,(1-x)\,g_s\,\mathcal{I}_1^{dip} = g_s\, 
\frac{P^+\,(m+xM)\,(1-x)^2}{\pi L_s^2(\Lambda_s^2)\,[{\bm p}_{\sT}^2+ 
L_s^2(\Lambda_s^2)]} \; .
\label{eq:J1sdip}
\end{equation} 

If the T-odd 
structures were deduced from the Drell--Yan amplitude, the $\mathcal{M}^{(1)}$ 
of Eq.~(\ref{eq:m-1loop}) would involve a $(l^++i\varepsilon)$ propagator, 
leading to the opposite sign in the cutting rule for $D_2$. In the spectator 
model, this is the origin of the predicted sign change for $f_{1T}^\perp$ and 
$h_1^\perp$ when extracting them in Drell--Yan spin asymmetries rather than in 
SIDIS ones~\cite{Collins:2002kn}. Analogously to Eq.~(\ref{eq:J1s}), we obtain
\begin{align} 
2\,\mathrm{Im}\,J_1^a &= -8P^+\,x\,(m+xM)\,g_a\,\mathcal{I}_1^{dip} = g_a\, 
\frac{2P^+\,x(1-x)\,(m+xM)}{\pi L_a^2(\Lambda_a^2)\,[{\bm p}_{\sT}^2+ 
L_a^2(\Lambda_a^2)]} \; . 
\label{eq:J1a} 
\end{align} 
By inserting these results in the model expressions of 
Eqs.~(\ref{eq:Siversspect2s}) to (\ref{eq:BoerMuldersspect2a}), we come to the
final form of the Sivers and Boer-Mulders functions with scalar and axial vector
diquarks:
\\ 
\begin{align}
f_{1T}^{\perp\,q(s)}(x,{\bm p}_{\sT}^2) &= -\frac{g_s^2}{4}\,
\frac{M\,e_c^2}{(2\pi)^4}\,\frac{(1-x)^3\,(m+xM)}{L_s^2(\Lambda_s^2)\,
[{\bm p}_{\sT}^2+L_s^2(\Lambda_s^2)]^3} 
\label{eq:Siversspect3s} 
\\
f_{1T}^{\perp\,q(a)}(x,{\bm p}_{\sT}^2) &= \frac{g_a^2}{4}\,
\frac{M\,e_c^2}{(2\pi)^4}\,\frac{(1-x)^2\,x\,(m+xM)}{L_a^2(\Lambda_a^2)\,
[{\bm p}_{\sT}^2+L_a^2(\Lambda_a^2)]^3}
\label{eq:Siversspect3a} 
\\ \nn \\
h_1^{\perp\,q(s)}(x,{\bm p}_{\sT}^2) &= f_{1T}^{\perp\,q(s)}(x,{\bm p}_{\sT}^2)
\label{eq:BoerMuldersspect3s}
\\
h_1^{\perp\,q(a)}(x,{\bm p}_{\sT}^2) &=
-\frac{1}{x}\,
f_{1T}^{\perp\,q(a)}(x,{\bm p}_{\sT}^2) \; .
\label{eq:BoerMuldersspect3a}
\end{align} 

To connect the ``Abelian'' version of the gluon interaction to the QCD color
interaction we shall apply the replacement~\cite{Brodsky:2002cx}
\begin{equation} 
e_c^2 \to 4 \pi C_F \alpha_s.
\end{equation} 

The Sivers and Boer-Mulders functions obtained in our model behave as 
$1/\bm{p}_{\sT}^6$ at high $\bm{p}_{\sT}^2$, similarly to the $f_1$ in 
Eq.~\eqref{eq:f1sspect}. As observed also in Ref.~\cite{Kotzinian:2008fe},  
this leads to a breaking of the
positivity bounds~\cite{Bacchetta:1999kz} for sufficiently high values of 
${\bm p}_{\sT}^2$. This problem is due to the fact
that the T-odd functions have been calculated at order $\alpha_S^1$, while the
T-even functions at order $\alpha_S^0$. At high ${\bm p}_{\sT}^2$, QCD 
radiative corrections generate a $1/\bm{p}_{\sT}^2$ tail for $f_1$ and a 
$1/\bm{p}_{\sT}^4$ tail for $f_{1\sT}^\perp$~\cite{Bacchetta:2008xw}. 
Our model is supposed to be valid for ${\bm p}_{\sT}^2 \sim M^2$ and for 
reasonable choices of the parameters no problems with positivity occurr 
in this region.

Often the following transverse-momentum moments of the Sivers and Boer-Mulders
functions are used:
\begin{align}
f_{1T}^{\perp\, (1)}(x) &= \int d{\bm p}_{\sT}\, \frac{{\bm p}_{\sT}^2}{2M^2}
\, f_{1T}^\perp (x,{\bm p}_{\sT}^2) \nn \\
f_{1T}^{\perp\, (1/2)}(x) &= \int d{\bm p}_{\sT}\, \frac{|{\bm p}_{\sT}|}{2M}
\, f_{1T}^\perp (x,{\bm p}_{\sT}^2)  \; .
\label{eq:Siversmom}
\end{align}
In our model, they turn out to be
\begin{align}
f_{1T}^{\perp\,q(s)\,(1)}(x) 
&= -\frac{g_s^2}{32}\,\frac{e_c^2}{(2\pi )^3\,M}\,
   \frac{(m+xM)\,(1-x)^3}{[L_s^2(\Lambda_s^2)]^2} 
\label{eq:Sivers1s} \\
f_{1T}^{\perp\,q(a)\,(1)}(x) 
&= \frac{g_a^2}{32}\,\frac{e_c^2}{(2\pi )^3\,M}\,
   \frac{x\,(m+xM)\,(1-x)^2}{[L_a^2(\Lambda_a^2)]^2},
\label{eq:Sivers1a} \\ \nn \\
h_1^{\perp\,q(s)\,(1)}(x) &= f_{1T}^{\perp\,q(s)\,(1)}(x) 
\label{eq:BoerMulders1s} \\
h_1^{\perp\,q(a)\,(1)}(x)
&= -\frac{1}{x}\,f_{1T}^{\perp\,q(a)\,(1)}(x).
\label{eq:BoerMulders1a} 
\end{align}

\begin{align}
f_{1T}^{\perp\,q(s)\,(1/2)}(x) 
&= 
-\frac{g_s^2}{256}\,\frac{e_c^2}{(2\pi )^2}\,
   \frac{(m+xM)\,(1-x)^3}{[L_s^2(\Lambda_s^2)]^{5/2}}, \\ 
f_{1T}^{\perp\,q(a)\,(1/2)}(x) 
&= \frac{g_a^2}{256}\,\frac{e_c^2}{(2\pi )^2}\,
   \frac{x\,(m+xM)\,(1-x)^2}{[L_a^2(\Lambda_a^2)]^{5/2}}, 
\\ \nn \\
h_1^{\perp\,q(s)\,(1/2)}(x) &= f_{1T}^{\perp\,q(s)\,(1/2)}(x) 
\label{eq:BoerMulders1-2s} \\
h_1^{\perp\,q(a)\,(1/2)}(x) 
&= -\frac{1}{x}\,f_{1T}^{\perp\,q(a)\,(1/2)}(x).
\label{eq:BoerMulders1-2a}
\end{align}


\subsection{T-odd functions: overlap representation}
\label{sec:LCtodd}

As already mentioned above, T-odd leading-twist parton distributions arise from
the interference of two channels leading to the same final state; for the case
considered here (and depicted in Fig.~\ref{fig:eikonal}), the two channels are
given by the tree-level and the single-gluon-exchange scattering amplitudes,
respectively. In Ref.~\cite{Burkardt:2002ks}, it was suggested that 
T-odd parton densities can also be represented by overlaps of 
LCWFs, as for their T-even partners, provided 
that a suitable operator is included to describe the final-state interactions 
(FSI) produced by the gluon rescattering. So far, this representation was fully
developed in a spectator model only for the Sivers function with scalar
diquarks~\cite{Burkardt:2003je,Lu:2006kt}. Here, we generalize it to the case of
axial-vector diquarks, as well as to the Boer-Mulders function. In this way,
all leading-twist (T-even and T-odd) parton densities can be given by overlaps
of LCWFs consistently within the model, contrary to the statement of 
Ref.~\cite{Meissner:2007rx}.

Following Ref.~\cite{Lu:2006kt}, for a nucleon transverse polarization state
described by Eqs.~\eqref{eq:Tup} and \eqref{eq:Tdown} along a generic direction 
$\hat{\bm{S}}_{\sT} = (\cos\phi_S, \sin\phi_S)$, and for an analogous quark
state described by Eqs.~\eqref{eq:Tupq} and \eqref{eq:Tdownq} along 
$\hat{\bm{S}}_{q\sT} = (\cos\phi_{S_q}, \sin\phi_{S_q})$, we can rewrite the 
Sivers~(\ref{eq:Sivers}) and Boer-Mulders~(\ref{eq:Boer-Mulders}) 
functions according to the Trento Conventions~\cite{Bacchetta:2004jz} 
(keeping in mind that $\lambda_X$ is absent 
for the scalar
diquark and $\lambda_X=\pm $ for the vector diquark)
as
\begin{align}
\frac{2\,(\hat{\bm{S}}_{\sT} \times {\bm p}_{\sT})\cdot\hat{\bm P}}{M}\,\, 
f_{1T}^{\perp}(x,{\bm p}_{\sT}^2) &= \int \frac{d{\bm p}_{\sT}'}{16\pi^3}
\, G(x,{\bm p}_{\sT},{\bm p}_{\sT}') \nn \\
& \quad  \sum_{\lambda_q, \lambda_X} \left[ 
\psi^{\uparrow\,*}_{\lambda_q\lambda_X}(x,{\bm p}_{\sT})\,
\psi^{\uparrow}_{\lambda_q\lambda_X}(x,{\bm p}_{\sT}') - 
\psi^{\downarrow\,*}_{\lambda_q\lambda_X}(x,{\bm p}_{\sT})\,
\psi^{\downarrow}_{\lambda_q\lambda_X}(x,{\bm p}_{\sT}') \right] + 
\mathrm{h.c.} \; ,  \\
\frac{(\hat{\bm{S}}_{q\sT} \times {\bm
    p}_{\sT})\cdot\hat{\bm P}}{M}\,\, 
h_1^{\perp}(x,{\bm p}_{\sT}^2) &= \int \frac{d{\bm p}_{\sT}'}{16\pi^3}\, 
G(x,{\bm p}_{\sT},{\bm p}_{\sT}') \nn \\
& \quad \frac{1}{2}\,\sum_{\lambda_N,\,\lambda_X} 
\left[ \psi^{\lambda_N\,*}_{\uparrow\lambda_X}(x,{\bm p}_{\sT})\,
\psi^{\lambda_N}_{\uparrow\lambda_X}(x,{\bm p}_{\sT}') - 
\psi^{\lambda_N\,*}_{\downarrow\lambda_X}(x,{\bm p}_{\sT})\,
\psi^{\lambda_N}_{\downarrow\lambda_X}(x,{\bm p}_{\sT}') \right] + 
\mathrm{h.c.} \; . \label{eq:Boer-MuldersLCWF}
\end{align}

The above equations should be considered as assumptions, since 
it is not known a priori if 
the FSI operator $G(x,{\bm p}_{\sT},{\bm p}_{\sT}')$ can be isolated and 
is the same for all
functions and all types of diquarks. 
In our model, it turns out to be actually the same in all
cases. In order to determine it, we must insert here 
above the expressions for the LCWFs of Sec.~\ref{sec:LCwf}, and 
compare the results with the ones from Eqs.~\eqref{eq:Siversspect2s} to
\eqref{eq:BoerMuldersspect2a}, after replacing
$\mathrm{Im}J_1^s$, $\mathrm{Im}J_1^a$ 
with Eqs.~(\ref{eq:J1s}-\ref{eq:J1a}), 
respectively, while keeping the definition of
$\mathcal{I}_1^{dip}$ (see App.~\ref{sec:C}). 
For the scalar diquark case, 
for example, we get
\begin{align}
f_{1T}^{\perp q(s)}(x,{\bm p}_{\sT}^2) 
&= \frac{g_s^2}{8\pi^3}\,
\frac{M\,(1-x)^3\,(m+xM)}{[{\bm p}_{\sT}^2+L_s^2(\Lambda_s^2)]^2} \int 
d{\bm p}_{\sT}'\, \frac{\mathrm{Im} G(x,{\bm p}_{\sT},{\bm p}_{\sT}')}
{[{\bm p}_{\sT}^{\prime\,2}+L_s^2(\Lambda_s^2)]^2}\, 
\frac{({\bm p}_{\sT}-{\bm p}_{\sT}')\cdot{\bm p}_{\sT}}{{\bm p}_{\sT}^2} \; .
\label{eq:G-s}
\end{align}
The above 
expression is identical to Eq.~(\ref{eq:Siversspect2s}), after inserting
Eq.~(\ref{eq:J1sdip}) and the definition~(\ref{eq:I1dip}) of $\mathcal{I}_1^{dip}$ 
(with the harmless substitution 
${\bm l}_{\sT}' \leftrightarrow -{\bm l}_{\sT}'$), provided that 
\begin{equation}
\mathrm{Im} G(x,{\bm p}_{\sT},{\bm p}_{\sT}') = -\frac{e_c^2}{2\,(2\pi)^2}\,
\frac{1}{({\bm p}_{\sT}-{\bm p}_{\sT}')^2} = -\frac{C_F\alpha_s}{2\pi}\,
\frac{1}{({\bm p}_{\sT}-{\bm p}_{\sT}')^2} \; ,
\label{eq:G-s2}
\end{equation}
in agreement with the expression of Ref.~\cite{Lu:2006kt}. Following similar
steps, we recover the same result~(\ref{eq:G-s2}) also for the Sivers function
with axial-vector diquarks, and for the Boer-Mulders function as well. The FSI
operator $G(x,{\bm p}_{\sT},{\bm p}_{\sT}')$ is indeed universal and describes
a rescattering via one gluon-exchange, which corresponds to the expansion at
first order of the gauge link operator of Eq.~(\ref{eq:link}).  Note that in
other versions of the model (see App.~\ref{sec:B}) the FSI 
cannot be as simple as Eq.~\eqref{eq:G-s2}, since we observe also a dependence
on the vector diquark anomalous chromomagnetic moment $\kappa_a$, which is
absent in the above equation. This does not imply that the FSI operator is not
universal, but simply that it could have additional parts that are not
interacting with scalar diquarks and transversely polarized vector diquarks.

We close this section by observing that in our model we can generalize the  
relation between the first ${\bm p}_{\sT}$-moment of the Sivers function and 
the nucleon anomalous magnetic moment $\kappa$, suggested in 
Ref.~\cite{Lu:2006kt} in the simple scalar diquark picture. In fact, we
define $\kappa$ in terms of the Dirac form factor using the overlap
representation for the nucleon matrix element of the spin-flip electromagnetic 
current operator~\cite{Brodsky:2000ii}:  
\begin{align}
\kappa &= \frac{e}{2M}\,F_2(0) \nn \\
&=  -\frac{1}{q_x-iq_y} \sum_{k,n,\lambda_n}\, e_n \int 
\frac{d{\bm p}_{\sT}\,dx}{16\pi^3}\,\Psi_k^{+\,*}(x,{\bm p}_{\sT}',\lambda_n)\, 
\Psi_k^{-}(x,{\bm p}_{\sT},\lambda_n)  \Bigg\vert_{{\bm p}_{_T}'={\bm p}_{\sT}} 
\; ,
\label{eq:kappa_def}
\end{align}
where the sum runs upon the number of Fock states $k$, the number of constituents
$n$ in each state $k$, and their helicities $\lambda_n$. Since in the diquark 
model of the nucleon initially at rest (${\bm P}_{\sT}=0$) there is only one 
Fock state with two constituents, and the kinematics of the diquark is 
constrained to the one of the valence quark, the wave functions $\Psi$
reduce to the usual LCWF~\cite{Lu:2006kt}. The momentum conservation for the
struck quark reads ${\bm p}_{\sT}'={\bm p}_{\sT}+(1-x)\,{\bm q}_{\sT}$. 
Distinguishing between $\kappa^{q(s)}$ and $\kappa^{q(a)}$ for scalar and axial-vector
diquarks, respectively, Eq.~(\ref{eq:kappa_def}) becomes
\begin{align}
\kappa^{q(s)} &=  -\frac{1}{q_x-iq_y}\,\int \frac{d{\bm p}_{\sT} dx}{16\pi^3} 
\, \sum_{\lambda_q}\,\left[ \psi_{\lambda_q}^{+\,*}(x,{\bm p}_{\sT}')\,
\psi_{\lambda_q}^-(x,{\bm p}_{\sT})\right]
\Bigg\vert_{{\bm p}_{_T}'={\bm p}_{_T}} \nn \\
&= \frac{g_s^2}{(2\pi )^2}\,\int_0^1 dx \, \frac{1-x}{12}\, 
\frac{(1-x)^3\,(m+xM)}{[L_s^2(\Lambda_s^2)]^3} 
\equiv \int_0^1 dx \, \kappa^{q(s)} (x) \nn \\
\kappa^{q(a)} &=  -\frac{1}{q_x-iq_y}\,\int \frac{d{\bm p}_{\sT} dx}{16\pi^3} 
\, \sum_{\lambda_q,\lambda_a}\,\left[ 
\psi_{\lambda_q\lambda_a}^{+\,*}(x,{\bm p}_{\sT}')\,
\psi_{\lambda_q\lambda_a}^-(x,{\bm p}_{\sT})\right] 
\Bigg\vert_{{\bm p}_{_T}'={\bm p}_{_T}} \nn \\
&= -\frac{g_a^2}{(2\pi )^2}\,\int_0^1 dx \, \frac{x}{12}\, 
\frac{(1-x)^3\,(m+xM)}{[L_a^2(\Lambda_a^2)]^3} 
\equiv \int_0^1 dx \, \kappa^{q(a)} (x) \; .
\label{eq:kappa}
\end{align}

By comparison with the first ${\bm p}_{\sT}$-moment of $f_{1T}^{\perp\,q(s)}$ 
and $f_{1T}^{\perp\,q(a)}$ in Eqs.~\eqref{eq:Siversspect3s} and \eqref{eq:Siversspect3a}, respectively, 
\begin{align}
f_{1T}^{\perp\,q(s)}(x) &= \int d{\bm p}_{\sT} \, 
f_{1T}^{\perp\,q(s)}(x,{\bm p}_{\sT}^2) \nn \\
&= -\frac{g_s^2}{(2\pi )^2}\, 
\frac{MC_F\alpha_s\,(1-x)^3\,(m+xM)}{[2\,L_s^2(\Lambda_s^2)]^3} \nn \\
f_{1T}^{\perp\,q(a)}(x) &= \int d{\bm p}_{\sT} \, 
f_{1T}^{\perp\,q(a)}(x,{\bm p}_{\sT}^2) \nn \\
&= \frac{g_a^2}{(2\pi )^2}\, 
\frac{MC_F\alpha_s\,x\,(1-x)^2\,(m+xM)}{[2\,L_a^2(\Lambda_a^2)]^3} \; ,
\label{eq:Sivers-pTmom}
\end{align}
we deduce the relation
\begin{align}
f_{1T}^{\perp\,q}(x) &= -\frac{3}{2}\,MC_F\alpha_s\,\frac{\kappa^{q} (x)}{1-x} ,
\label{eq:Sivers-kappa_x}
\end{align}
valid for both types of diquarks, 
from which we have
\begin{align}
\int_0^1 dx\, (1-x)\, f_{1T}^{\perp\,q}(x) &= -\frac{3}{2}\,MC_F\alpha_s\, 
\kappa^{q} 
\label{eq:Sivers-kappa}
\end{align}
that generalize the findings of Ref.~\cite{Burkardt:2003je,Lu:2006kt}.


\section{Numerical results and comparison with various parametrizations}
\label{sec:fit}

In this section, after fixing the parameters of the model by fitting some
known distribution functions, we show the numerical results of our model 
for a few selected TMDs.


\subsection{Choice of model parameters}
\label{sec:fitf1g1}

In order to fix the parameters of the model, we try to reproduce the
parametrizations of parton
distribution functions extracted from experimental data. When doing this,
however, we have to face the problem of choosing a scale $Q^2$ at which
our model can be compared to the parametrization. In principle, this scale
should be considered as a further parameter of the model. However, we checked
that the lowest possible value of $Q^2$ is always 
preferred by the fit. This is not
surprising, since probably the model is applicable to a very low scale, beyond
the limit of applicability of the perturbative QCD evolution equations.
Therefore, we
have decided to compare it to a parametrization at the lowest 
possible value of $Q^2$. 

For the unpolarized distribution functions $f_1^u$ and $f_1^d$, 
we have chosen the parametrization of the ZEUS
collaboration~\cite{Chekanov:2002pv} (ZEUS2002) at $Q_0^2= 0.3$ GeV$^2$. This set of PDFs
gives also an estimate of the errors, which is important to perform a $\chi^2$
fit. Other parametrizations either do not reach such low $Q^2$ or provide no
error estimate. 

For the helicity distributions $g_1^u$ and $g_1^d$, we chose the leading-order
(LO) version 
from Ref.~\cite{Gluck:2000dy} (GRSV2000) at 
$Q^2 = 0.26$ GeV$^2$. Since this parametrization comes with no error estimate,
we assigned a fixed relative error of 10\% and 25\% to the up and down quark
distributions, respectively, which is reasonably similar to the error
estimates of other parametrizations at higher $Q^2$ (see, e.g.,
Ref.~\cite{Hirai:2003pm}). 

Finally, in order to perform the fit we arbitrarily chose to select from each parametrization 25
equally spaced points in the range $x=0.1$ to 0.7.

The free parameters of the model include the quark mass $m$, 
the nucleon-quark-diquark coupling $g_{X}$, the diquark mass $M_X$, and the 
cutoff $\Lambda_X$ in the nucleon-quark-diquark form factor, for $X=s,a$
scalar and axial-vector diquarks. 
It turns out that in order to achieve a good fit 
we need also to make a distinction between the two isospin states of the 
vector diquark. Hence, we will use $g_{a}$  
$M_a$, and $\Lambda_{a}$, for the coupling, mass and cutoff of the 
vector isoscalar diquark with $I_3=0$ (corresponding to the $ud$ system), 
and $g'_{a}$, $M_a'$, and $\Lambda'_{a}$, 
for the normalization, mass and cutoff of 
the vector isovector diquark with $I_3=1$ (corresponding to the $uu$ 
system). 

In order to reduce the number of free parameters, we decided to
fix the value of the constituent quark mass to $m=0.3$ GeV. We checked that
the results are not very sensitive to the value of this parameter.

To perform the fit, we 
need to discuss the relation between the functions $f_1^{q(s)}$, $f_1^{q(a)}$ and
$f_1^{q(a')}$,  
computed in the model, and the functions $f_1^u$ and $f_1^d$ of the
global fits. 
For ease of interpretation, it is better to use
normalized versions of the $f_1^{q(X)}$. 
Therefore, we
write 
 $f_{1 \,\rm{norm}}^{q(X)}= \left(N_X^2/g_X^2\right) f_1^{q(X)}$  
where $N_X$ are normalization constants determined by imposing
\begin{equation}
\pi \int_0^1 dx \int_0^{\infty} d\bm{p}_{\sT}^2 \,
f_{1 \,\rm{norm}}^{q(X)}(x,\bm{p}_{\sT}^2) = 1. 
\end{equation} 
Quite generally, the relation between quark flavors and diquark types can be
written as
\begin{align}
f_1^u &= c_{s}^2\, f_{1 \,\rm{norm}}^{u(s)} + c_{a}^2\, f_{1 \,\rm{norm}}^{u(a)} 
\label{eq:uX}
\\
f_1^d &= c_a^{\prime 2}\, f_{1 \,\rm{norm}}^{d(a')}  \; .
\label{eq:dX}
\end{align} 
We will refer to the coefficients $c_{X}$ as ``couplings'', although they differ
from the original couplings $g_{X}$ by the normalization constants $N_{X}$. 
They are free
parameters of the model.

In past versions of the spectator diquark model~\cite{Jakob:1997wg}, 
the quarks were assumed to occupy the 
lowest-energy available orbital with positive parity $(J^P={\half}^+ )$; in 
this case, the proton wave function assumes an SU(4)=SU(2)$\otimes$SU(2) 
spin-isospin symmetry, leading to probabilistic weights 3:1:2 among the scalar
isoscalar (quark $u$ with diquark $s$), vector isoscalar (quark $u$ 
with diquark $a$), and vector isovector (quark $d$ with diquark 
$a'$) configurations. Moreover, the overall size of the couplings was
adjusted to give a total number of three quarks.
These choices led to the relations~\cite{Jakob:1997wg}
\begin{align}
f_1^u &= \frac{3}{2}\,f_{1 \,\rm{norm}}^{u(s)} + \frac{1}{2}\, f_{1 \,\rm{norm}}^{u(a)} 
\\
f_1^d &= f_{1 \,\rm{norm}}^{d(a')}  \; .  
\end{align} 
There are two reasons to criticize this choice. First of all, 
in the present work the quark-diquark system can have a 
nonvanishing relative orbital angular momentum, as shown in the previous 
section. Thus, the proton wave function no longer displays an SU(4) symmetry.
Secondly, strictly speaking the SU(4) decomposition gives coefficients that
are three times smaller then the ones in the above relation. 
This is because the total
number of quarks ``seen'' in the spectator model is only one, since the other
two are always hidden inside the diquark.
This is actually a fundamental limitation of the spectator model, it is
independent of the SU(4) choice, and in our
opinion it has not been sufficiently stressed in the literature. 
The only possible way out
is to consider the diquark not as an elementary particle, but as formed by two
quarks that can be also probed by the photon (see, e.g.,
Ref.~\cite{Cloet:2007em}). 

A different way to see this problem is by considering the (longitudinal)
momentum sum
rule. Since also the diquarks can carry momentum, they should be included in
the corresponding sum rule.\footnote{A similar approach has been used in
Ref.~\cite{Goeke:2006ef} to verify in the spectator model the validity of the so-called
Burkardt sum rule~\cite{Burkardt:2004ur}, 
which is related to transverse-momentum conservation.}
 Using the handbag diagram 
of Fig.~\ref{fig:DISdiquark}, we calculated the corresponding diquark 
distribution function $f_1^{X(q)}$ for the active diquark in the state $X$ and 
the spectator quark with flavor $q$, again using the first choice 
in Eq.~(\ref{eq:lc}) (independently of the choice of form
factor). 
We found the 
remarkable property
\begin{equation} 
f_1^{X(q)}(x) = f_1^{q(X)}(1-x).
\label{eq:pdfsymm}
\end{equation} 
By splitting the total proton momentum sum rule into the contributions of
quarks, $P_q$, and of diquarks, $P_X$,
using the 
symmetry property~(\ref{eq:pdfsymm}) 
we get 
\begin{equation} \begin{split} 
P_q + P_X &= 
\int_0^1 dx \,x \left[ c_{s}^2\, f_{1\,\rm{norm}}^{u(s)}(x) + 
c_{a}^2\, f_{1\,\rm{norm}}^{u(a)}(x) + c_a^{\prime 2}\, f_{1\,\rm{norm}}^{d(a')}(x) \right] 
\\
& \quad + \int_0^1 dx \,x \left[ c_{s}^2\, f_{1\,\rm{norm}}^{s(u)}(x) + 
c_{a}^2\, f_{1\,\rm{norm}}^{a(u)}(x) + c_a^{\prime 2}\, f_{1\,\rm{norm}}^{a'(d)}(x) \right] 
\\
&= \int_0^1 dx \, \left[ c_{s}^2\, f_{1\,\rm{norm}}^{u(s)}(x) + 
c_{a}^2\, f_{1\,\rm{norm}}^{u(a)}(x) + c_a^{\prime 2}\, f_{1\,\rm{norm}}^{d(a')}(x) \right] 
\\
&=  c_{s}^2 + c_{a}^2 + c_a^{\prime 2} \;. 
\label{eq:momsumrule2}
\end{split} \end{equation} 
It is therefore impossible in our spectator model to fulfill at the same time 
the momentum sum rule and the quark number sum rule. 

Although from the fundamental point of view it is more important to satisfy
the momentum sum rule, from the phenomenological point of view it is impossible
to reproduce the parametrizations in a satisfactory way. We decided therefore
to avoid imposing the momentum sum rule and let the fit choose the values of
the parameters $c_X$.

\begin{figure}[h]
\begin{center}
\includegraphics[height=5cm]{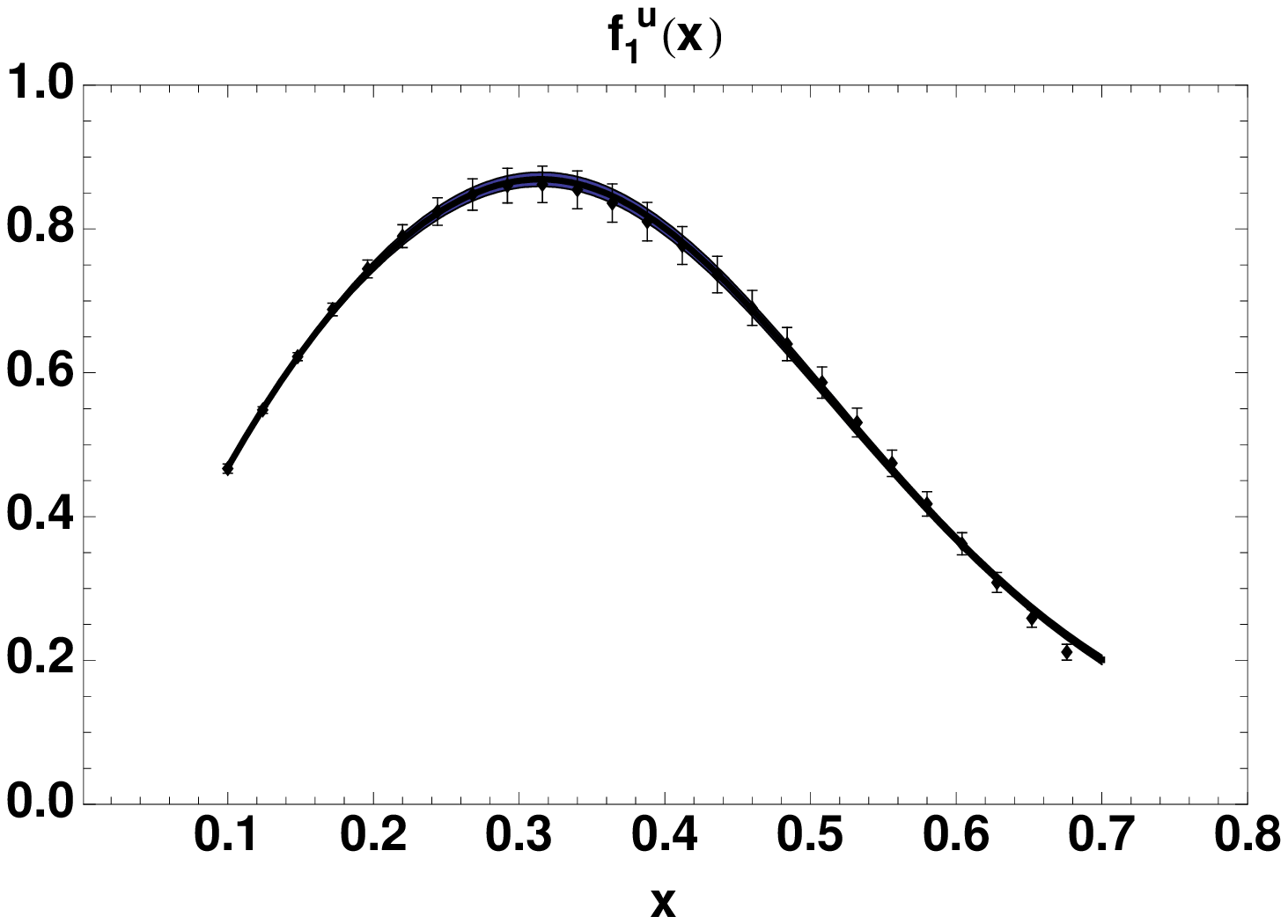} \hspace{1cm}
\includegraphics[height=5cm]{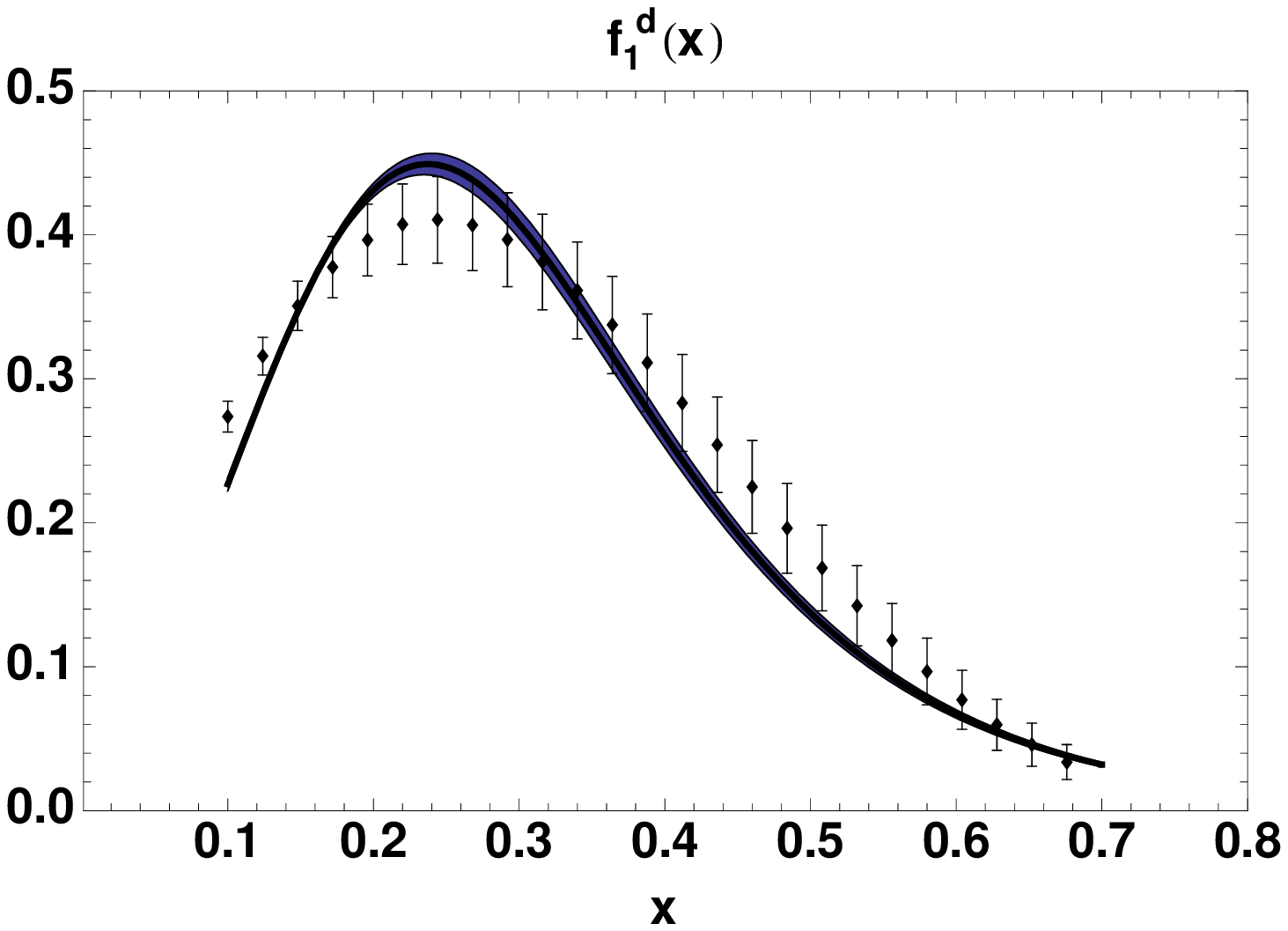} \\
\includegraphics[height=5cm]{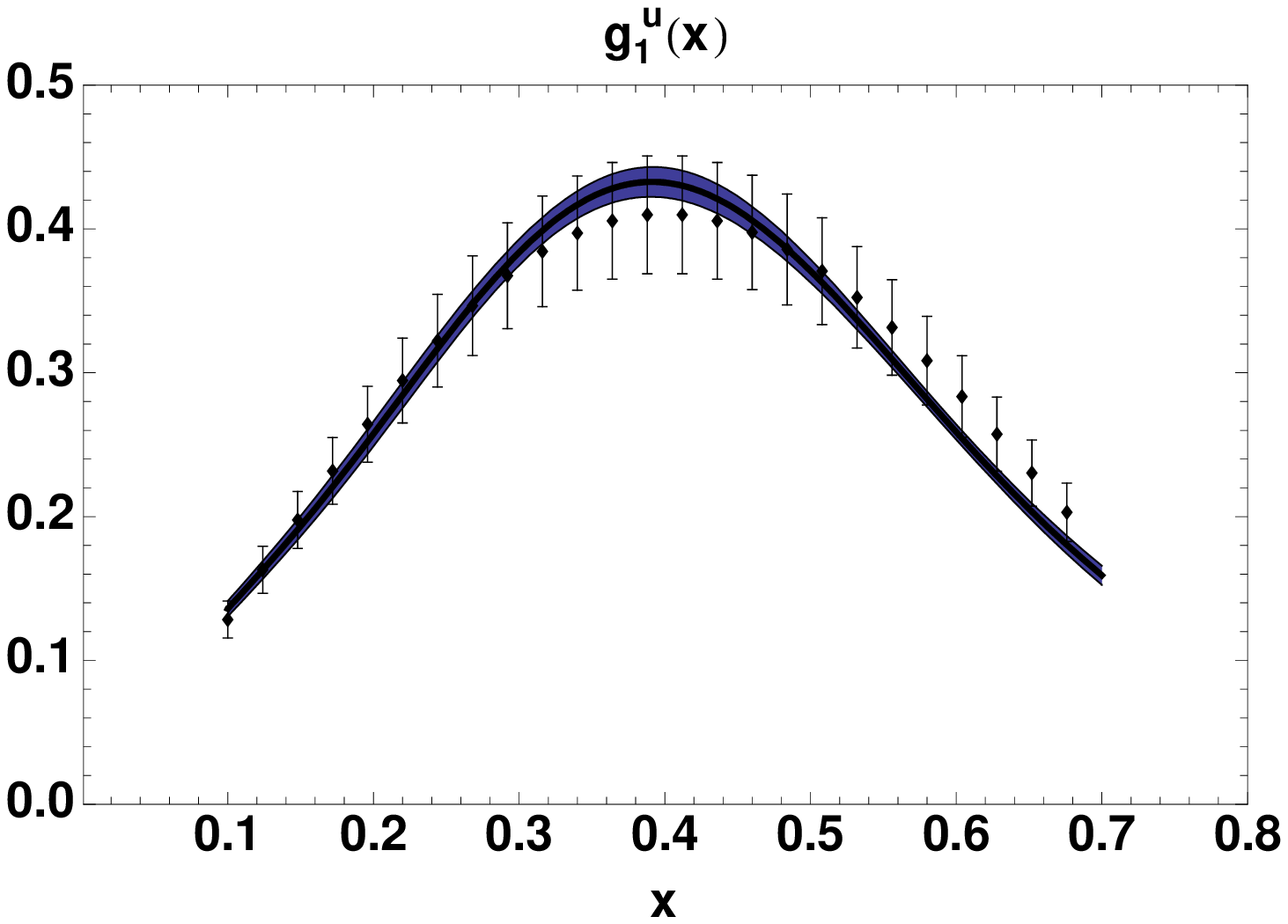} \hspace{1cm}
\includegraphics[height=5cm]{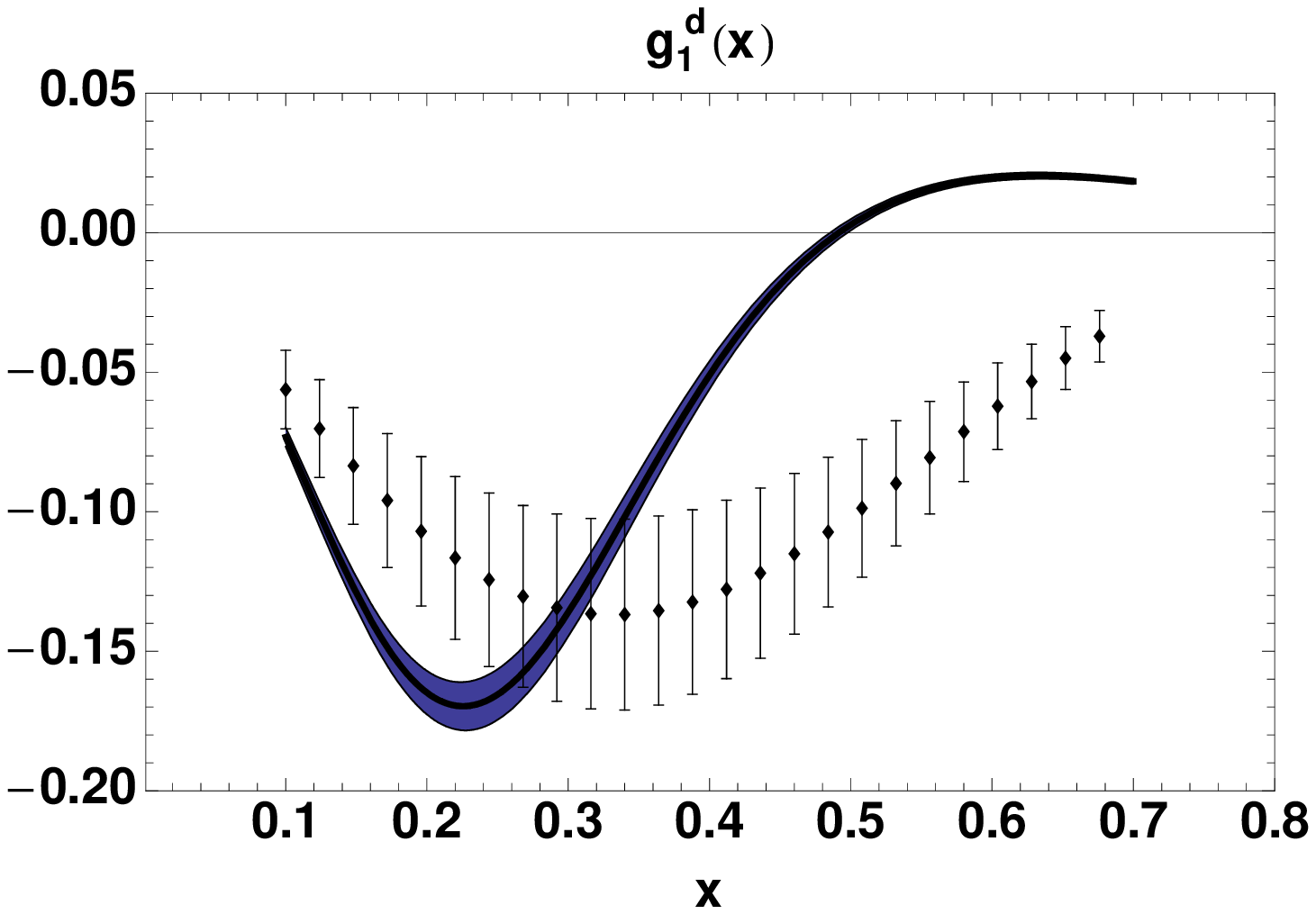} 
\end{center}
\caption{The distribution functions $f_1(x)$ (above) and $g_1(x)$ (below) for
the up quark (left panel) and the down quark (right panel). Data are a 
selection of 25 equidistant points in $0.1 \leq x \leq 0.7$
from the parametrizations of Ref.~\protect{\cite{Chekanov:2002pv}} (ZEUS2002) and
Ref.~\protect{\cite{Gluck:2000dy}} (GRSV2000) at LO, respectively (we assigned a constant
relative error
of 10\% to $g_1^u$ and 25\% to $g_1^d$ based on comparisons with similar
fits~\protect{\cite{Hirai:2003pm}}). The curves represent the best fit
($\chi^2/\rm{d.o.f.}=3.88$) obtained with 
our spectator model. The statistical uncertainty bands correspond to $\Delta
\chi^2 = 1$.}
\label{fig:fit}
\end{figure}

In summary, we have 9 free parameters for the model.
We fix them
by fitting at the same time $f_1^u,\, f_1^d$ at $Q^2=0.3$ GeV$^2$ 
from Ref.~\cite{Chekanov:2002pv}, and $g_1^u,\, g_1^d$ at $Q^2=0.26$ GeV$^2$ 
from Ref.~\cite{Gluck:2000dy} at LO. The fit was performed using the MINUIT 
program. A $\chi^2$/d.o.f. = 3.88 was 
reached. The results are shown in Fig.~\ref{fig:fit}. In spite of the very
high $\chi^2$, the agreement is acceptable, except perhaps for the down quark
helicity distribution. 
The error 
band is deduced from 
the covariance matrix given by MINUIT and represents the standard
1-$\sigma$ statistical uncertainty ($\Delta \chi^2 = 1$). 
The corresponding values for the 
various model parameters are listed in Tab.~\ref{tab:fit}.

\begin{table}[h]
\begin{tabular}{|c||c|c|c|}
\hline
\textit{Diquark} & $M_X$ (GeV) & $\Lambda_X$ (GeV) & $c_X$ \\
\hline
Scalar $s$ $(uu)$ & 0.822 $\pm$ 0.053 & 0.609 $\pm$ 0.038 & 0.847 $\pm$
0.111 \\
\hline
Axial-vector $a$ $(ud)$ & 1.492 $\pm$ 0.173 & 0.716 $\pm$ 0.074 &
1.061 $\pm$ 0.085 \\
\hline
Axial-vector $a'$ $(uu)$ & 0.890 $\pm$ 0.008 & 0.376 $\pm$ 0.005 &
0.880 $\pm$ 0.008 \\
\hline
\end{tabular}
\caption{Results for the model parameters with dipolar nucleon-quark-diquark 
form factor and light-cone transverse polarizations of the vector diquark: 
the diquark masses $M_X$, 
the 
cutoffs $\Lambda_X$ in the form factors, and the $c_X$ couplings for 
$X=s,a,a'$ scalar isoscalar, vector isoscalar, and vector isovector 
diquarks. The fit was performed using the MINUIT program on the parametrization 
of $f_1(x)$ from Ref.~\protect{\cite{Chekanov:2002pv}} (ZEUS2002), and of $g_1(x)$ from
Ref.~\protect{\cite{Gluck:2000dy}} (GRSV2000) at LO, reaching a $\chi^2$/d.o.f. = 3.88.} 
\label{tab:fit} 
\end{table} 

\begin{figure}[h]
\begin{center}
\includegraphics[height=6cm]{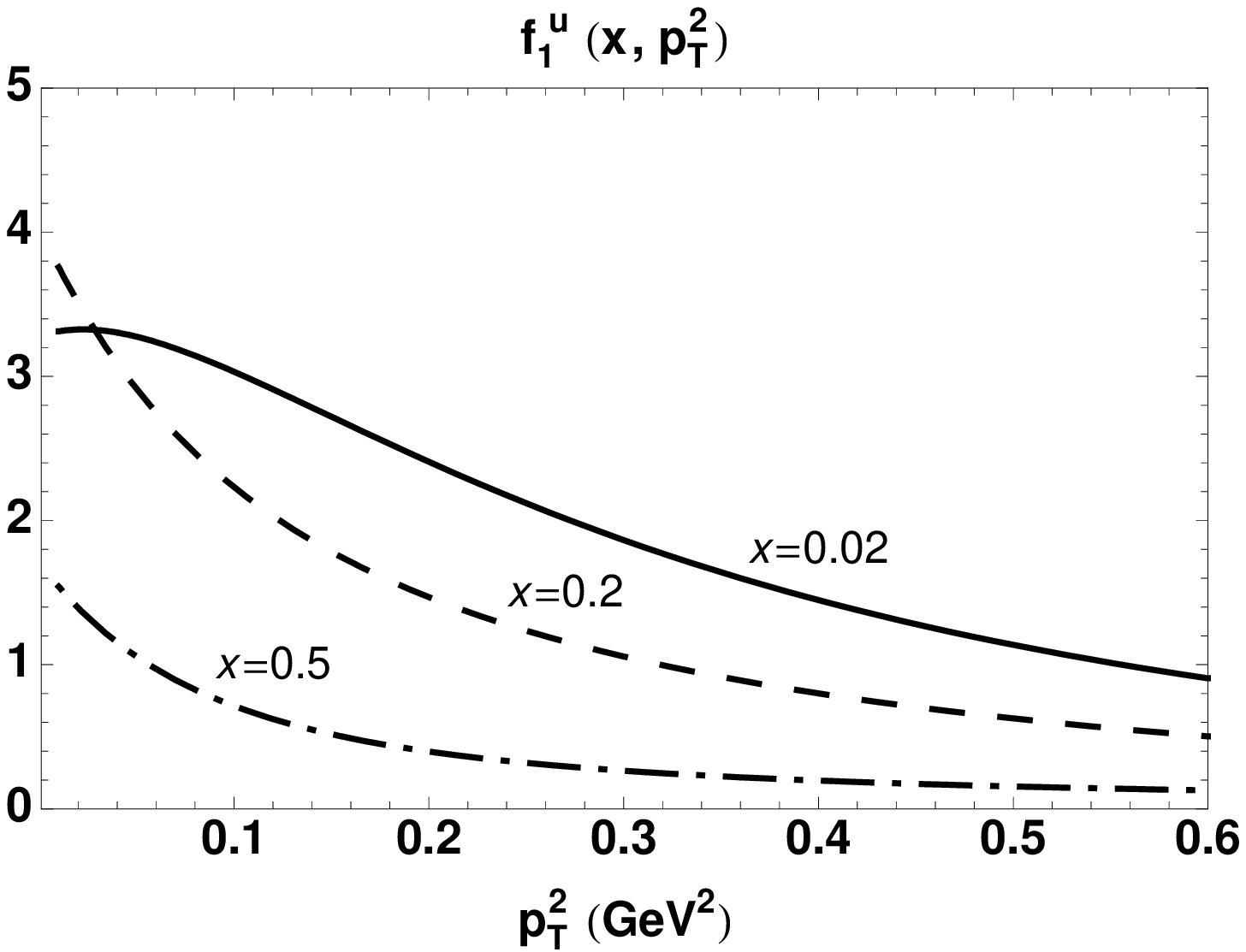} \hspace{0.2cm}
\includegraphics[height=6cm]{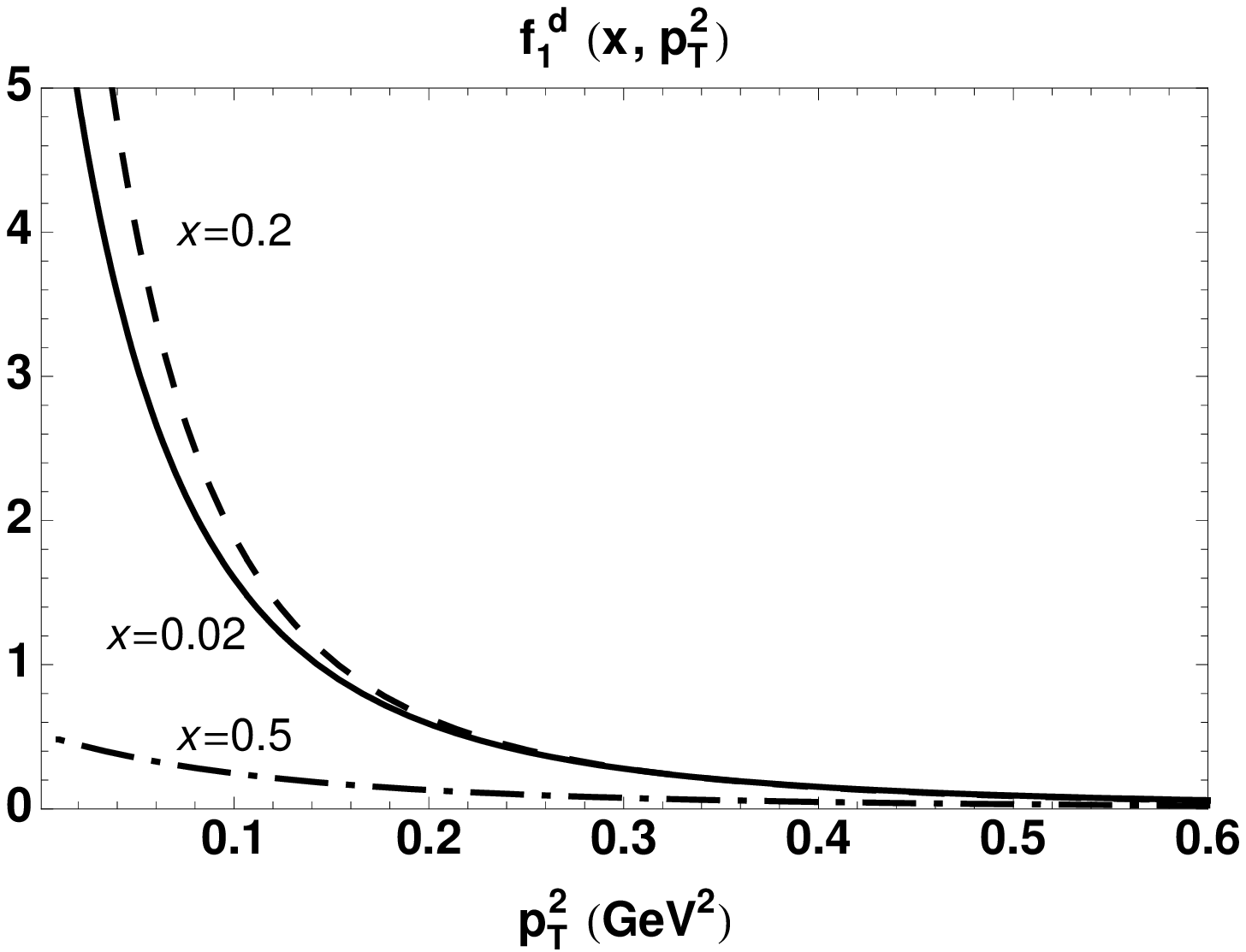}
\end{center}
\caption{The $\bm{p}_{\sT}^2$ dependence of the 
unpolarized distribution $f_1(x,\bm{p}_{\sT}^2)$ 
for up (left panel) and down
quark (right panel). Different lines correspond to 
different values of $x$. The downturn of the function $f_1^u$ at relatively
small $x$ is due to wavefunctions with nonzero orbital angular
momentum.} 
\label{fig:f1_xpt}
\end{figure}

\subsection{Unpolarized parton densities}

With the above model parameters, 
the proton momentum fraction $P_q$ carried by 
valence quarks, is
\begin{align}
P_q &= \int_0^1 dx \,x \left[ c_{s}^2\, f_{1\,\rm{norm}}^{u(s)}(x) + 
c_{a}^2\, f_{1\,\rm{norm}}^{u(a)}(x) + c_a^{\prime 2}\,
f_{1\,\rm{norm}}^{d(a')}(x) \right] 
\nn \\
&= \int_0^1 dx\, x\, \left[ f_1^u(x) + f_1^d(x) \right] \approx 
0.584 \pm 0.010  \;  , 
\label{eq:momsumrule}
\end{align}
which is consistent with the ZEUS result of 0.55~\cite{Chekanov:2002pv}.

While for $f_1^d$ only the vector-isovector diquark plays a role,
for $f_1^u$ it turns out that 
the contributions from the scalar and vector diquark have
about the same size. 
The vector diquark is always dominant at high $x$. However, we know that the
model is not reliable in the limit $x \to 1$. In fact, 
the behavior at high $x$ does
not follow the predictions of Ref.~\cite{Avakian:2007xa}, 
since our model does not
correctly take into account the dominant dynamics in that region.

We consider now the $\bm{p}_{\sT}^2$ dependence of the unpolarized distribution 
function obtained in our model. In Fig.~\ref{fig:f1_xpt} we show the behavior of the 
up and down components as functions of $\bm{p}_{\sT}^2$ for a few values of the
variable $x$. 

First of
all, we observe that $f_1^{u}$ displays a nonmonotonic behavior at $x \le
0.02$. This is due to the contribution from
LCWFs with nonzero orbital angular momentum. Although the details of
where and how this feature occurs is model-dependent, it is generally true
that the contribution of LCWFs with one unit of 
orbital angular momentum falls linearly with $\bm{p}_{\sT}^2$ for $\bm{p}_{\sT}^2
\to 0$. This behavior is sharply different from the contribution of 
LCWFs with no orbital angular momentum. This simple example shows how 
the study of the
$\bm{p}_{\sT}^2$ dependence of unpolarized TMDs can therefore already expose
some effects due to orbital angular momentum.

Finally, we observe that in our model the average quark transverse momentum
decreases as $x$ increases, and that down quarks on
average carry less transverse momentum than up quarks.
Although this is just a
model result, a general message can be derived:  
the widely used assumption of
a flavor-independent quark transverse momentum distribution is already
falsified in a relatively simple model (see also Ref.~\cite{Mkrtchyan:2007sr}).

\subsection{Longitudinally polarized parton densities}

The model parameters of Tab.~\ref{tab:fit} produce the axial charge 
\begin{equation} 
g_A = \int_0^1 dx\, \left[ g_1^u(x) - g_1^d(x)\right] = 0.966 \pm 0.038 \; ,
\label{eq:ga}
\end{equation}
in excellent agreement with the value $0.969 \pm 0.096$ deduced from the GRSV 
parametrization~\cite{Gluck:2000dy}. 

It is, however,
evident from Fig.~\ref{fig:fit} that our description of the down quark
helicity distribution is in bad disagreement with the GRSV parametrization at
large $x$. Nevertheless, 
we point out that  there is a qualitative agreement with
the parametrization of the so-called BBS model of Ref.~\cite{Leader:1997kw} 
and the analogous parametrization of Ref.~\cite{Avakian:2007xa}. In
particular, our model shows the same feature highlighted in this latter
reference, namely that the contribution of the LCWFs with nonvanishing orbital
angular momentum is dominant at high $x$. This is true in all distribution
functions, but becomes particularly evident for the down helicity
distribution, since the contribution from the LCWFs $\psi^+_{++}$ and
$\psi^+_{+-}$ (carrying nonzero orbital angular momentum) are positive and
make the distribution positive at $x>0.5$.

\begin{figure}[h]
\begin{center}
\includegraphics[height=6cm]{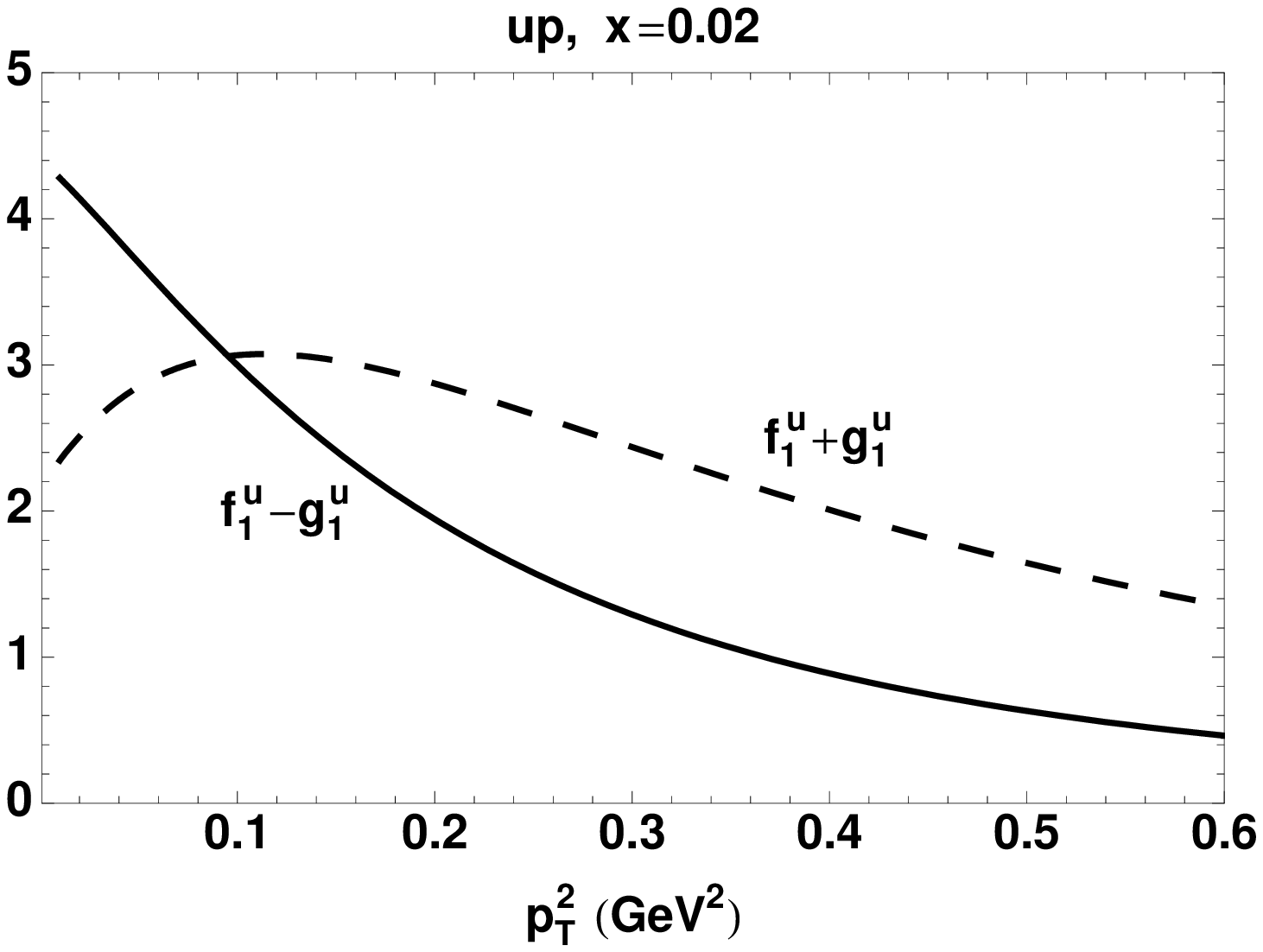} \hspace{0.2cm}
\includegraphics[height=6cm]{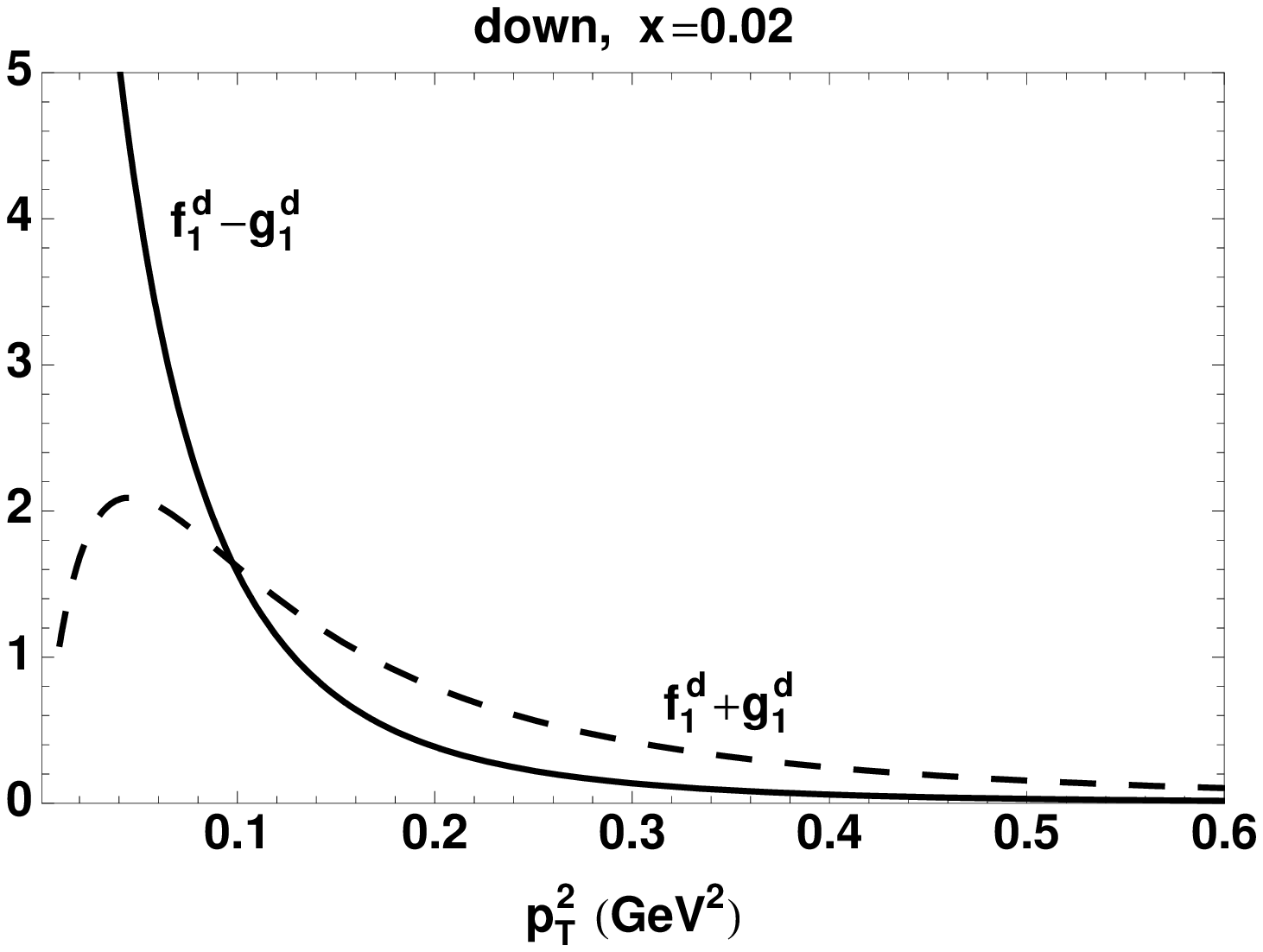}
\end{center}
\caption{The $\bm{p}_{\sT}^2$ dependence of the  distributions
$f_1(x,\bm{p}_{\sT}^2)-g_1(x,\bm{p}_{\sT}^2)$  (solid line) and 
  $f_1(x,\bm{p}_{\sT}^2)+g_1(x,\bm{p}_{\sT}^2)$  (dashed line) 
for up (left panel) and down
quark (right panel), at $x=0.02$.
The difference in their behavior is due to the different role played in the
two combinations by 
wavefunctions with nonzero orbital angular
momentum.} 
\label{fig:f1g1_pt}
\end{figure}

The effect of orbital angular momentum becomes even more evident when
considering the $\bm{p}_{\sT}^2$ behavior of the helicity distribution
function. As an illustration, we show in Fig.~\ref{fig:f1g1_pt} the behavior
of the combinations 
$f_1(x,\bm{p}_{\sT}^2)-g_1(x,\bm{p}_{\sT}^2)$  and 
$f_1(x,\bm{p}_{\sT}^2)+g_1(x,\bm{p}_{\sT}^2)$. In the case of the scalar diquark,
LCWFs with one unit of
orbital angular momentum are filtered by the first combination. 
In the case of the vector diquark, the situation
is opposite. The down quark distribution is entirely given by the vector
diquark, therefore the $f_1(x,\bm{p}_{\sT}^2)+g_1(x,\bm{p}_{\sT}^2)$ sum clearly 
turns down to
zero for $\bm{p}_{\sT}^2\to 0 $.
For the up quark, the situation is less clear due to the simultaneous presence
of 
scalar and vector diquark contributions. However, at $x=0.02$ 
the vector diquark is responsible for the nontrivial shape of 
$f_1(x,\bm{p}_{\sT}^2)+g_1(x,\bm{p}_{\sT}^2)$.

\begin{figure}[h]
\begin{center}
\includegraphics[height=6cm]{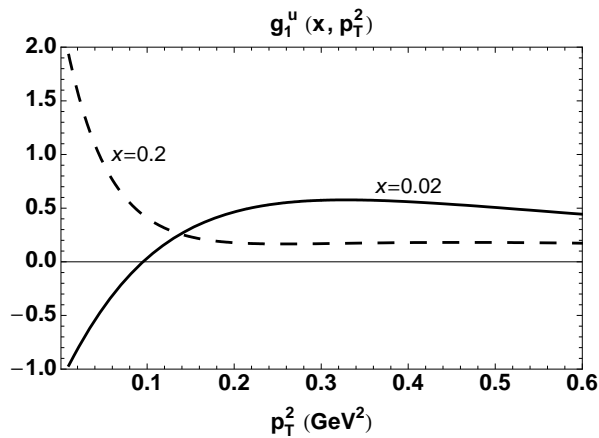} 
\end{center}
\caption{The $\bm{p}_{\sT}^2$ dependence of the 
helicity distribution $g_1^u(x,\bm{p}_{\sT}^2)$. Different lines correspond to 
different values of $x$. 
} 
\label{fig:g1_pt}
\end{figure}

It is interesting also to investigate the $\bm{p}_{\sT}^2$ behavior of $g_1^u$
alone, shown in Fig.~\ref{fig:g1_pt}. There is a dramatic
change of behavior for different values of $x$, due to the difference between 
the scalar and vector diquark components of the function. If
the spectator is a scalar diquark, for
${\bm p}_{\sT}=0$, where the LCWFs with orbital angular momentum vanish, the
spin of the up quark has to be parallel to that of the proton, 
 thus $g_1^{u(s)}(x,0) \geq 0$. At high transverse
momentum, where LCWFs with $L_z=1$ dominate, the spin of the up quark has to
be antiparallel to that of the proton,  thus $g_1^{u(s)}(x,\infty)
\leq 0$. The situation is exactly reversed in the case of the vector diquark.  
As is already visible in
Eqs.~\eqref{eq:g1Ls} and \eqref{eq:g1La}, at high transverse momentum
the vector diquark always dominates and gives a positive result. At low
transverse momentum, the relative size of the functions $L_X^2(\Lambda_X^2)$ in the 
denominator determines which contribution is dominant. At higher $x$ the
scalar diquark dominates and gives a positive $g_1^u(x,0)$, while at lower $x$
the 
vector diquark dominates and gives a negative $g_1^u(x,0)$.   

Once again, apart from the details specific to our model, these examples show
that the exploration of the $\bm{p}_{\sT}^2$ dependence of the unpolarized and
helicity distribution functions 
can expose very interesting features of the inner
structure of the nucleon, related in particular to orbital angular momentum.


\begin{figure}[h]
\begin{center}
\includegraphics[height=6cm]{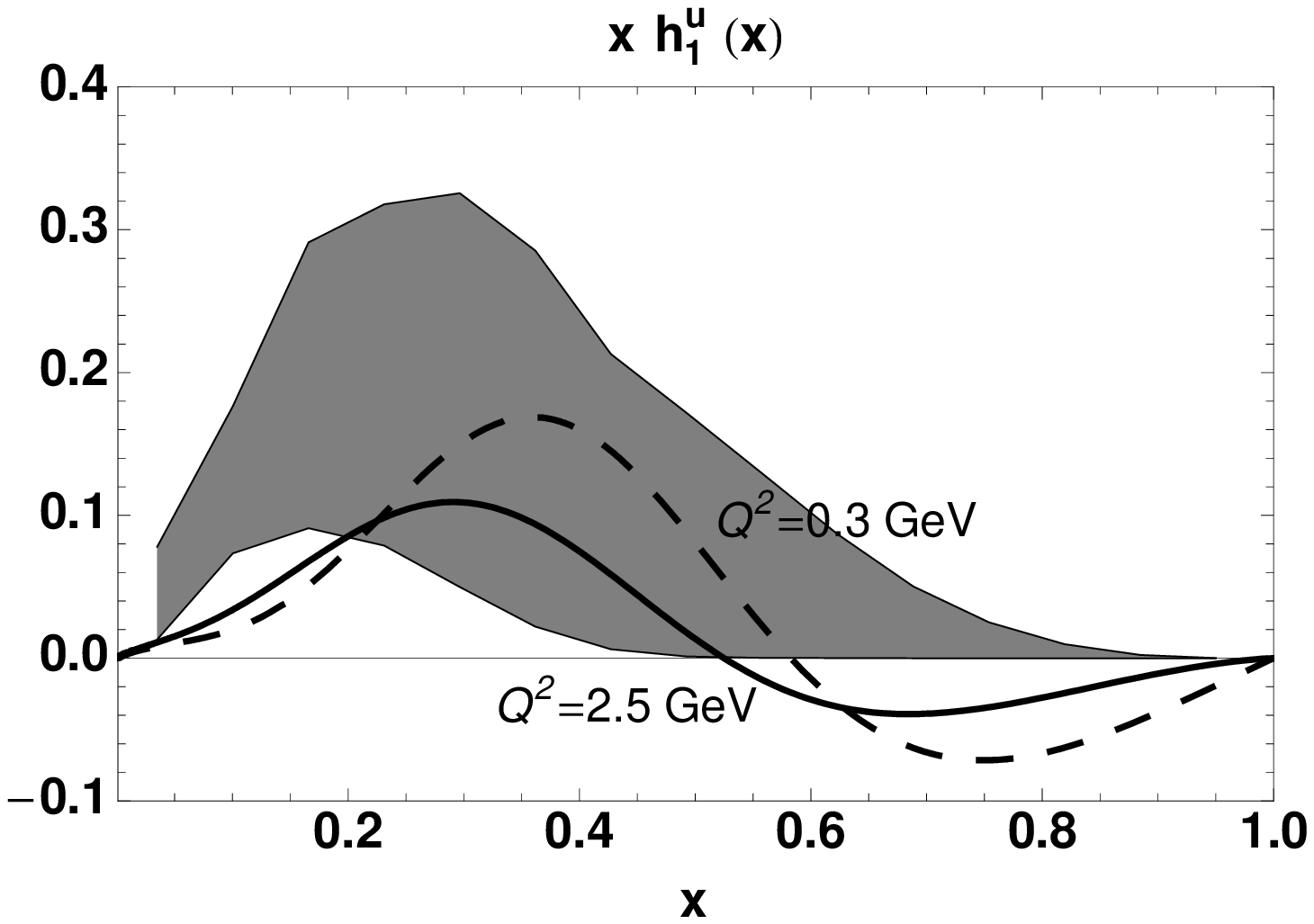} \hspace{0.2cm}
\includegraphics[height=6cm]{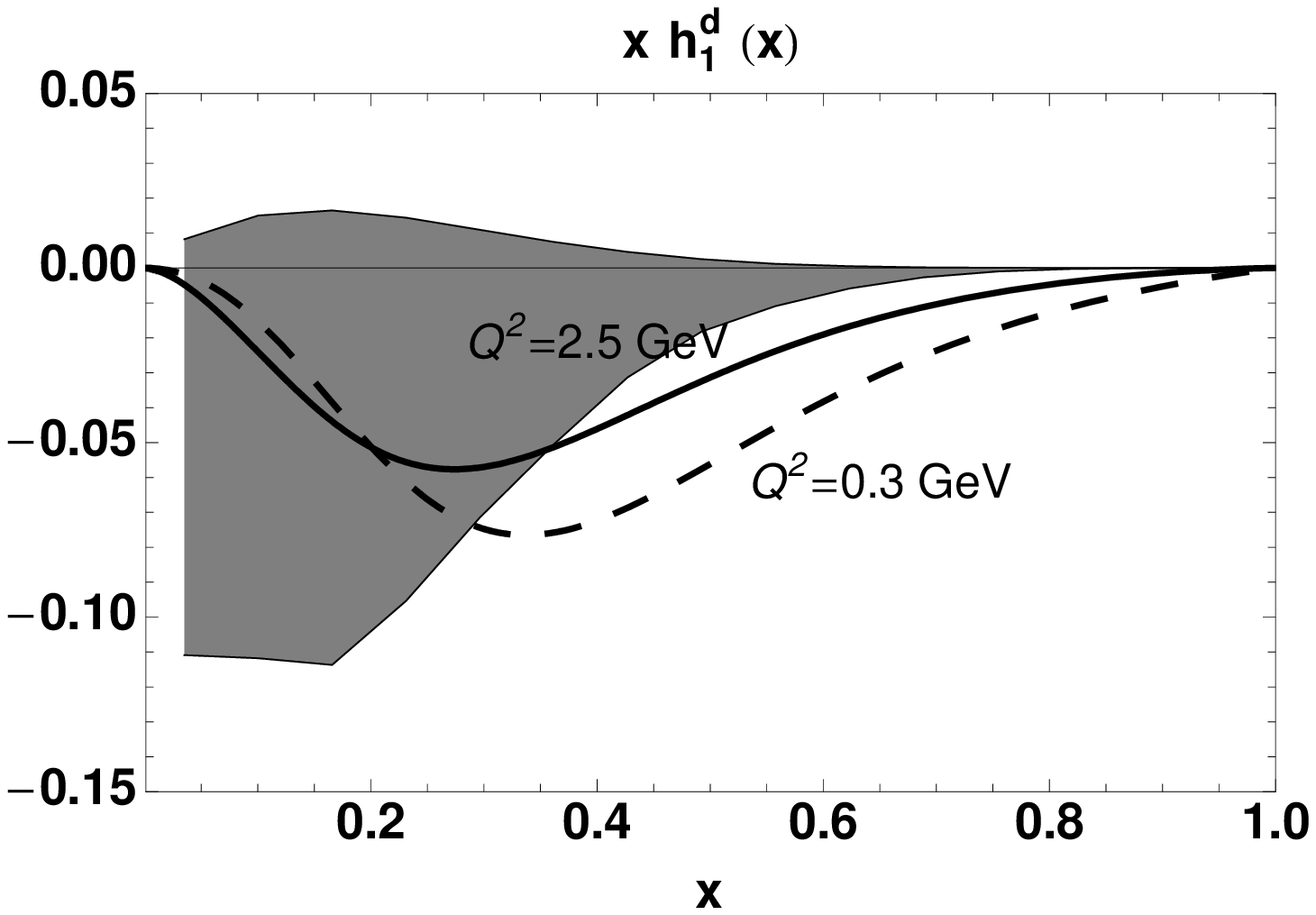}
\end{center}
\caption{The transversity distribution $x h_1(x)$ for up (left panel) and down
quark (right panel). Dashed (solid) line for the model result before (after) the
evolution at LO using the code of Ref.~\protect{\cite{Hirai:1997mm}} up to the
scale of the parametrization from Ref.~\protect{\cite{Anselmino:2007fs}}, whose 
uncertainty band due to errors in the fit parameter is represented by the shaded
area.}
\label{fig:h1x}
\end{figure}

\subsection{Transversity}
\label{sec:h1}

In Fig.~\ref{fig:h1x}, the predictions of the spectator diquark model for the
transversity distribution are compared with the only available parametrization 
of Ref.~\cite{Anselmino:2007fs}. In the left panel, 
$x h_1^u(x)$ is shown,
whereas $x h_1^d(x)$ is shown in the right panel. 
All 
the model results at the assumed original scale $Q_0^2=0.3$ GeV$^2$ are 
represented by the dashed line. The solid line indicates the result after 
applying the DGLAP evolution at LO up to the scale $Q^2=2.5$ GeV$^2$ using the 
code from Ref.~\cite{Hirai:1997mm}. The latter scale pertains the 
parametrization of Ref.~\cite{Anselmino:2007fs}, whose errors in the fit 
parameters produce the uncertainty band represented by the shaded area. The 
model is in reasonable agreement with the parametrization, with the maxima in 
the correct position and a somewhat too small result for the up quark at small 
$x$. It should also be kept in mind that the present data reach at most 
$x \approx 0.4$~\cite{Airapetian:2004tw,Ageev:2006da} and, moreover, the ansatz 
of Ref.~\cite{Anselmino:2007fs} does not allow for a sign change.\footnote{We
  point out that new fits of the transversity distribution functions have 
  been presented at some
  conferences~\protect{\cite{Anselmino:2008sj}} but
  not published yet.}

Interestingly, for the up quark 
the model predicts a change of sign at $x\sim 0.5$. To our knowledge, no other
model of transversity displays this feature (see Ref.~\cite{Barone:2001sp} and
references therein; see also recent calculations in
Refs.~\cite{Pasquini:2005dk,Wakamatsu:2007nc,Cloet:2007em}).  
The reason for this sign change is that the
contribution of the vector diquark is negative, as evident from
Eq.\eqref{eq:h1xspecta}. In our model, at moderate $x$ the
scalar diquark contribution is dominant, whereas at sufficiently high $x$
the contribution of the vector diquark becomes in absolute size bigger, 
thus leading to the sign change. 
Other versions of the diquark model, even with
vector diquarks, may not show this property. This is already evident from
inspecting the results (listed in the appendices) for
different choices of the diquark polarization sum.
We don't think that our model calculation should be trusted more than
others. Nevertheless, it might be interesting to contemplate the possibility
of a sign change when choosing a form for the parametrization of the
transversity function in ``global fits.'' 

\begin{figure}[h]
\begin{center}
\includegraphics[height=6cm]{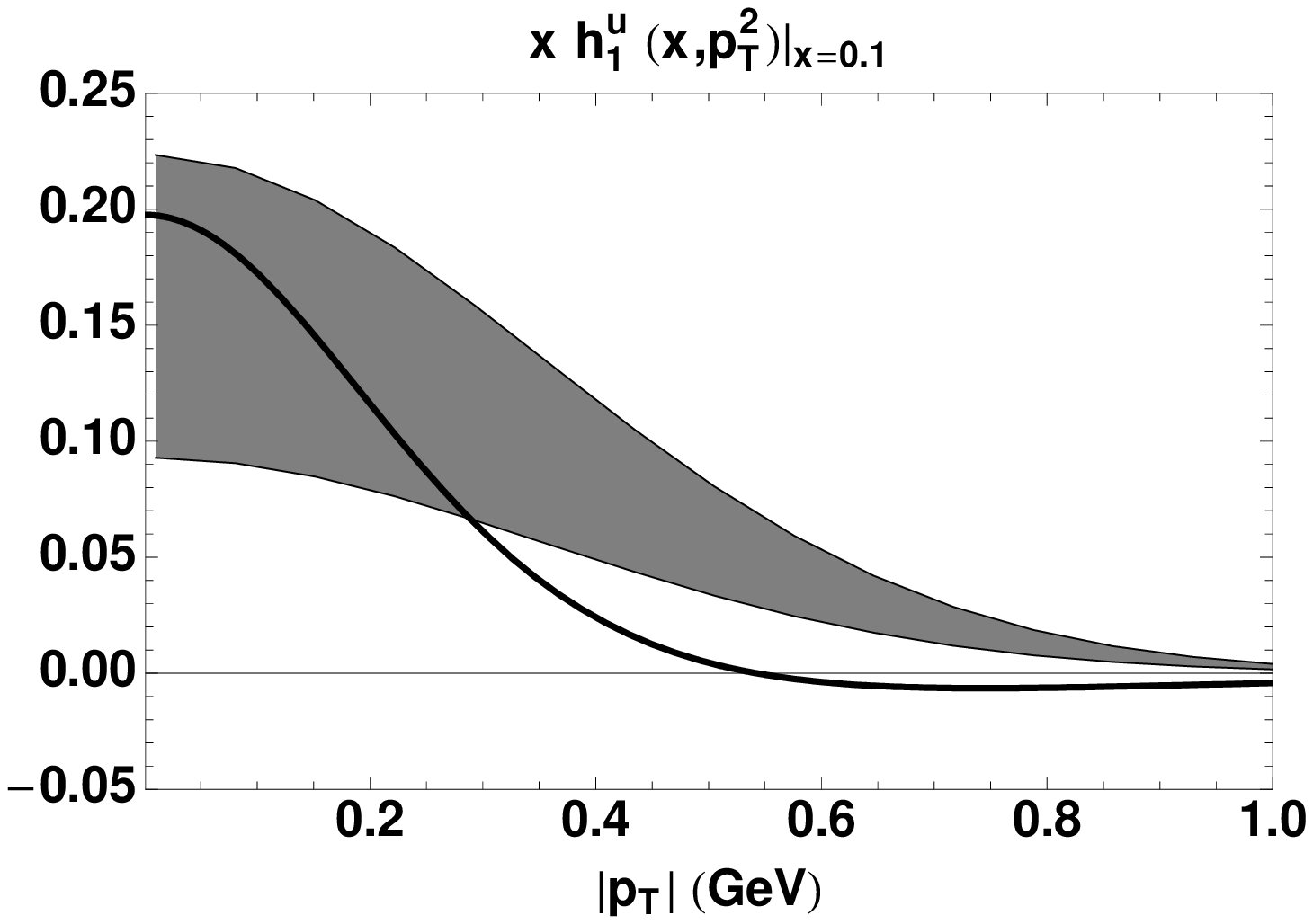} \hspace{0.2cm}
\includegraphics[height=6cm]{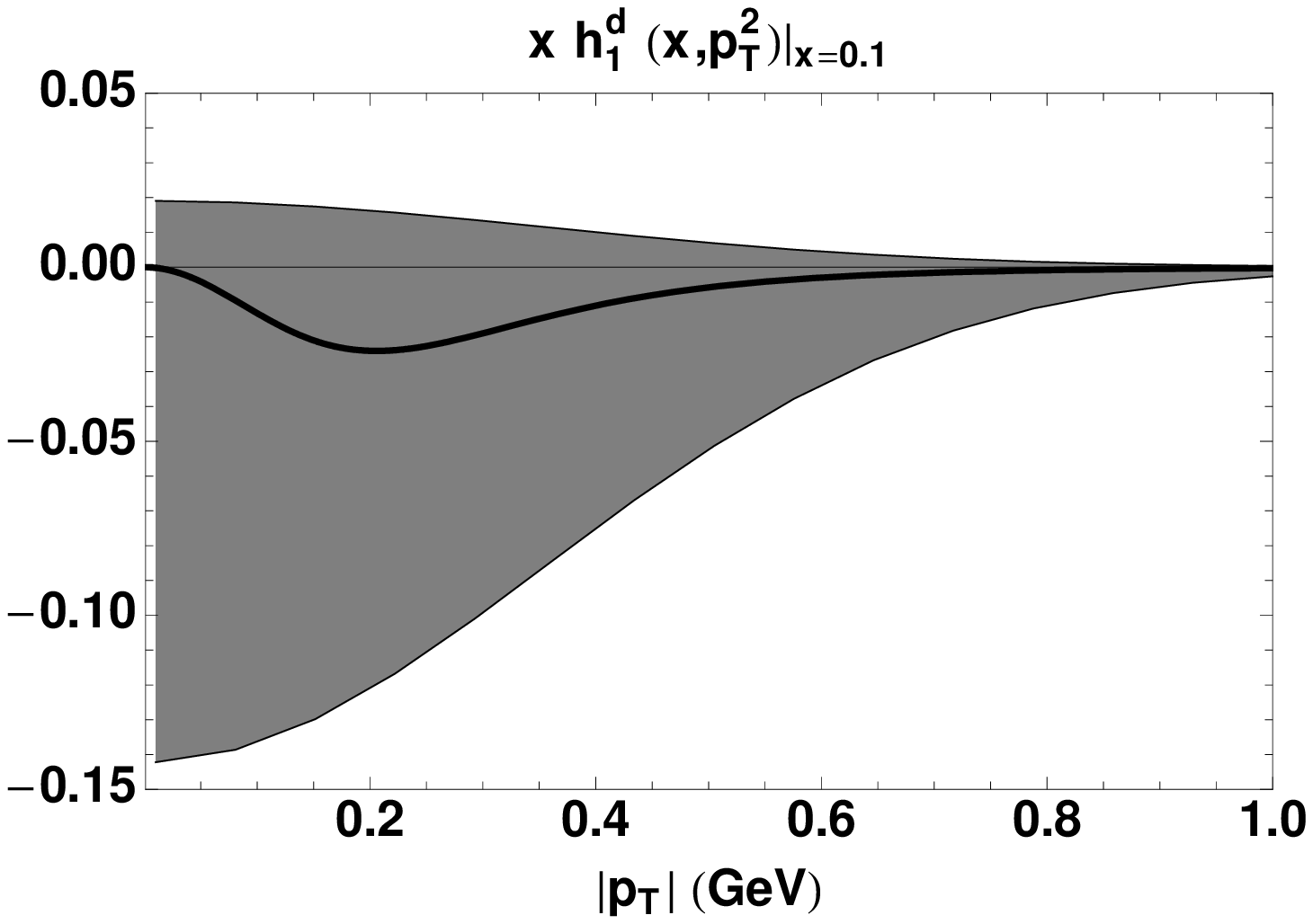}
\end{center}
\caption{Same as in the previous figure, but for the ${\bm p}_{\sT}$ dependence 
of transversity at $x=0.1$.}
\label{fig:h1pt}
\end{figure}

In Fig.~\ref{fig:h1pt}, the same comparison is performed as in the previous 
figure, but for the ${\bm p}_{\sT}$ dependence of the transversity at $x=0.1$, 
as it is deduced from Eqs.~(\ref{eq:h1spects},\ref{eq:h1spect}). Again, there is a 
reasonable agreement between model predictions and parametrizations but for the 
trend of the result for the up quark at $|{\bm p}_{\sT}| > 0.3$ GeV/$c$.  
However, we stress that the comparison may be affected by the different 
scale of the model results ($Q^2=0.3$ GeV$^2$) and the one at which the 
parametrizations are extracted ($Q^2=2.5$ GeV$^2$). The proper evolution of
the TMDs has not been considered yet.
It is interesting to point out  that in our model $h_1^u(x,{\bm p}_{\sT}^2)$
changes sign at
${\bm p}_{\sT}\sim 0.5$ GeV. This is due to the fact that the vector diquark
contribution is always negative and dominant at high ${\bm p}_{\sT}$. For
the down quark, we note that $h_1^d(x,0)=0$, because the vector-diquark
contribution to $h_1$ is entirely given by LCWFs with nonvanishing orbital
angular momentum.  

\begin{figure}[h]
\begin{center}
\includegraphics[height=6cm]{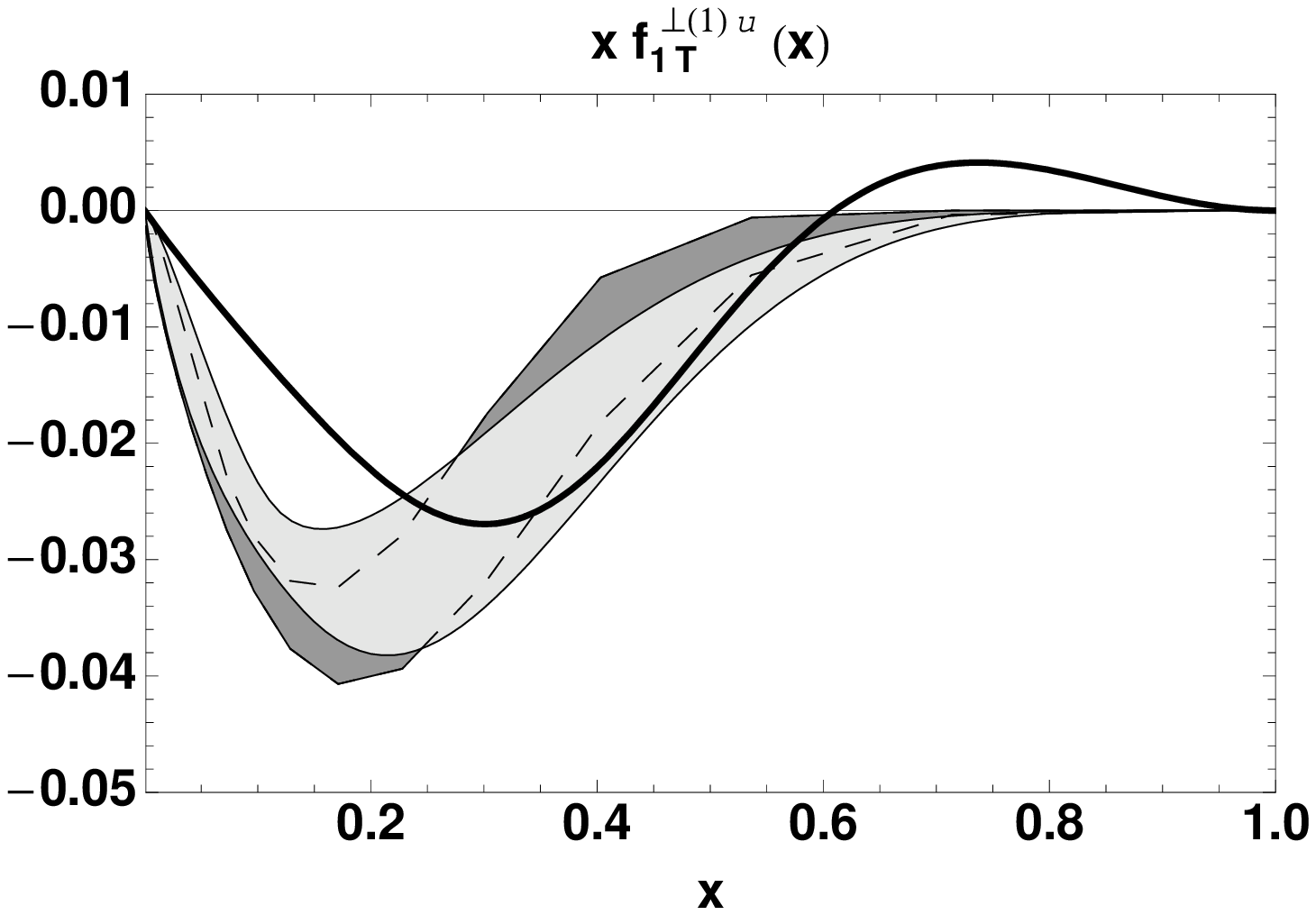}
\hspace{0.2cm}
\includegraphics[height=6cm]{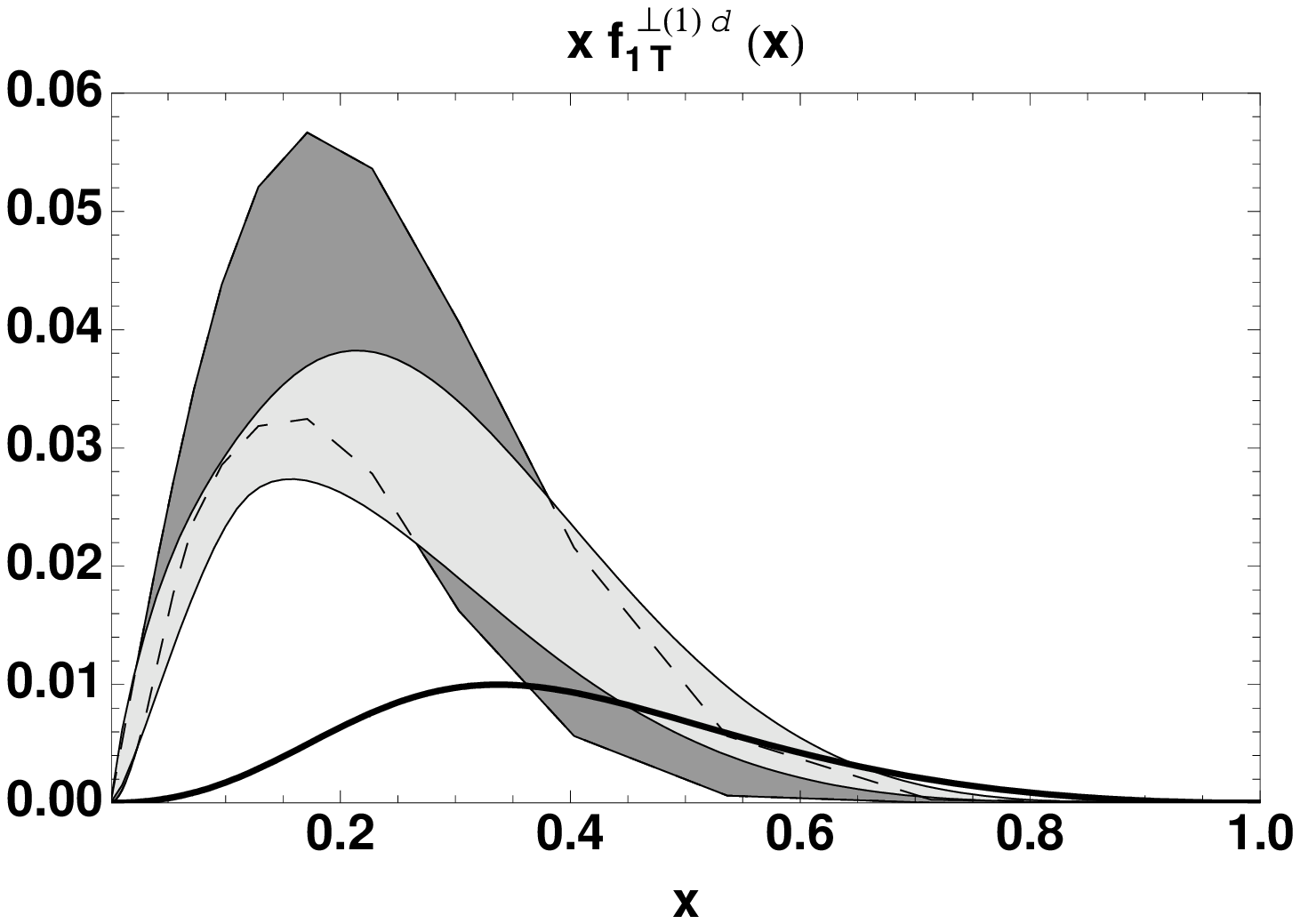} 
\end{center}
\caption{The first ${\bm p}_{\sT}$-moment $x f_{1T}^{\perp\, (1)}(x)$ of the 
Sivers function; left (right) panel for up (down) quark. Solid line for the   
results of the spectator diquark model. Darker shaded area for the uncertainty 
band due to the statistical error of the quark parametrizations from 
Ref.~\protect{\cite{Anselmino:2008sg}}, lighter one from 
Ref.~\protect{\cite{Collins:2005wb}}.}
\label{fig:sivers1}
\end{figure}


\subsection{Sivers function}
\label{sec:Toddresults}

In Fig.~\ref{fig:sivers1}, the $x f_{1T}^{\perp\,(1)}(x)$ moment of the 
Sivers function, predicted using 
Eqs.~(\ref{eq:Sivers1s}) and (\ref{eq:Sivers1a}), is given by the 
solid line and it is compared with two different parametrizations of the same 
observable. The darker shaded area represents the uncertainty due to the 
statistical errors in the parametrization of Ref.~\cite{Anselmino:2008sg}, 
while the lighter shaded area corresponds to the same for 
Ref.~\cite{Collins:2005wb}. Left panel refers to the up quark, right panel to 
down quark. First of all, we observe the agreement between the signs of the 
various flavor components, which also agree with the findings from calculations 
on the lattice~\cite{Gockeler:2006zu}. Also the maxima are reached at 
approximately the same $x\sim 0.3$ as the parametrizations. Instead, 
the ``strength'' of the asymmetry (related to the modulus of each moment) is 
too much weaker for the down quark, while it seems reasonable for the up one. 
Again, it must be stressed that no evolution was applied in the displayed 
model results.

\begin{figure}[h]
\begin{center}
\includegraphics[width=8cm]{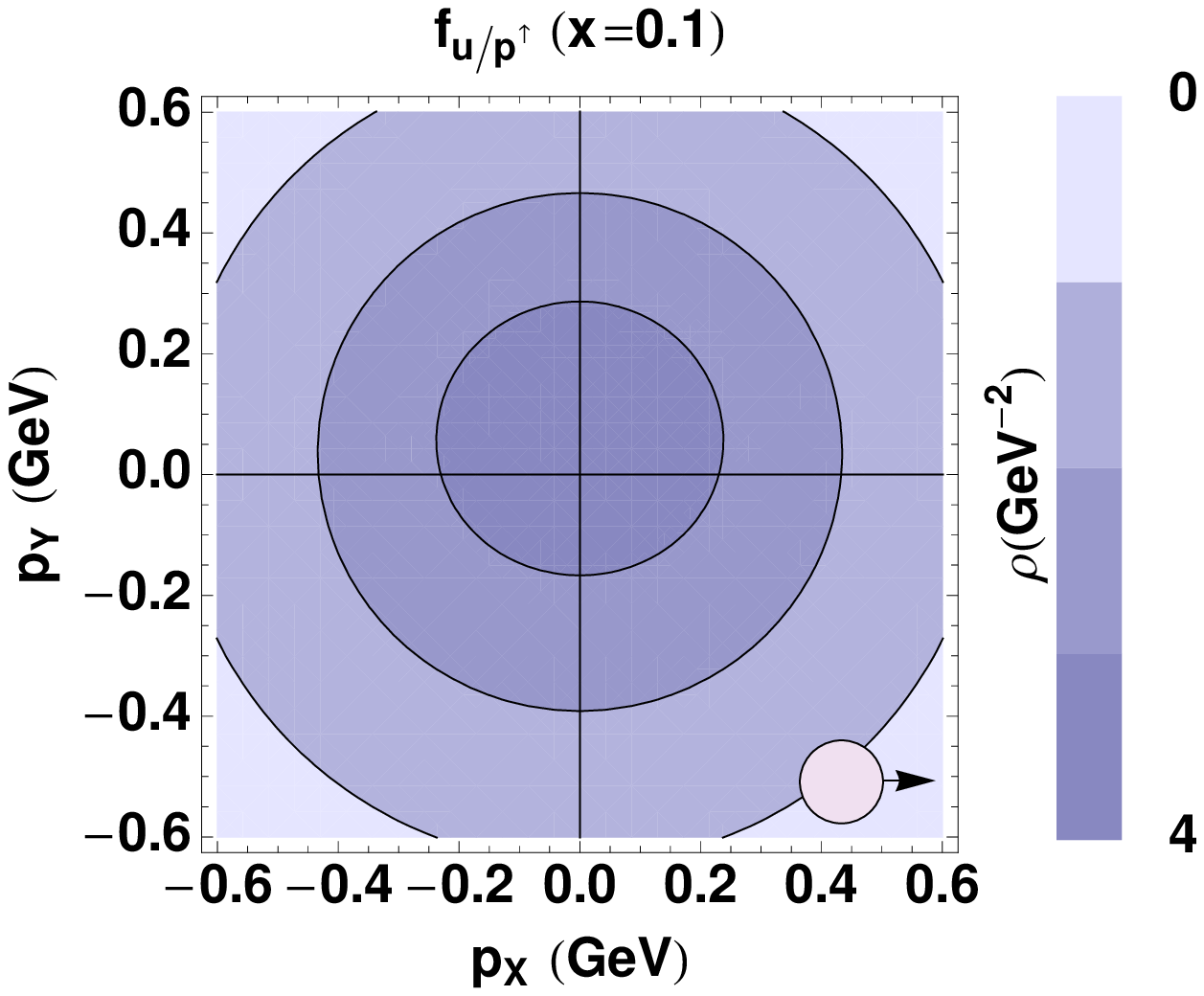} 
\hspace{0.2cm}
\includegraphics[width=8cm]{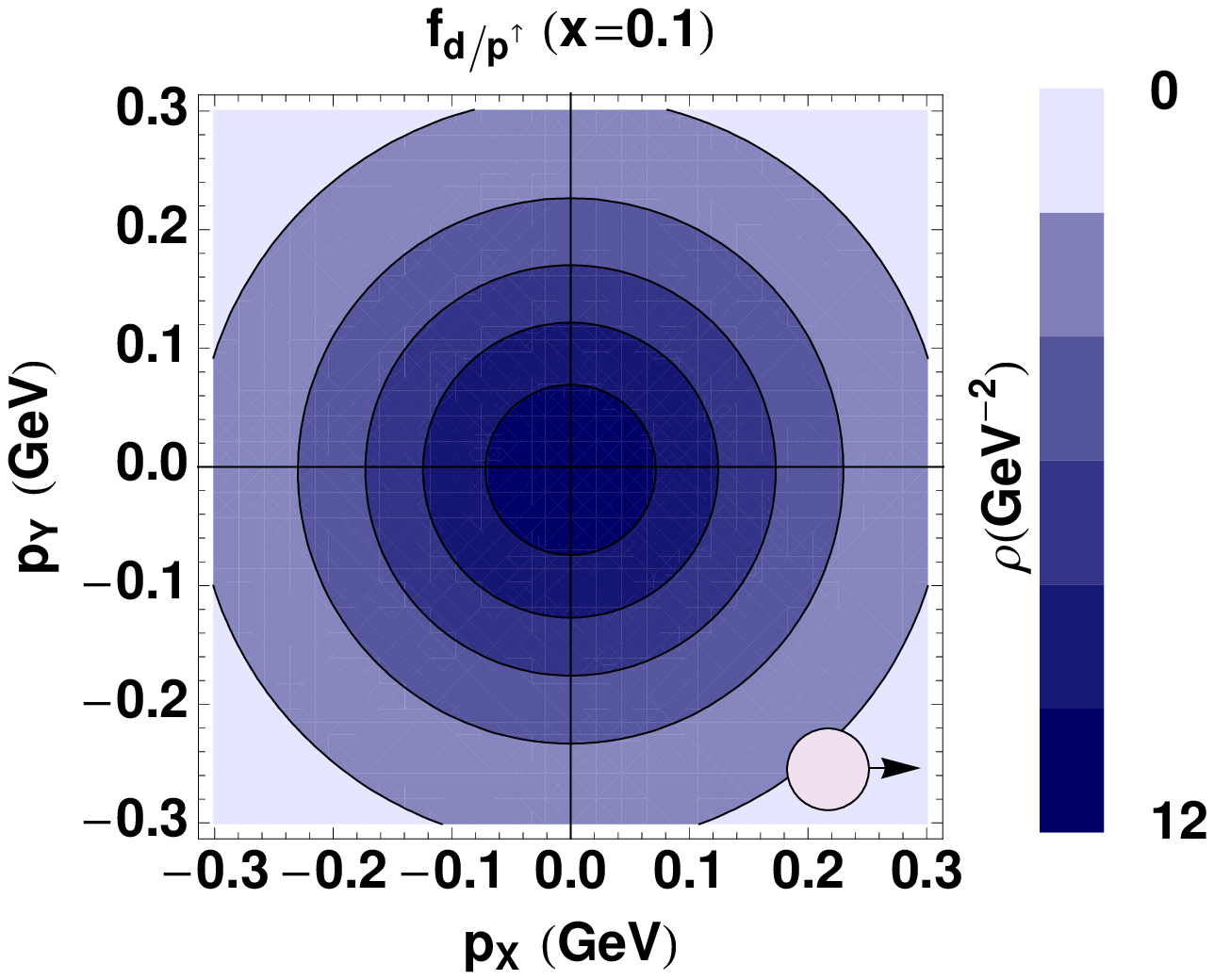} 
\end{center}
\caption{The model result for the spin density of unpolarized quarks in 
transversely polarized protons (see text for the precise definition) in 
$\bm{p}_{\sT}$ space at $x=0.1$. Left panel for up quark, right panel 
for down quark. The circle with the arrow indicates the direction of the
proton polarization.}
\label{fig:plotSiv}
\end{figure}

According to the Trento conventions~\cite{Bacchetta:2004jz}, we define the spin 
density of unpolarized quarks with flavor $q$ in transversely polarized 
protons as
\begin{equation} 
f_{q/p^\uparrow}(x,\bm{p}_{\sT}) = f_1^q(x,\bm{p}_{\sT}^2) - 
f_{1T}^{\perp\,q}(x,\bm{p}_{\sT}^2)\, 
\frac{(\hat{\bm{P}}\times \bm{p}_{\sT})\,\cdot\,\bm{S}}{M} \; .
\label{eq:spindens}
\end{equation}

In a SIDIS experiment, typically $\hat{\bm{P}}$ is antialigned to the 
$\hat{\bm{z}}$ axis that points in the direction of the momentum transfer 
$\bm{q}$. Hence, if the proton polarization is chosen along the $\hat{\bm{x}}$ 
axis, the spin density~(\ref{eq:spindens}) shows an asymmetry in momentum space 
along the $p_y$ direction, whose size is driven by the Sivers function. In 
Fig.~\ref{fig:plotSiv}, we show $f_{q/p^\uparrow}(0.1,\bm{p}_{\sT})$ for $q=u$ 
(left panel) and $q=d$ (right panel). Since the Sivers function for the up 
(down) quark is negative (positive), the density is deformed towards positive
(negative) values of $p_y$. This feature is in agreement with the lattice 
results of Ref.~\cite{Gockeler:2006zu} and with the signs of the anomalous
magnetic moments $\kappa^q$~\cite{Burkardt:2002ks}.



\subsection{Boer-Mulders function}

\begin{figure}[h]
\begin{center}
\includegraphics[width=7.5cm]{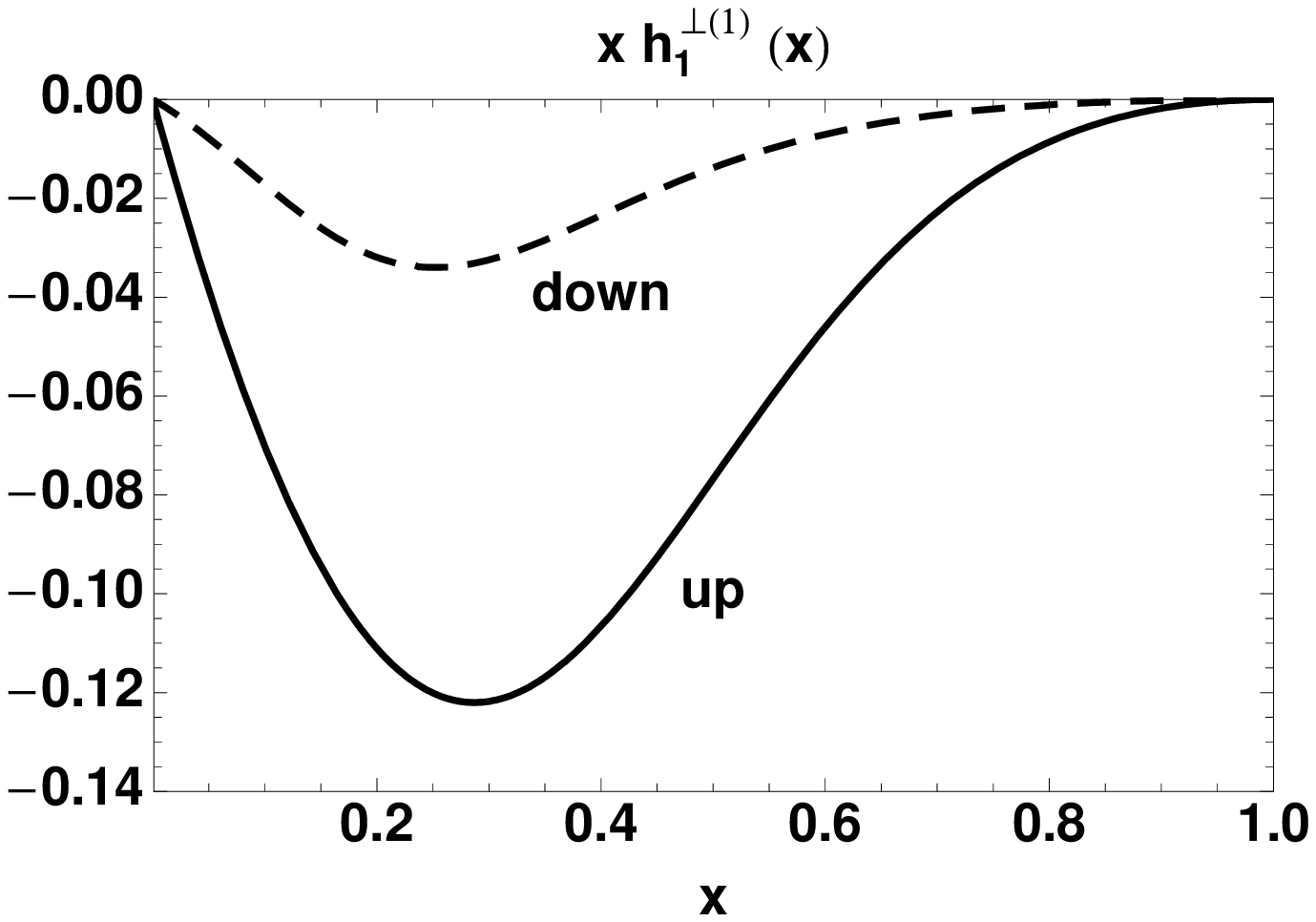} 
\hspace{0.5cm}
\includegraphics[width=7.5cm]{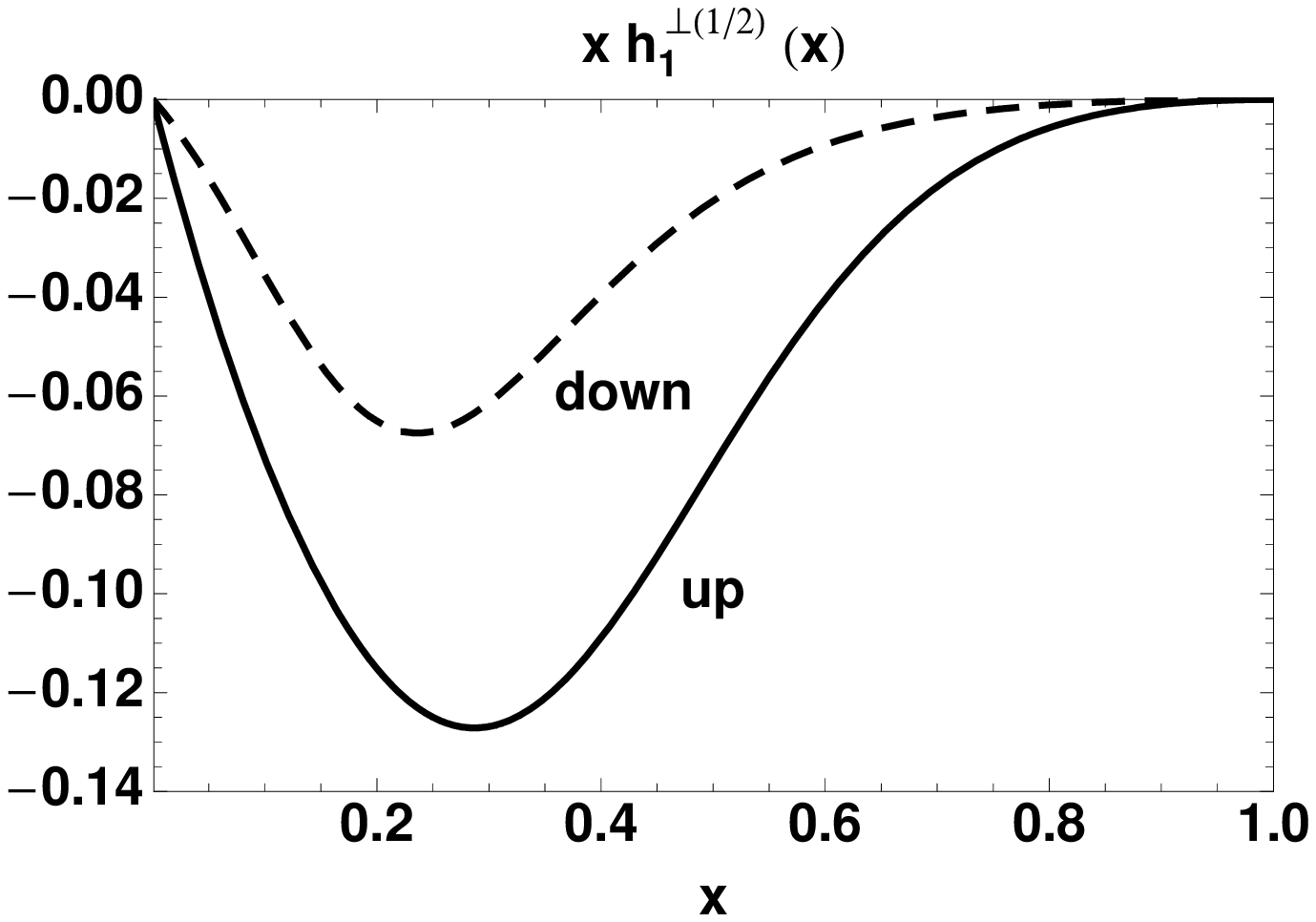} 
\end{center}
\caption{The $x h_1^{\perp\, (1)}(x)$ (left) and $x h_1^{\perp\, (1/2)}(x)$ 
(right) moments of the Boer-Mulders function. Solid and dashed lines for up and
down quarks, respectively.}
\label{fig:BM}
\end{figure}

In Fig.~\ref{fig:BM}, the $x h_1^{\perp\,(1)}(x)$ and $x h_1^{\perp\,(1/2)}(x)$ 
moments of the Boer-Mulders function, as deduced from 
Eqs.~(\ref{eq:BoerMulders1s},\ref{eq:BoerMulders1a}) and 
(\ref{eq:BoerMulders1-2s},\ref{eq:BoerMulders1-2a}), 
are displayed in the left and right panel, 
respectively. The solid lines correspond to the results for the up quark; 
dashed lines for the down quark. 
For the Boer-Mulders function, the only available parametrization appeared
recently in \cite{Zhang:2008nu}, but the overall normalization depends on a
parameter $\omega$ that cannot be fixed with available experimental
information. Our result agrees in sign and shape with that extraction. The
absolute values of our functions correspond to $\omega \approx 0.3$.
We remark that 
there is full agreement between the sign of the $u$ and $d$ components and the 
aforementioned lattice calculations~\cite{Gockeler:2006zu}, as observed also in
a different version of the spectator model~\cite{Gamberg:2007wm} and in the
bag model~\cite{Yuan:2003wk}. This
agreement seems to be a general feature, as argued in
Ref.~\cite{Burkardt:2007xm}.~\footnote{A different result for the sign of the
  down quark Boer-Mulders function was obtained in
  Ref.~\cite{Bacchetta:2003rz},
  probably due to a mistake in that calculation (see
  App.~\ref{sec:toddtimelike}).}    

\begin{figure}[h]
\begin{center}
\includegraphics[width=8cm]{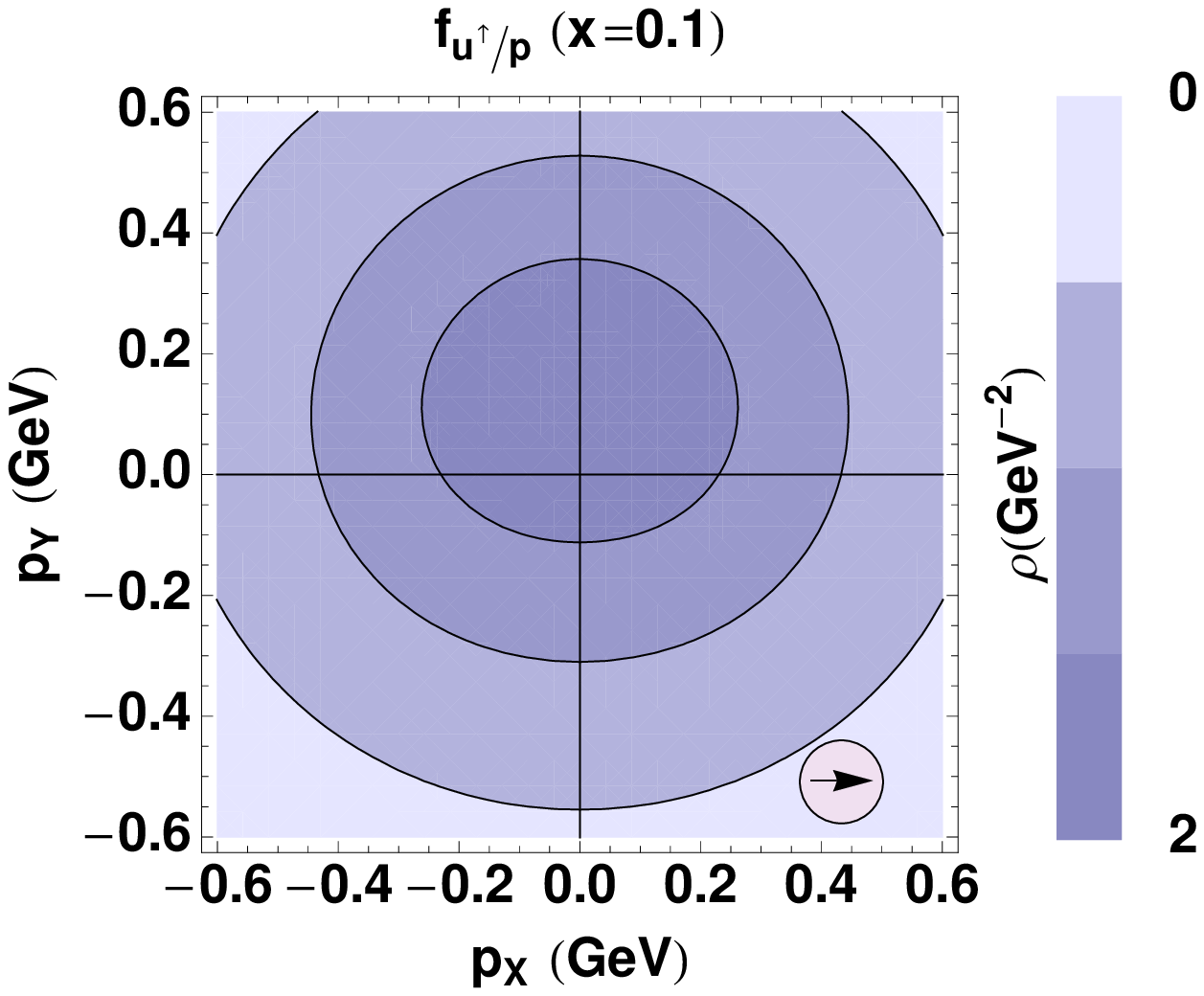} 
\hspace{0.2cm}
\includegraphics[width=8cm]{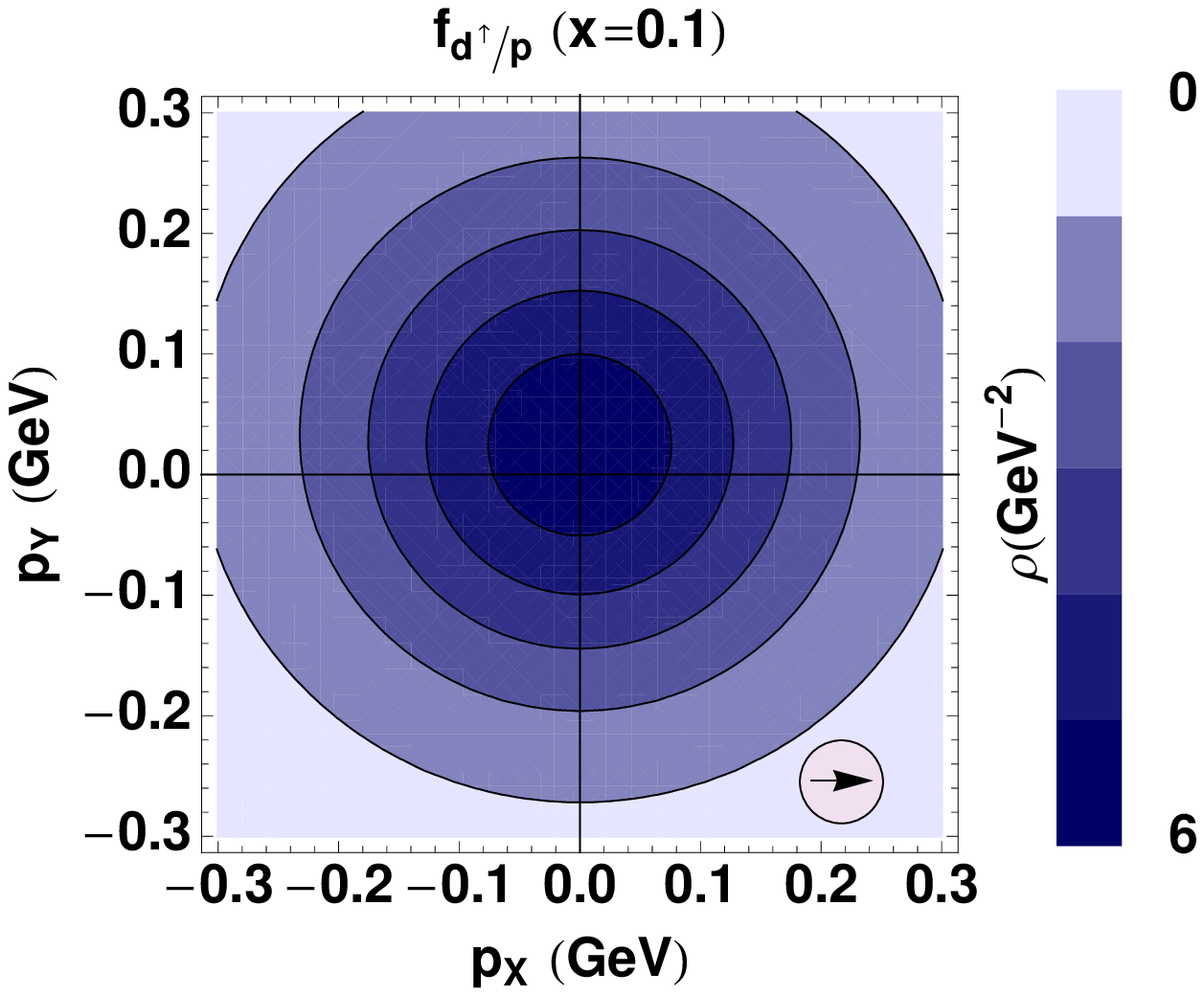} 
\end{center}
\caption{The model result for the spin density of transversely polarized quarks 
in unpolarized protons (see text for the precise definition) in 
$\bm{p}_{\sT}$ space at $x=0.1$. Left panel for up quark, right panel 
for down quark. The arrow inside the circle indicates the direction of the
quark polarization.}
\label{fig:plotBM}
\end{figure}

In Fig.~\ref{fig:plotBM}, we show, again at $x=0.1$, the spin density of 
transversely polarized quarks with flavor $q$ in unpolarized protons, related
to  the Boer-Mulders effect by~\cite{Bacchetta:2004jz} 
\begin{equation} 
f_{q^\uparrow/p}(x,\bm{p}_{\sT}) = \frac{1}{2}\, \left[ 
f_1^q(x,\bm{p}_{\sT}^2) - h_1^{\perp\,q}(x,\bm{p}_{\sT}^2) \, 
\frac{(\hat{\bm{P}}\times \bm{p}_{\sT})\,\cdot\,\bm{S}_q}{M} 
\right] \; , 
\label{eq:spindens2}
\end{equation}
where now the quark polarization $\bm{S}_q$ points along $\hat{\bm{x}}$. 
Since the Boer-Mulders function is negative for both flavors (see 
Fig.~\ref{fig:BM}), the related spin density is always deformed towards 
positive $p_y$, again in agreement with the lattice 
results~\cite{Gockeler:2006zu}.

\section{Conclusions}
\label{sec:end}

We have presented a systematic calculation of all leading-twist parton 
distributions in the nucleon in a diquark spectator model. 
We have generated the relative phase necessary to 
produce T-odd structures by approximating the gauge link operator with a one 
gluon-exchange interaction. All results have been presented in analytic form and
interpreted in terms  
of overlaps of light-cone wavefunctions. 

We tried to extend and improve the spectator model
calculations presented in 
Refs.~\cite{Jakob:1997wg,Bacchetta:2003rz,Gamberg:2007wm} 
by considering several choices of
the 
axial-vector diquark polarization states and of the nucleon-quark-diquark
form factor. We listed the analytic expressions for all possible choices
in the appendices. We critically reconsidered some of the limits of the model
and the choice of model parameters 
used in the past literature. In particular, we showed that the spectator diquark 
model is not able to reproduce both the quark number and momentum sum rule at the 
same time, because the diquark is considered as a charged parton, hence active in 
the sum rules. We argued 
that the proton wave function does not show the usual SU(4)=SU(2)$\otimes$SU(2) 
symmetry~\cite{Jakob:1997wg}, since 
the quark-diquark system in its ground state 
can have a nonvanishing relative orbital angular momentum. 

For numerical studies, we chose the version of the model that in our opinion
is more sensible and practical, i.e., the one where only light-cone transverse
polarizations of the diquark are present and a dipolar form factor is used.
We identified nine free parameters of the model and we 
fixed them by reproducing the phenomenological 
parametrization of unpolarized~\cite{Chekanov:2002pv} and longitudinally 
polarized~\cite{Gluck:2000dy} parton distributions at the lowest available 
scale, i.e.\ $Q^2=0.3$ and $0.26$ GeV$^2$, respectively. 

Whenever possible, results have been compared with available parametrizations. 
For the chiral-odd transversity distribution, there is only one available 
from Ref.~\cite{Anselmino:2007fs}, which was deduced from SSA data at $Q^2=2.5$ 
GeV$^2$. The $\bm{p}_{\sT}$-integrated model result, once evolved to this scale 
using the code from Ref.~\cite{Hirai:1997mm} at LO, displays a satisfactory 
overall agreement. 
The $f_{1T}^{\perp\,(1)}(x)$ moment of the chiral-even T-odd Sivers function 
$f_{1T}^{\perp}$ was compared with the parametrizations of 
Refs.~\cite{Anselmino:2008sg,Collins:2005wb}. There is 
agreement between the signs of the various flavor components and between the 
positions of the maxima in $x$, but the absolute value of the function is
somewhat too small for the down quark. 
The comparison is affected by the 
difference of the scales, since evolution equations for the Sivers function
have not been used. We also plotted the $h_1^{\perp\,(1)}(x)$ and 
$h_1^{\perp\,(1/2)}(x)$ moments of the chiral-odd T-odd Boer-Mulders function 
$h_1^{\perp}$.
We have also shown the quark spin densities defined in the Trento
conventions~\cite{Bacchetta:2004jz}, as produced in turn by the Sivers or 
the Boer-Mulders effects. For unpolarized quarks in transversely polarized 
protons, the spin density $f_{q/p^\uparrow}$ is linked to $f_{1T}^{\perp}$, 
while for transversely polarized quarks in unpolarized protons the  
$f_{q^\uparrow/p}$ is linked to $h_1^{\perp}$. For transverse polarizations 
along the $\hat{\bm{x}}$ axis, the contour plot in the quark momentum space of 
such densities at $x=0.1$ displays a distortion in the $p_y$ direction,
whose sign is consistent with the lattice findings for the corresponding spin 
densities in impact parameter space~\cite{Gockeler:2006zu}. 

Using the model parton densities discussed above, various spin, 
beam, and azimuthal asymmetries in semi-inclusive hadronic reactions can be 
predicted, which are of interest for several experimental collaborations. 
Model calculations can be useful to interpret experimental measurements,
helping us to explore spin-orbit 
parton correlations inside hadrons and shed light on the 
well-known puzzle of the proton spin sum rule.


\section*{Acknowledgments}

F.~C. and M.~R. would like to thank B.~Pasquini for useful discussions. 

This work is part of the European Integrated Infrastructure Initiative in
Hadronic Physics project under Contract No. RII3-CT-2004-506078.

Authored by Jefferson Science Associates, LLC under U.S. DOE Contract No. 
DE-AC05-06OR23177. The U.S. Government retains a non-exclusive, paid-up, 
irrevocable, world-wide license to publish or reproduce this manuscript for 
U.S. Government purposes.


\appendix

\section{T-even functions in different variations of the model}
\label{sec:A}

In this Appendix we list the leading-twist T-even parton densities obtained in
the context of our spectator diquark model, for all the choices of
axial-vector diquark polarization sum and
nucleon-quark-diquark vertex. To avoid overloading the notation, we will use
the same ones for the parameters involved ($g_X$, $M_X$,
$\Lambda_X$). However, it must be kept in mind that the numerical value of
these parameters can be different in the various versions of the model.

\subsection{Scalar diquark}

The results for the scalar diquark are
\begin{align}
f_1^{q(s)}(x,{\bm p}_{\sT}^2) &= \frac{g_s^2}{(2\pi)^3}\,
\frac{[(m+xM)^2+{\bm p}_{\sT}^2]\,(1-x)}{2\,[{\bm p}_{\sT}^2+L_s^2(m^2)]^2}
\\[0.2cm]
g_{1L}^{q(s)}(x,{\bm p}_{\sT}^2) &= \frac{g_s^2}{(2\pi)^3}\,
\frac{[(m+xM)^2-{\bm p}_{\sT}^2]\,(1-x)}{2\,[{\bm p}_{\sT}^2+L_s^2(m^2)]^2}
\; ,  \\[0.2cm]
g_{1T}^{q(s)} (x,{\bm p}_{\sT}^2) &= \frac{g_s^2}{(2\pi)^3}
\, \frac{M\,(m+xM)\,(1-x)}{[{\bm p}_{\sT}^2 + L_s^2(m^2)]^2} \; ,
\\[0,2cm]
h_{1L}^{\perp\,q(s)}(x,{\bm p}_{\sT}^2) &= - g_{1T}^{q(s)} (x,{\bm p}_{\sT}^2) \; ,
\\[0.2cm]
h_{1T}^{q(s)}(x,{\bm p}_{\sT}^2) &= f_1^{q(s)} (x,{\bm p}_{\sT}^2) \; ,  \\[0.2cm]
h_{1T}^{\perp\,q(s)}(x,{\bm p}_{\sT}^2) &= -\frac{g_s^2}{(2\pi)^3}\,
\frac{M^2\,(1-x)}{[{\bm p}_{\sT}^2+L_s^2(m^2)]^2} \; ,
\label{eq:pTTeven-s-pl}
\end{align}
where we recall that $M$ is the nucleon mass and $m$ is the mass of the
parton.
From the latter two densities, we construct the contribution
of the scalar diquark to the transversity:
\begin{equation} \begin{split}
h_{1}^{q(s)}(x,{\bm p}_{\sT}^2) &= h_{1T}^{q(s)}(x,{\bm p}_{\sT}^2) +
\frac{{\bm p}_{\sT}^2}{2M^2}\, h_{1T}^{\perp\,q(s)}(x,{\bm p}_{\sT}^2)  \\
&= \frac{g_s^2}{(2\pi)^3}
\, \frac{(m+xM)^2\,(1-x)}{2\, [{\bm p}_{\sT}^2 + L_s^2(m^2)]^2}
=\frac{1}{2} \, \left( f_1^{q(s)}(x,{\bm p}_{\sT}^2) + g_1^{q(s)}(x,{\bm
    p}_{\sT}^2) \right)
\; .
\label{eq:pTh1s-pl}
\end{split} \end{equation}

The above results are valid for a point-like nucleon-quark-diquark
coupling. For the other form factors it is sufficient to apply the replacements
\begin{align}
g_s^2 &\to g_s^2 (1-x)^2 \frac{[{\bm p}_{\sT}^2 + L_s^2(m^2)]^2}{[{\bm
    p}_{\sT}^2 + L_s^2(\Lambda_s^2)]^4}
&&\text{dipolar form factor,}
\label{eq:repdip}
\\
g_s^2 &\to g_s^2  e^{\, -[{\bm p}_{\sT}^2 + L_X^2(m^2)]/[(1-x)\,
\Lambda_X^2]}
&&\text{exponential form factor.}
\label{eq:repexp}
\end{align}

The integrated results are obviously different for the three form-factor
choices. In all cases
the transversity function saturates the Soffer bound, i.e.,
\begin{equation}
h_1^{q(s)}(x) = \frac{1}{2} \, \left( f_1^{q(s)}(x) + g_1^{q(s)}(x) \right).
\end{equation}

\begin{itemize}
\item{Point-like coupling (to avoid divergences
we assume that the $\bm{p}_{\sT}^2$
integration is extended up to a finite cutoff $\Lambda_s^2$)
\begin{align}
f_1^{q(s)}(x) &= \frac{g_s^2\, (1-x)}{(2\pi)^2}\,
\frac{(m+x M)^2\,\Lambda_s^2 - L_s^2(m^2)\,\Lambda_s^2
+ L_s^2(m^2)\,[\Lambda_s^2+L_s^2(m^2)]\, \log \left(
\frac{\Lambda_s^2}{L_s^2(m^2)}+1 \right)}{4\,L_s^2(m^2)\,[\Lambda_s^2+L_s^2(m^2)]} \; ,
\\
g_1^{q(s)}(x) &= \frac{g_s^2\, (1-x)}{(2\pi)^2}\,
\frac{(m+x M)^2\,\Lambda_s^2 + L_s^2(m^2)\,\Lambda_s^2
- L_s^2(m^2)\,[\Lambda_s^2+L_s^2(m^2)]\, \log \left(
\frac{\Lambda_s^2}{L_s^2(m^2)}+1 \right)}{4\,L_s^2(m^2)\,[\Lambda_s^2+L_s^2(m^2)]} \; ,
\end{align}
}
\item{Dipolar form factor [same as Eqs.~\eqref{eq:f1sdip_int} and \eqref{eq:g1sdip_int}]
\begin{align}
f_1^{q(s)} (x) &=\frac{g_s^2}{(2\pi)^2}\,
\frac{[2\,(m+xM)^2 + L_s^2(\Lambda_s^2) ]\,(1-x)^3}{24\,L_s^6(\Lambda_s^2)}
\\
g_1^{q(s)}(x) &=\frac{g_s^2}{(2\pi)^2}\,
\frac{[2\,(m+xM)^2-L_s^2(\Lambda_s^2)]\,(1-x)^3}{24\,L_s^6(\Lambda_s^2)}
\end{align}
}
\item{Exponential form factor
\begin{align}
\begin{split}
f_1^{q(s)}(x) &= \frac{g_s^2}{(2\pi)^2}\, \frac{1}{4}\, \Bigg\{
e^{\, - 2\, L_s^2(m^2)/[(1-x)\,\Lambda_s^2]} \, \frac{1-x}{L_s^2(m^2)}\,
\bigl[(m+xM)^2-L_s^2(m^2)\bigr]
\\
& \hspace{3cm} - \Gamma \left( 0,
\frac{2 L_s^2(m^2)}{(1-x)\,\Lambda_s^2} \right) \,
\frac{2\, \bigl[(m+xM)^2-L_s^2(m^2)\bigr] - (1-x)\,\Lambda_s^2}{\Lambda_s^2}
\Bigg\} \; ,
\end{split}
\\
\begin{split}
g_1^{q(s)}(x) &= \frac{g_s^2}{(2\pi)^2}\, \frac{1}{4}\, \Bigg\{
e^{\, - 2\, L_s^2(m^2)/[(1-x)\,\Lambda_s^2]} \,
\frac{1-x}{L_s^2(m^2)}\, \bigl[ (m+xM)^2 + L_s^2(m^2) \bigr]
\\
& \hspace{2cm} - \Gamma \left( 0,
\frac{2\, L_s^2(m^2)}{(1-x)\,\Lambda_s^2} \right) \,
\frac{ 2\, [(m+xM)^2+L_s^2(m^2)] + \Lambda_s^2\,(1-x)}{\Lambda_s^2}
\Bigg\}\; ,
\end{split}
\end{align}
where $\Gamma$ is the incomplete $\Gamma$ function, defined as
\begin{equation}
\Gamma (a,z) = \int_z^{\infty} t^{a-1}\,e^{-t} dt \; .
\label{eq:specfunct}
\end{equation}
}
\end{itemize}

\subsection{Axial-vector diquark with light-cone transverse polarization only}

The unintegrated parton densities are
\begin{align}
f_1^{q(a)}(x,{\bm p}_{\sT}^2) &=\frac{g_a^2}{(2\pi)^3}\,
\frac{{\bm p}_{\sT}^2\,(1+x^2)+(m+xM)^2\,(1-x)^2}
{2\,[{\bm p}_{\sT}^2+L_a^2(m^2)]^2\,(1-x)}
\nn \\[0.2cm]
g_{1L}^{q(a)}(x,{\bm p}_{\sT}^2) &=\frac{g_a^2}{(2\pi)^3}\,
\frac{{\bm p}_{\sT}^2\,(1+x^2)-(m+xM)^2\,(1-x)^2}
{2\,[{\bm p}_{\sT}^2+L_a^2(m^2)]^2\,(1-x)} \; , \nn \\[0.2cm]
g_{1T}^{q(a)} (x,{\bm p}_{\sT}^2) &=\frac{g_a^2}{(2\pi)^3}
\, \frac{M\,x \,(m+xM)}{[{\bm p}_{\sT}^2 +L_a^2(m^2)]^2} \; , \nn
\\[0,2cm]
h_{1L}^{\perp\,q(a)}(x,{\bm p}_{\sT}^2) &=g_{1T}^{q(a)} (x,{\bm
p}_{\sT}^2)/x
\; , \nn \\[0.2cm]
h_{1T}^{q(a)}(x,{\bm p}_{\sT}^2) &=-\frac{g_a^2}{(2\pi)^3} \,
\frac{x\,{\bm p}_{\sT}^2}{[{\bm p}_{\sT}^2 + L_a^2(m^2)]^2\,(1-x)}
\; , \nn
\\[0.2cm]
h_{1T}^{\perp\,q(a)}(x,{\bm p}_{\sT}^2) &=0 \; , \nn \\[0.2cm]
h_1^{q(a)} (x,{\bm p}_{\sT}^2) &=h_{1T}^{q(a)}(x,{\bm p}_{\sT}^2)
\; . \label{eq:pTTeven-a-pl-lc}
\end{align}
The above results are valid for a point-like nucleon-quark-diquark
coupling. For the other form factors it is sufficient to apply the replacements
in Eqs.~\eqref{eq:repdip} and \eqref{eq:repexp}.

The integrated results are
\begin{itemize}
\item{Point-like coupling (to avoid divergences
we assume that the $\bm{p}_{\sT}^2$
integration is extended up to a finite cutoff $\Lambda_a^2$)
\begin{align}
f_1^{q(a)}(x) &=\frac{g_a^2}{(2\pi)^2}\,
\frac{1}{4\, L_a^2(m^2)\,[\Lambda_a^2+L_a^2(m^2)]\, (1-x)}\, \Bigg[
x\,\Lambda_a^2\,\left[ (M^2-m^2)\,(1-x^2)+2mM\,(1-x)^2-M_a^2\,(1+x^2) \right]
\nn \\
&\quad + L_a^2(m^2)\,[\Lambda_a^2+L_a^2(m^2)]\,(1+x^2)\, \log
        \left( \frac{\Lambda_a^2}{L_a^2(m^2)}+1 \right) \Bigg] \; , \nn
\\[0.2cm]
g_1^{q(a)}(x) &=\frac{g_a^2}{(2\pi)^2}\,
\frac{1}{4\, L_a^2(m^2)\,[\Lambda_a^2+L_a^2(m^2)]\,(1-x)} \, \Bigg\{
L_a^2(m^2)\,[\Lambda_a^2+L_a^2(m^2)]\,(1+x^2)\,
\log \left( \frac{\Lambda_a^2}{L_a^2(m^2)}+1 \right) \nn \\
&\quad - \Lambda_a^2\, \Big[ (1-x)\,mM\,[x(1-x)\,(2M-m)+2m] + x(1-x)\,M^3\,
(x-2x^2-1) + x(1+x^2)\,M M_a^2 \Big] \Bigg\} \; , \nn \\[0.2cm]
h_1^{q(a)}(x) &=-\frac{g_a^2}{(2\pi)^2}\,
\frac{x\left[ \Lambda_a^2 [(1-x)\,(xM^2-m^2)-xM_a^2] + L_a^2(m^2)\,
[\Lambda_a^2+L_a^2(m^2)]\,\log \left( \frac{\Lambda_a^2}{L_a^2(m^2)}+1 \right)\right]}
{2\, L_a^2(m^2)\,[\Lambda_a^2+L_a^2(m^2)]\,(1-x)} \; .
\label{eq:Teven-a-pl-lc}
\end{align}
}
\item{Dipolar form factor
[same as Eqs.~(\ref{eq:f1adip_int},\ref{eq:g1adip_int},
\ref{eq:h1xspecta})]
\begin{align}
f_1^{q(a)} (x) &=\frac{g_a^2}{(2\pi)^2}\,
\frac{[2\,(m+xM)^2\,(1-x)^2+(1+x^2)\,L_a^2(\Lambda_a^2) ]\,(1-x)}
     {24\,L_a^6(\Lambda_a^2)}  \;, \nn
\\
g_1^{q(a)}(x) &=-\frac{g_a^2}{(2\pi)^2}\,
\frac{[2\,(m+xM)^2\,(1-x)^2-(1+x^2)\,L_a^2(\Lambda_a^2)]\,(1-x)}
     {24\,L_a^6(\Lambda_a^2)} \;, \nn
\\
h_1^{q(a)}(x) &=-\frac{g_a^2}{(2\pi)^2}\, \frac{x(1-x)}{12\,L_a^4(\Lambda_a^2)} \;.
\end{align}
}
\item{Exponential form factor
\begin{align}
f_1^{q(a)}(x) &=\frac{g_a^2}{(2\pi)^2}\,
\frac{1}{4\Lambda_a^2\,L_a^2(m^2)\,(1-x)^2}\, \Bigg\{
\Lambda_a^2\,L_a^2(m^2)\, (1-x)\, (1+x^2)\, \Gamma \left( 0, 2\,
\frac{L_a^2(m^2)}{(1-x)\,\Lambda_a^2} \right) \nn \\
&\qquad \quad + \left[ (m+xM)^2\,(1-x)^2-L_a^2(m^2)\,(1+x^2) \right] \nn \\
&\qquad \qquad \times \left[ (1-x)\, \Lambda_a^2\,
e^{\, - 2\, L_a^2(m^2)/[(1-x)\,\Lambda_a^2]} - 2L_a^2(m^2)\,
\Gamma \left( 0, 2\, \frac{L_a^2(m^2)}{(1-x)\,\Lambda_a^2} \right)
\right] \Bigg\} \; , \nn \\[0.2cm]
g_1^{q(a)}(x) &=\frac{g_a^2}{(2\pi)^2}\,
\frac{1}{4\Lambda_a^2\,L_a^2(m^2)\,(1-x)^2}\, \Bigg\{
\Lambda_a^2\,L_a^2(m^2)\, (1-x)\, (1+x^2)\, \Gamma \left( 0, 2\,
\frac{L_a^2(m^2)}{(1-x)\,\Lambda_a^2} \right) \nn \\
&\qquad \quad + \left[ (m+xM)^2\,(1-x)^2+L_a^2(m^2)\,(1+x^2) \right] \nn \\
&\qquad \qquad \times \left[ 2L_a^2(m^2)\,
\Gamma \left( 0, 2\, \frac{L_a^2(m^2)}{(1-x)\,\Lambda_a^2} \right)
- (1-x)\, \Lambda_a^2\, e^{\, - 2\, L_a^2(m^2)/[(1-x)\,\Lambda_a^2]}
\right] \Bigg\}\; , \nn \\[0.2cm]
h_1^{q(a)}(x) &=\frac{g_a^2}{(2\pi)^2}\,
\frac{x}{2\Lambda_a^2\,(1-x)^2}\, \Bigg\{
\Lambda_a^2\,(1-x)\, \left[ e^{\, -2\,L_a^2(m^2)/[(1-x)\,\Lambda_a^2]}
- \Gamma \left( 0, 2\, \frac{L_a^2(m^2)}{(1-x)\,\Lambda_a^2} \right)
\right] \nn \\
&\hspace{4cm} - 2L_a^2(m^2)\, \Gamma \left( 0, 2\,
\frac{L_a^2(m^2)}{(1-x)\,\Lambda_a^2} \right) \Bigg\}  \; .
\label{eq:Teven-a-exp-lc}
\end{align}
}
\end{itemize}

\subsection{Axial-vector diquark including also longitudinal polarization}

The unintegrated parton densities are
\begin{align}
f_1^{q(a)}(x,{\bm p}_{\sT}^2) &=\frac{g_a^2}{(2\pi)^3}\,
\frac{1}{4\,[{\bm p}_{\sT}^2+L_a^2(m^2)]^2\,M_a^2\,(1-x)} \, \Bigg[
{\bm p}_{\sT}^4 + xM_a^2\,(2{\bm p}_{\sT}^2+xM_a^2) \nn \\
&\qquad + (1-x)^2\,\left[ {\bm p}_{\sT}^2 \, (M^2+m^2+2M_a^2) +
2m^2M_a^2+6xmMM_a^2+2x^2M^2M_a^2+m^2M^2\,(1-x)^2 \right] \Bigg]\; ,
\nn \\[0.2cm]
g_{1L}^{q(a)}(x,{\bm p}_{\sT}^2) &=\frac{g_a^2}{(2\pi)^3}\,
\frac{1}{4\,[{\bm p}_{\sT}^2+L_a^2(m^2)]^2\,M_a^2\,(1-x)} \, \Bigg[
{\bm p}_{\sT}^4 + xM_a^2\,(2{\bm p}_{\sT}^2+xM_a^2) \nn \\
&+ (1-x)^2\,\left[ {\bm p}_{\sT}^2 \, (2M_a^2-m^2-M^2-4mM) -
2m^2M_a^2-2xmMM_a^2-2x^2M^2M_a^2+m^2M^2\,(1-x)^2 \right] \Bigg]\; ,
\nn\\[0.2cm]
g_{1T}^{q(a)} (x,{\bm p}_{\sT}^2) &=
\frac{g_a^2\, M}{(2\pi)^3}\,
\frac{(m+M)\, {\bm p}_{\sT}^2-mM\,(m+M)\, (1-x)^2+xM_a^2\,[M(2x-1)+m]}
{2\,[{\bm p}_{\sT}^2+L_a^2(m^2)]^2\,M_a^2} \; , \nn \\[0,2cm]
h_{1L}^{\perp\,q(a)}(x,{\bm p}_{\sT}^2) &=\frac{g_a^2\,
M}{(2\pi)^3}\, \frac{(m+M)\,{\bm p}_{\sT}^2-mM\,(m+M)\,
(1-x)^2+xMM_a^2-mM_a^2(x-2)} {2\,[{\bm
p}_{\sT}^2+L_a^2(m^2)]^2\,M_a^2}
\; , \nn \\[0.2cm]
h_{1T}^{q(a)}(x,{\bm p}_{\sT}^2) &=-\frac{g_a^2}{(2\pi)^3}
\, \frac{{\bm p}_{\sT}^4 +[2xM_a^2+(m^2+M^2)\,(1-x)^2]\,
{\bm p}_{\sT}^2+[xM_a^2+mM\, (1-x)^2]^2}
{4\,[{\bm p}_{\sT}^2+L_a^2(m^2)]^2\,M_a^2\, (1-x)}\; , \nn \\[0.2cm]
h_{1T}^{\perp\,q(a)}(x,{\bm p}_{\sT}^2) &=
\frac{g_a^2}{(2\pi)^3}\,
\frac{M^2\, (m+M)^2\, (1-x)}{2\,[{\bm p}_{\sT}^2+L_a^2(m^2)]^2\,M_a^2}
\; , \nn \\[0.2cm]
h_1^{q(a)} (x,{\bm p}_{\sT}^2) &=-\frac{g_a^2}{(2\pi)^3}\,
\frac{{\bm p}_{\sT}^4 +[2xM_a^2-2mM\,(1-x)^2]\, {\bm p}_{\sT}^2+
[xM_a^2+mM\, (1-x)^2]^2}
{4M_a^2\, (1-x)\, [{\bm p}_{\sT}^2 +xM_a^2 +(1-x)\, (m^2-xM^2)]^2}\; .
\label{eq:pTTeven-a-pl-covar}
\end{align}
The above results are valid for a point-like nucleon-quark-diquark
coupling. For the other form factors it is sufficient to apply the replacements
in Eqs.~\eqref{eq:repdip} and \eqref{eq:repexp}.

The integrated results are
\begin{itemize}
\item{Point-like coupling (to avoid divergences
we assume that the $\bm{p}_{\sT}^2$
integration is extended up to a finite cutoff $\Lambda_a^2$)
\begin{align}
f_1^{q(a)}(x) &=\frac{g_a^2}{(2\pi)^2}\,
\frac{1}{8\,L_a^2(m^2)\,[\Lambda_a^2+L_a^2(m^2)]\, M_a^2\, (1-x)} \, \Bigg[
\Lambda_a^2\, [ L_a^4(m^2) + \Lambda_a^2\, L_a^2(m^2)] \nn \\
&\quad + (1-x)^2\,\left[ 2M_a^2\, [m^2-L_a^2(m^2)]+
x\, [(m^2-M^2)^2 -M_a^2\,(m^2-6mM+M^2)] + x^2\,2M^2M_a^2 \right] \nn \\
&\quad + L_a^2(m^2)\, [\Lambda_a^2+L_a^2(m^2)]\, (1-x)\,
\log \left( \frac{\Lambda_a^2}{L_a^2(m^2)}+1 \right) \, [(1+x)\, (m^2-M^2)-
2M_a^2\, (1-x)] \Bigg] \; , \nn \\[0.2cm]
g_1^{q(a)}(x) &=\frac{g_a^2}{(2\pi)^2}\,
\frac{1}{8\, L_a^2(m^2)\,[\Lambda_a^2+L_a^2(m^2)]\,MM_a^2\,x\,(1-x)}\, \Bigg\{
xM L_a^2(m^2)\,\Lambda_a^2\,[\Lambda_a^2+L_a^2(m^2)]\, + \Lambda_a^2\,(1-x)^2\, \nn \\
&\quad \times \Bigg[ L_a^2(m^2)\, \left[ mM^2\,(2x-1)+m^2M\,(x-2)-
m^3+xM\,(M^2-2M_a^2) \right] \nn \\
&\qquad + m^5\,(1-x)-m^4M\,(x-2)+m^3M_a^2x + m^3M^2\,(1-x^2) +
m^2M^3x\,(x^2-2x-1) \nn \\
&\qquad + mM^2M_a^2x+mM^4x\,(2x^2-x-1)+ x^3M^3\,(M^2-2M_a^2) \Bigg] \nn \\
&\quad + (1-x)\, L_a^2(m^2)\,[\Lambda_a^2+L_a^2(m^2)]\, \log \left(
\frac{\Lambda_a^2}{L_a^2(m^2)}+1 \right) \nn \\
&\qquad \times \Bigg[ L_a^2(m^2)\,(m+M)-m^3\,(1-x)+m^2M\,(x^2-2x-1)
-xm\,(3M^2\,(1-x)+M_a^2) \nn \\
&\qquad \quad +xM\,(M_a^2(1-2x)+2xM^2) \Bigg] \Bigg\} \; , \nn
\\[0.2cm]
h_1^{q(a)}(x) &=-\frac{g_a^2}{(2\pi)^2}\,
\frac{1}{8\,L_a^2(m^2)\,[\Lambda_a^2+L_a^2(m^2)]\,M_a^2\,(1-x)} \,
\Bigg[ L_a^2(m^2)\,\Lambda_a^2\,(L_a^2(m^2)+\Lambda_a^2) \nn \\
&\quad +(1-x)^2\, \Lambda_a^2\, [(m+M)^2\,L_a^2(m^2)+x\,(m-M)^2\,((m+M)^2-
M_a^2)] \nn \\
&\quad -2L_a^2(m^2)\,(L_a^2(m^2)+\Lambda_a^2)\,(m+M)\,(m-xM)\,(1-x)\,
\log \left( \frac{\Lambda_a^2}{L_a^2(m^2)}+1 \right) \Bigg] \; .
\label{eq:Teven-a-pl-covar}
\end{align}
}
\item{Dipolar form factor
\begin{align}
f_1^{q(a)}(x) &=\frac{g_a^2}{(2\pi)^2}\,
\frac{1-x}{48\,L_a^6(\Lambda_a^2)\, M_a^2} \, \Bigg[
2 L_a^4(\Lambda_a^2) + 2xM_a^2\,(L_a^2(\Lambda_a^2)+xM_a^2) \nn \\
&\quad + (1-x)^2\,\left[ (2M_a^2+m^2+M^2)\,L_a^2(\Lambda_a^2)+
4m^2M_a^2+x\, 12mMM_a^2 + x^2\,4M^2M_a^2 + 2m^2M^2\,(1-x)^2 \right]
\Bigg]\; ,\nn\\[0.2cm]
g_1^{q(a)}(x) &=\frac{g_a^2}{(2\pi)^2}\,
\frac{1-x}{48\, L_a^6(\Lambda_a^2)\,M_a^2}\, \Bigg[
2 L_a^4(\Lambda_a^2) + 2xM_a^2\,(L_a^2(\Lambda_a^2)+xM_a^2) \nn \\
&\quad + (1-x)^2\,\left[ (2M_a^2-m^2-M^2-4mM)\,L_a^2(\Lambda_a^2)-
4m^2M_a^2-x\, 4mMM_a^2 - x^2\,4M^2M_a^2 \right.
\nn \\
&\left. \quad \qquad \qquad + 2m^2M^2\,(1-x)^2 \right] \Bigg]\; ,\nn\\[0.2cm]
h_1^{q(a)}(x) &=-\frac{g_a^2}{(2\pi)^2}\,
\frac{1-x}{24\,L_a^6(\Lambda_a^2)\,M_a^2} \, \left[ L_a^4(\Lambda_a^2)
+[xM_a^2-mM\,(1-x)^2]\,L_a^2(\Lambda_a^2)+[xM_a^2+mM\,(1-x)^2]^2
\right] \; .
\label{eq:Teven-a-dip-covar}
\end{align}
}
\item{Exponential form factor
\begin{align}
f_1^{q(a)}(x) &=\frac{g_a^2}{(2\pi)^2}\,
\frac{1}{16\, L_a^2(m^2)\, \Lambda_a^2\, M_a^2\, (1-x)^2} \nn \\
&\; \times \Bigg\{
2\, \left[ \Lambda_a^2\, (1-x)\, e^{\, - 2\,[{\bm p}_{\sT}^2 + L_a^2(m^2)]
/[(1-x)\,\Lambda_a^2]} - 2\, L_a^2(m^2)\,\Gamma \left( 0,
\frac{2\, L_a^2(m^2)}{(1-x)\,\Lambda_a^2} \right) \right] \nn \\
&\quad \times \Bigg[ (1-x)^2\, \Big[ m^2\, (2M_a^2+M^2\,(1-x)^2) -
L_a^2(m^2)\, (2M_a^2+m^2+M^2) + 2xMM_a^2\,(xM+3m) \Big] \nn \\
&\qquad + (1-x)\, xM_a^2\, (xM^2-m^2) + L_a^2(m^2)\,
(L_a^2(m^2)-xM_a^2) \Bigg] \nn \\
&\; + \Lambda_a^2\, L_a^2(m^2)\, (1-x) \, \Bigg[ (1-x)^2\, 2\Gamma
\left( 0, \frac{2\, L_a^2(m^2)}{(1-x)\,\Lambda_a^2} \right) \,
(m^2+M^2+2M_a^2) \nn \\
& \qquad \quad + (1-x)\, \Lambda_a^2 \,
e^{\, - 2\, [{\bm p}_{\sT}^2 + L_a^2(m^2)]/[(1-x)\, \Lambda_a^2]} + 4\,
\Gamma \left( 0, \frac{2\, L_a^2(m^2)}{(1-x)\,\Lambda_a^2} \right) \,
[xM_a^2 - L_a^2(m^2)] \Bigg] \Bigg\} \nn \\[0.2cm]
g_1^{q(a)}(x) &=\frac{g_a^2}{(2\pi)^2}\,
\frac{1}{16\, L_a^2(m^2)\, \Lambda_a^2\, M_a^2\, (1-x)^2} \nn \\
&\; \times \Bigg\{
\left[ \Lambda_a^2\, (1-x)\, e^{\, - 2\,[{\bm p}_{\sT}^2 +L_a^2(m^2)]
/[(1-x)\,\Lambda_a^2]} - 2\, L_a^2(m^2)\,\Gamma \left( 0,
\frac{2\, L_a^2(m^2)}{(1-x)\,\Lambda_a^2} \right) \right] \nn \\
&\quad \times \Bigg[ (1-x)^4\, 2m^2M^2 -(1-x)^2\, \left[ 2mM
[xM_a^2-L_a^2(m^2)] -L_a^2(m^2)\, (m+M)^2 \right] \nn \\
&\qquad + (1-x)\, \left[ \Lambda_a^2 \, L_a^2(m^2) + 2M_a^2 \,
(m^2\, (x-2) + (2x-1)\, x^2 M^2) \right] \nn \\
&\qquad + 2L_a^2(m^2)\, \left[ L_a^2(m^2) - xM_a^2 - 2 M_a^2\,
(1-x)^2\right] \Bigg] \nn \\
&\; - 2L_a^2(m^2)\, \Lambda_a^2 \, (1-x)\, \Gamma \left( 0,
\frac{2\, L_a^2(m^2)}{(1-x)\,\Lambda_a^2} \right) \, \left[
L_a^2(m^2) -2xM_a^2 + (1-x)^2\, [(m+M)^2+2mM-2M_a^2] \right] \Bigg\}
\; , \nn \\[0.2cm]
h_1^{q(a)}(x) &=\frac{g_a^2}{(2\pi)^2}\,
\frac{1}{16\,L_a^2(m^2)\, \Lambda_a^2\, M_a^2\, (1-x)^2}\, \Bigg\{
\Bigg[ \Lambda_a^2\, (1-x)\, \left[ \sinh \left( 2\,
\frac{{\bm p}_{\sT}^2 + L_a^2(m^2)}{(1-x)\,\Lambda_a^2} \right)-
\cosh \left( 2\, \frac{{\bm p}_{\sT}^2 + L_a^2(m^2)}
{(1-x)\,\Lambda_a^2} \right) \right] \nn \\
&\qquad \quad + 2\, L_a^2(m^2)\,\Gamma \left( 0,
\frac{2\, L_a^2(m^2)}{(1-x)\,\Lambda_a^2} \right) \Bigg] \; \Bigg[
(1-x)^4\, 2m^2M^2 +(1-x)^2\, 4mM\, [xM_a^2+L_a^2(m^2)] \nn \\
&\qquad \qquad \quad + (1-x)\, \left[ \Lambda_a^2 \, L_a^2(m^2) +
2xM_a^2 \, (xM^2-m^2) \right] + 2L_a^2(m^2)\, (L_a^2(m^2) - xM_a^2)
\Bigg] \nn \\
&\; + 2L_a^2(m^2)\, \Lambda_a^2 \, (1-x)\, \Gamma \left( 0,
\frac{2\, L_a^2(m^2)}{(1-x)\,\Lambda_a^2} \right) \, \left[
L_a^2(m^2) -2xM_a^2 + (1-x)^2\, 2mM \right] \Bigg\} \; .
\label{eq:Teven-a-exp-covar}
\end{align}
}
\end{itemize}

\subsection{Axial-vector diquark including also time-like polarization}

The unintegrated parton densities are
\begin{align}
f_1^{q(a)}(x,{\bm p}_{\sT}^2) &=\frac{g_a^2}{(2\pi)^3}\,
\frac{[{\bm p}_{\sT}^2+(m+xM)^2+2mMx]\,(1-x)}{2\,[{\bm p}_{\sT}^2+
L_a^2(m^2)]^2} \nn \\[0.2cm]
g_{1L}^{q(a)}(x,{\bm p}_{\sT}^2) &=-\frac{g_a^2}{(2\pi)^3}\,
\frac{[-{\bm p}_{\sT}^2+m^2+x^2 M^2]\,(1-x)}{2\,[{\bm p}_{\sT}^2+
L_a^2(m^2)]^2} \; , \nn \\[0.2cm]
g_{1T}^{q(a)} (x,{\bm p}_{\sT}^2) &=-\frac{g_a^2}{(2\pi)^3}
\, \frac{M^2\,x (1-x)}{[{\bm p}_{\sT}^2 + L_a^2(m^2)]^2} \; , \nn \\[0,2cm]
h_{1L}^{\perp\,q(a)}(x,{\bm p}_{\sT}^2) &=\frac{g_a^2}{(2\pi)^3} \,
\frac{mM\,(1-x)}{[{\bm p}_{\sT}^2 + L_a^2(m^2)]^2} \; , \nn
\\[0.2cm]
h_{1T}^{q(a)}(x,{\bm p}_{\sT}^2) &=-x\,h_{1L}^{\perp\,q(a)}(x,{\bm p}_{\sT}^2)
\; , \nn \\[0.2cm]
h_{1T}^{\perp\,q(a)}(x,{\bm p}_{\sT}^2) &=0 \; , \nn \\[0.2cm]
h_1^{q(a)} (x,{\bm p}_{\sT}^2) &\equiv h_{1T}^{q(a)}(x,{\bm p}_{\sT}^2) \; .
\label{eq:pTTeven-a-pl-feyn}
\end{align}
The above results are valid for a point-like nucleon-quark-diquark
coupling. For the other form factors it is sufficient to apply the replacements
in Eqs.~\eqref{eq:repdip} and \eqref{eq:repexp}. The result for $f_1$ with
dipolar form factor
corresponds to that obtained in Ref.~\cite{Bacchetta:2003rz}.

The integrated results are
\begin{itemize}
\item{Point-like coupling (to avoid divergences
we assume that the $\bm{p}_{\sT}^2$
integration is extended up to a finite cutoff $\Lambda_a^2$)
\begin{align}
f_1^{q(a)}(x) &=\frac{g_a^2\, (1-x)}{(2\pi)^2}\,
\frac{[(m+M)^2+2mM-M_a^2]\,x\,\Lambda_a^2 + L_a^2(m^2)\,[\Lambda_a^2+L_a^2(m^2)]\, \log
\left( \frac{\Lambda_a^2}{L_a^2(m^2)}+1 \right)}{4\,L_a^2(m^2)\,[\Lambda_a^2+L_a^2(m^2)]}
\; , \nn \\
g_1^{q(a)}(x) &=-\frac{g_a^2\, (1-x)}{(2\pi)^2}\,
\frac{\Lambda_a^2\, (L_a^2(m^2)+m^2+x^2M^2) -L_a^2(m^2)\,[\Lambda_a^2+L_a^2(m^2)]\,
\log \left(
 \frac{\Lambda_a^2}{L_a^2(m^2)}+1 \right)}{4\,L_a^2(m^2)\,[\Lambda_a^2+L_a^2(m^2)]}
\; , \nn \\
h_1^{q(a)}(x) &=-\frac{g_a^2\, (1-x)\, \Lambda_a^2}{(2\pi)^2}\,
\frac{x m M}{2\,L_a^2(m^2)\,[\Lambda_a^2+L_a^2(m^2)]} \; .
\label{eq:Teven-a-pl-feyn}
\end{align}
}
\item{Dipolar form factor
\begin{align}
f_1^{q(a)}(x) &=\frac{g_a^2}{(2\pi)^2}\,
\frac{\left[ L_a^2(\Lambda_a^2)+2\,[(m+xM)^2+2xmM]\right] \, (1-x)^3}
{24\,L_a^6(\Lambda_a^2)}  \; , \nn \\
g_1^{q(a)}(x) &=\frac{g_a^2}{(2\pi)^2}\,
\frac{[ L_a^2(\Lambda_a^2) -2\,(m^2+x^2M^2)]\,(1-x)^3}
{24\,L_a^6(\Lambda_a^2)}  \; , \nn \\
h_1^{q(a)}(x) &=-\frac{g_a^2}{(2\pi)^2}\,
\frac{mM\,x\,(1-x)^3}{6\,L_a^6(\Lambda_a^2)} \; .
\label{eq:Teven-a-dip-feyn}
\end{align}
}
\item{Exponential form factor
\begin{align}
f_1^{q(a)}(x) &=\frac{g_a^2}{(2\pi)^2}\, \frac{1}{4}\,
\Bigg\{ e^{\, - 2\, L_a^2(m^2)/[(1-x)\,\Lambda_a^2]} \, \;
\frac{[(m+xM)^2+2mxM-L_a^2(m^2)]\,(1-x)}{L_a^2(m^2)} \nn \\
&\hspace{2cm} - \Gamma \left( 0, 2\,
\frac{L_a^2(m^2)}{(1-x)\,\Lambda_a^2} \right) \,
\frac{2\, [(m+xM)^2+2mxM-L_a^2(m^2)] - (1-x)\,\Lambda_a^2}{\Lambda_a^2}
\Bigg\} \; , \nn \\[0.2cm]
g_1^{q(a)}(x) &=\frac{g_a^2}{(2\pi)^2}\, \frac{1}{4}\,
\Bigg\{ \Gamma \left( 0, 2\, \frac{L_a^2(m^2)}{(1-x)\,\Lambda_a^2}
\right) \,
\frac{2\,[m^2+x^2 M^2+L_a^2(m^2)] + (1-x)\,\Lambda_a^2}{\Lambda_a^2}
\nn \\
&\hspace{2cm} - e^{\, - 2\, L_a^2(m^2)/[(1-x)\,\Lambda_a^2]} \, \;
\frac{(1-x)\, [m^2+x^2 M^2+L_a^2(m^2)]}{L_a^2(m^2)}\Bigg\} \; ,
\nn \\[0.2cm]
h_1^{q(a)}(x) &=\frac{g_a^2}{(2\pi)^2}\, \frac{mxM}{2}\,
\left\{ \frac{2}{\Lambda_a^2}\, \Gamma \left( 0,
\frac{2\, L_a^2(m^2)}{(1-x)\,\Lambda_a^2} \right) -
e^{\, - 2\, L_a^2(m^2)/[(1-x)\,\Lambda_a^2]} \,
\frac{1-x}{L_a^2(m^2)} \right\} \; .
\label{eq:Teven-a-exp-feyn}
\end{align}
}
\end{itemize}


\section{T-odd functions in different variations of the model}
\label{sec:B}

As a continuation of App.~\ref{sec:A}, here we list the Sivers and
Boer-Mulders functions, namely the leading-twist T-odd parton
densities obtained in the context of our spectator diquark model,
again for all the combinations of
diquark propagators and nucleon-quark-diquark vertices.

\subsection{Scalar diquark}
\label{sec:B_scalar}

For scalar diquarks, we have
\begin{align}
f_{1T}^{\perp\,q(s)}(x,{\bm p}_{\sT}^2) &=-\frac{g_s(p^2)}{4}\,
\frac{1}{(2\pi)^3}\, \frac{M\,e_c^2}{2(1-x)P^+}\,
\frac{2\,\mathrm{Im}\,J_1^{s}}{p^2-m^2} \nn \\
h_1^{\perp\,q(s)}(x,{\bm p}_{\sT}^2) &=f_{1T}^{\perp\,q(s)}(x,
{\bm p}_{\sT}^2) \; ,
\label{eq:SivBM1-s-pl}
\end{align}
where the $J_1^{s}$ integral is defined as
\begin{equation} 
\begin{split} 
\left(\varepsilon_{\sT}^{ij}p_{\sT i}S_{\sT j}\right)J_1^{s}&=\intl\,
\frac{g_s\bigl((p-l)^2\bigr)}{(D_1+i\varepsilon )\,(D_2-i \varepsilon )\,
(D_3+i \varepsilon )\,(D_4+i \varepsilon )}\nn \\
&\mbox{\hspace{2cm}} \mathrm{Tr}\bigg[
(\pslash-\lslash+m)\,(\Pslash+M)\,\g_5\,\Sslash\,(\pslash+m)\,(2P-2p+l)_\rho\,
n_-^\rho\;\g^+\bigg] \\
&=\intl\,\frac{g_s\bigl((p-l)^2\bigr)}{(D_1+i \varepsilon )\,
(D_2-i \varepsilon )\,(D_3+i \varepsilon )\,(D_4+i \varepsilon )}\;
4 i\Big(l^++2(1-x)P^+\Big) \nn \\
&\mbox{\hspace{2cm}} \Big(l^+M\,\varepsilon_{\sT}^{ij}p_{\sT i}S_{\sT j}
-P^+(m+xM)\,\varepsilon_{\sT}^{ij}\,l_{\sT i}S_{\sT j}\Big),
\label{eq:J1s-expl}
\end{split} 
\end{equation} 
with $D_1,D_2,D_3,D_4$ defined in Eq.~(\ref{eq:denoms}). The imaginary part of 
$J_1^{s}$ can be extracted by using the Cutkosky cutting rules on the loop 
diagram of Fig.~\ref{fig:eikonal}, which in the present case amount to put on 
shell the eikonalized virtual quark propagator $D_2$ and the virtual scalar 
diquark propagator $D_4$. The resulting $\delta$ functions (see below) reduce 
the integral in Eq.~(\ref{eq:J1s-expl}) to a bidimensional integral in 
$d^2 {\bm l}_{\sT}$. In general, for a $n$-dimensional integral 
$\int d^nl \, l_{\rho} f(l,p)$ the term $l_{\rho}$ can be replaced by the 
expression $p_{\rho} (l \cdot p)/p^2$. For the present case $n=2$ and with the 
identification $l_{\rho}=l_{\sT i},\, p_{\rho}=p_{\sT i}$, we finally can write
\begin{equation} 
\begin{split} 
2\,\mathrm{Im}\,J_1^{s} &=\intl\,
\frac{g_s\bigl((p-l)^2\bigr)}{D_1\,D_3}\,4\,\left( l^+ +2(1-x)P^+ \right) \, 
\left(l^+M-P^+(m+xM)\,\frac{{\bm l}_{\sT}\cdot {\bm p}_{\sT}}{{\bm p}_{\sT}^2} 
\right)\, (2\pi i)\, \delta(D_2)\,(-2\pi i)\, \delta(D_4)  \\
&=-4P^+\,(m+xM)\,(1-x)\,g_s\,\mathcal{I}_1\; .
\end{split} 
\end{equation} 
The explicit expression of $\mathcal{I}_1$ clearly depends on the choice of the
nucleon-quark-scalar diquark vertex form factor.

The Boer-Mulders calculation gives exactly the same results as the
Sivers one, in the scalar diquark framework, since the relevant
trace over Dirac-Lorentz structures is now given by
\begin{equation}
\mathrm{Tr}\Bigl[
(\pslash-\lslash+m)\,(\Pslash+M)\,(\pslash+m)\,(2P-2p+l)_\rho\,n_-^\rho 
i\sigma^{i+}\gamma_5 \Bigr]
=-4 i\Bigl(l^++2(1-x)P^+\Big)\Big(l^+M\,\varepsilon_{\sT}^{ij}p_{\sT j}
-P^+(m+xM)\,\varepsilon_{\sT}^{ij}\,l_{\sT j}\Bigr) \; .
\end{equation}

\begin{itemize}
\item{Point-like coupling
\begin{equation} 
\begin{split} 
2\,\mathrm{Im}\,J_1^{s} &=g_s \intl\,\frac{1}{D_1\,D_3}\,4\,\left( l^+ +
2(1-x)P^+ \right) \, \left(l^+M-P^+(m+xM)\,
\frac{{\bm l}_{\sT}\cdot {\bm p}_{\sT}}{{\bm p}_{\sT}^2} \right)\,
(2\pi i)\, \delta(D_2)\,(-2\pi i)\, \delta(D_4) \\
&=-4P^+\,(m+xM)\,(1-x)\,g_s\,\mathcal{I}_1^{p.l.} =
-g_s \, \frac{P^+\,(m+xM)\,(1-x)}{\pi {\bm p}_{\sT}^2}\, \log \left(
\frac{L_s^2(m^2) + {\bm p}_{\sT}^2}{L_s^2(m^2)} \right) \; ,
\label{eq:B-J1s}
\end{split} \end{equation} 
where $\mathcal{I}_1^{p.l.}$ is calculated in App.~\ref{sec:C}.

Using Eq.~(\ref{eq:offshell}), the final result is then
\begin{align}
f_{1T}^{\perp\,q(s)}(x,{\bm p}_{\sT}^2) &=-\frac{g_s^2}{4}\,
\frac{M\,e_c^2}{(2\pi)^4}\,
\frac{(m+xM)\,(1-x)}{{\bm p}_{\sT}^2\, [L_s^2(m^2) + {\bm p}_{\sT}^2)]}
\, \log \left( \frac{L_s^2(m^2) + {\bm p}_{\sT}^2}{L_s^2(m^2)} \right)
\nn \\
h_1^{\perp\,q(s)}(x,{\bm p}_{\sT}^2) &=f_{1T}^{\perp\,q(s)}(x,
{\bm p}_{\sT}^2) \; .
\label{eq:SivBM2-s-pl}
\end{align}
}
\item{Dipolar form factor.
The final results, already given in Eq.~\eqref{eq:Siversspect3s} and
\eqref{eq:BoerMuldersspect3s}, are
\begin{align}
f_{1T}^{\perp\,q(s)}(x,{\bm p}_{\sT}^2) &= -\frac{g_s^2}{4}\,
\frac{M\,e_c^2}{(2\pi)^4}\,\frac{(1-x)^3\,(m+xM)}{L_s^2(\Lambda_s^2)\,
[{\bm p}_{\sT}^2+L_s^2(\Lambda_s^2)]^3}
\nn \\
h_1^{\perp\,q(s)}(x,{\bm p}_{\sT}^2) &= f_{1T}^{\perp\,q(s)}(x,{\bm p}_{\sT}^2)
\end{align}
}
\item{Exponential form factor
\begin{equation} \begin{split} 
2\,\mathrm{Im}\,J_1^s &= g_s \intl\,
\frac{e^{[(p-l)^2-m^2]/\Lambda_s^2}}{D_1\,D_3}\,4\left( l^+ +
2(1-x)P^+ \right) \left(l^+M-P^+(m+xM)\,
\frac{{\bm l}_{\sT}\cdot {\bm p}_{\sT}}{{\bm p}_{\sT}^2} \right)\,
(2\pi i)\, \delta(D_2)\,(-2\pi i)\, \delta(D_4)  \\
&=-4P^+\,(m+xM)\,(1-x)\,g_s\,\mathcal{I}_1^{exp} 
\nn \\ &
=-g_s\, \frac{P^+\,(m+xM)\,(1-x)}{\pi {\bm p}_{\sT}^2}\, \left[ \Gamma
\left( 0,\frac{L_s^2(m^2)}{(1-x)\,\Lambda_s^2} \right) - \Gamma
\left( 0, \frac{L_s^2(m^2)+{\bm p}_{\sT}^2}{(1-x)\,\Lambda_s^2}
\right) \right]\; ,
\label{eq:B-J1s-exp}
\end{split} \end{equation} 
and $\mathcal{I}_1^{exp}$ is calculated in App.~\ref{sec:C}.

The final results are then
\begin{align}
f_{1T}^{\perp\,q(s)}(x,{\bm p}_{\sT}^2) &=-\frac{g_s^2}{4}\,
\frac{M\,e_c^2}{(2\pi)^4}\,
\frac{(m+xM)\,(1-x)}{{\bm p}_{\sT}^2\, [L_s^2(m^2) + {\bm p}_{\sT}^2]}
\, e^{-[{\bm p}_{\sT}^2 + L_s^2(m^2)]/[(1-x)\,\Lambda_s^2]} \nn \\[0.2cm]
&\qquad \times
\left[ \Gamma \left( 0,\frac{L_s^2(m^2)}{(1-x)\,\Lambda_s^2} \right) -
\Gamma \left( 0, \frac{L_s^2(m^2)+{\bm p}_{\sT}^2}{(1-x)\,\Lambda_s^2}
 \right) \right] \; , \nn \\
h_1^{\perp\,q(s)}(x,{\bm p}_{\sT}^2) &=f_{1T}^{\perp\,q(s)}(x,
{\bm p}_{\sT}^2) \; .
\label{eq:SivBM2-s-exp}
\end{align}
}
\end{itemize}

\subsection{Axial-vector diquark with light-cone transverse polarization only}

We have
\begin{align}
f_{1T}^{\perp\,q(a)}(x,{\bm p}_{\sT}^2)
&=\frac{g_a(p^2)}{4}\,\frac{1}{(2\pi)^3}\,
\frac{M\,e_c^2}{4(1-x)P^+}\,\frac{2\,\mathrm{Im}\,J_1^a}{p^2-m^2}
\nn \\
h_1^{\perp\,q(a)}(x,{\bm p}_{\sT}^2) &=\frac{g_a(p^2)}{4}\,
\frac{1}{(2\pi)^3}\, \frac{M\,e_c^2}{4(1-x)P^+}\,
\frac{2\,\mathrm{Im}\,J_1^{\prime(a)}}{p^2-m^2} \; ,
\label{eq:SivBM1-a-pl-lc}
\end{align}
where now the $J_1^a$ and $J_1^{\prime\,a}$ integrals are defined as
\begin{eqnarray}
\left(\varepsilon_{\sT}^{ij}p_{\sT i}S_{\sT j}\right)J_1^{a}&=&\intl\,
\frac{g_a\bigl((p-l)^2\bigr)}{(D_1+i \varepsilon )\,(D_2-i \varepsilon )\,
(D_3+i\varepsilon )\,(D_4+i \varepsilon )}\nn \\
& &\qquad  
\mathrm{Tr}\bigg[(\pslash-\lslash+m)\,\gamma^\mu\,\gamma_5\,(\Pslash+M)\,\g_5\,
\Sslash\,\gamma^\alpha\,\gamma_5\,(\pslash+m)\,d_{\mu\nu}(p-l-P)\,
d_{\sigma\alpha}(P-p)\nn\\
& &\qquad \left[ (2P-2p+l)_\rho\, g^{\nu \sigma}
-(P-p+(1+\kappa_a)l)^\sigma g_\rho^\nu-(P-p-\kappa_a l)^\nu \,
g_\rho^\sigma \right]n_-^\rho\;\g^+\bigg]
\label{eq:J1a-expl}
\end{eqnarray} 
and
\begin{eqnarray} 
\left(-\varepsilon_{\sT}^{ij}p_{\sT j}\right)J_1^{\prime\, a}&=&\intl\,
\frac{g_a\bigl((p-l)^2\bigr)}{(D_1+i\varepsilon )\,(D_2-i\varepsilon )\,
(D_3+i\varepsilon )\,(D_4+i\varepsilon )}\nn \\
& &\qquad 
\mathrm{Tr}\bigg[(\pslash-\lslash+m)\,\gamma^\mu\,\gamma_5\,(\Pslash+M)\,
\gamma^\alpha\,\gamma_5\,(\pslash+m)\,d_{\mu\nu}(p-l-P)\,d_{\sigma\alpha}(P-p)
\nn\\
& &\qquad \left[ (2P-2p+l)_\rho\, g^{\nu \sigma}
-(P-p+(1+\kappa_a)l)^\sigma g_\rho^\nu-(P-p-\kappa_a l)^\nu \,
g_\rho^\sigma \right]n_-^\rho\;i\sigma^{i+}\gamma_5\bigg] \; ,
\label{eq:J1primea-expl}
\end{eqnarray} 
and the explicit expressions of the $d_{\mu\nu}(p-l-P)$ and
$d_{\sigma\alpha}(P-p)$ structures are those expressed in the first
line in Eq.~(\ref{eq:lc}).

\begin{itemize}
\item{Point-like coupling (to avoid divergences
we assume that the $\bm{p}_{\sT}^2$
integration is extended up to a finite cutoff $\Lambda_a^2$)
\begin{align}
2\,\mathrm{Im}\,J_1^a &=-8P^+x\,(m+xM)\,g_a\,\mathcal{I}_1^{p.l.} =
-g_a \, \frac{2P^+\,x\,(m+xM)}{\pi {\bm p}_{\sT}^2}\, \log \left(
\frac{L_a^2(m^2) + {\bm p}_{\sT}^2}{L_a^2(m^2)} \right) \nn \\[0.2cm]
2\,\mathrm{Im}\,J_1^{\prime\,a} &=8P^+\,(m+xM)\,g_a\,
\mathcal{I}_1^{p.l.} = g_a\, \frac{2P^+\,(m+xM)}{\pi {\bm p}_{\sT}^2}\,
\log \left( \frac{L_a^2(m^2) + {\bm p}_{\sT}^2}{L_a^2(m^2)} \right)
\; , \label{eq:B-J1a-pl-lc}
\end{align}
where $\mathcal{I}_1^{p.l.}$ is the same integral as in Eq.~(\ref{eq:B-J1s}) but
with the substitution $L_s(m^2) \leftrightarrow L_a(m^2)$.

Using again Eq.~(\ref{eq:offshell}), the final result is
\begin{align}
f_{1T}^{\perp\,q(a)}(x,{\bm p}_{\sT}^2) &=
\frac{g_a^2}{4}\,\frac{M\,e_c^2}{(2\pi)^4}\,
\frac{x\,(m+xM)}{{\bm p}_{\sT}^2\, [L_a^2(m^2) + {\bm p}_{\sT}^2]}\,
\log \left( \frac{L_a^2(m^2) + {\bm p}_{\sT}^2}{L_a^2(m^2)} \right)
\nn \\
h_1^{\perp\,q(a)}(x,{\bm p}_{\sT}^2) &=-\frac{1}{x}\, 
f_{1T}^{\perp\,q(a)}(x,{\bm p}_{\sT}^2)
\; .
\label{eq:SivBM2-a-pl-lc}
\end{align}
}
\item{Dipolar form factor.
The final results, already given in Eq.~\eqref{eq:Siversspect3a} and
\eqref{eq:BoerMuldersspect3a}, are
\begin{align}
f_{1T}^{\perp\,q(a)}(x,{\bm p}_{\sT}^2) &= \frac{g_a^2}{4}\,
\frac{M\,e_c^2}{(2\pi)^4}\,\frac{(1-x)^2\,x\,(m+xM)}{L_a^2(\Lambda_a^2)\,
[{\bm p}_{\sT}^2+L_a^2(\Lambda_a^2)]^3}
\nn \\
h_1^{\perp\,q(a)}(x,{\bm p}_{\sT}^2) &= -\frac{1}{x}\, 
f_{1T}^{\perp\,q(a)}(x,{\bm p}_{\sT}^2)\; .
\end{align}
}
\item{Exponential form factor
\begin{align}
2\,\mathrm{Im}\,J_1^a &=-8P^+x\,(m+xM)\,g_a\, \mathcal{I}_1^{exp} \nn \\
&=-g_a\, \frac{2P^+\,x\,(m+xM)}
{\pi {\bm p}_{\sT}^2}\, \left[ \Gamma \left( 0,
\frac{L_a^2(m^2)}{(1-x)\,\Lambda_a^2} \right) - \Gamma \left( 0,
\frac{L_a^2(m^2)+{\bm p}_{\sT}^2}{(1-x)\,\Lambda_a^2} \right) \right]
\; , \nn \\
2\,\mathrm{Im}\,J_1^{\prime\,a} &=8P^+\,(m+xM)\,g_a\,\mathcal{I}_1^{exp}
\nn \\
&=g_a\, \frac{2P^+\,(m+xM)}
{\pi {\bm p}_{\sT}^2}\, \left[ \Gamma \left( 0,
\frac{L_a^2(m^2)}{(1-x)\,\Lambda_a^2} \right) - \Gamma \left( 0,
\frac{L_a^2(m^2)+{\bm p}_{\sT}^2}{(1-x)\,\Lambda_a^2} \right) \right]
\; .
\label{eq:B-J1a-exp-lc}
\end{align}

The final result is, then,
\begin{align}
f_{1T}^{\perp\,q(a)}(x,{\bm p}_{\sT}^2) &=
\frac{g_a^2}{4}\,\frac{M\,e_c^2}{(2\pi)^4}\,
\frac{x\,(m+xM)}{{\bm p}_{\sT}^2\, [L_a^2(m^2) + {\bm p}_{\sT}^2]}\,
e^{-[{\bm p}_{\sT}^2 + L_a^2(m^2)]/[(1-x)\,\Lambda_a^2]} \nn \\
& \qquad \times
\left[ \Gamma \left( 0,\frac{L_a^2(m^2)}{(1-x)\,\Lambda_a^2} \right) -
\Gamma \left( 0, \frac{L_a^2(m^2)+{\bm p}_{\sT}^2}{(1-x)\,\Lambda_a^2}
 \right) \right] \; , \nn \\
h_1^{\perp\,q(a)}(x,{\bm p}_{\sT}^2) &=-\frac{1}{x}\, 
f_{1T}^{\perp\,q(a)}(x,{\bm p}_{\sT}^2) \; .
\label{eq:SivBM2-a-exp-lc}
\end{align}
}
\end{itemize}

\subsection{Axial-vector diquark including also longitudinal polarization}

We have
\begin{align}
f_{1T}^{\perp\,q(a)}(x,{\bm p}_{\sT}^2) &=
\frac{g_a(p^2)}{4}\,\frac{1}{(2\pi)^3}\,
\frac{M\,e_c^2}{4(1-x)P^+}\,\frac{2\,\mathrm{Im}\,J_1^a}{p^2-m^2}
\nn \\
h_1^{\perp\,q(a)}(x,{\bm p}_{\sT}^2) &=\frac{g_a(p^2)}{4}\,
\frac{1}{(2\pi)^3}\, \frac{M\,e_c^2}{4(1-x)P^+}\,
\frac{2\,\mathrm{Im}\,J_1^{\prime\,a}}{p^2-m^2} \; ,
\label{eq:SivBM1-a-pl-cov}
\end{align}
where the $J_1^a$ and $J_1^{\prime\,a}$ integrals are defined as in
Eqs.~(\ref{eq:J1a-expl}) and (\ref{eq:J1primea-expl}),
respectively, but now the second line in
Eq.~(\ref{eq:lc}) is employed for the $d_{\mu\nu}(p-l-P)$ and
$d_{\sigma\alpha}(P-p)$ Lorentz structures.

\begin{itemize}
\item{Point-like coupling (to avoid divergences
we assume that the $\bm{p}_{\sT}^2$
integration is extended up to a finite cutoff $\Lambda_a^2$)
\begin{align}
2\,\mathrm{Im}\,J_1^a &=g_a \intl\,\frac{1}{D_1\,D_3}\,(2\pi i)\,
\delta(D_2)\,(-2\pi i)\, \delta(D_4) \nn \\
&\times \Bigg\{ X(x,{\bm p}_{\sT}^2)\,\left( 1+
\frac{{\bm l}_{\sT}^2}{2M_a^2}\right) \, {\bm l}_{\sT}\cdot \left(
{\bm p}_{\sT}-\frac{1}{2}\,{\bm l}_{\sT}\right) + \nn \\
&\qquad \left( \frac{{\bm l}_{\sT}\cdot {\bm p}_{\sT}} {{\bm
p}_{\sT}^2} \right) \, \bigg[ Y_1(x,{\bm p}_{\sT}^2)\, + Y_2(x,{\bm
p}_{\sT}^2) \, ({\bm l}_{\sT}\cdot {\bm p}_{\sT}) +
Y_3(x,{\bm p}_{\sT}^2)\, {\bm l}_{\sT}^2 \bigg] + \nn \\
&\qquad W(x,{\bm p}_{\sT}^2)\, \left(\frac{{\bm l}_{\sT}\cdot {\bm
S}_{\sT}} {{\bm S}_{\sT}^2}\right) {\bm S}_{\sT} \cdot \left[ {\bm
l}_{\sT} -
2{\bm p}_{\sT}\right] \Bigg\} \; , \nn \\
2\,\mathrm{Im}\,J_1^{\prime\,a} &=g_a \intl\,\frac{1}{D_1\,D_3}\,
(2\pi i)\, \delta(D_2)\,(-2\pi i)\, \delta(D_4) \nn \\
&\times \Bigg\{ -X(x,{\bm p}_{\sT}^2)\, \left( 1+
\frac{{\bm l}_{\sT}^2}{2M_a^2}\right) \, {\bm l}_{\sT}\cdot \left(
{\bm p}_{\sT}-\frac{1}{2}\,{\bm l}_{\sT}\right) + \nn \\
&\qquad \left(
\frac{{\bm l}_{\sT}\cdot {\bm p}_{\sT}}
{{\bm p}_{\sT}^2} \right) \, \Bigg[ Y_1(x,{\bm p}_{\sT}^2)\, -
Y_2(x,{\bm p}_{\sT}^2) \, ({\bm l}_{\sT}\cdot {\bm p}_{\sT}) -
Y_3(x,{\bm p}_{\sT}^2)\, {\bm l}_{\sT}^2 + \nn \\
&\qquad \qquad +2M_a^2\,[(m+xM)\,(1-x)-xm]
+M_a^2\,[m\,(\kappa_a-1)-M(1+\kappa_a)\,x]\, {\bm l}_{\sT}^2 \Bigg]
\Bigg\} \; , \label{eq:B-J1a-pl-cov}
\end{align}
where the integrals $X, W, Y_i,\, i=1-3$ are listed in
App.~\ref{sec:C}. Unfortunately, most of the above combinations are
divergent under the $d{\bm l}_{\sT}$ integration. This is a typical
pathology when choosing the point-like form factor for the
nucleon-quark-diquark vertex, without any {\it ad-hoc} cut-off.
}
\item{Dipolar form factor
\begin{align}
2\,\mathrm{Im}\,J_1^a &=g_a \intl\,\frac{1}{D_1\,(D_3)^2}\,(2\pi i)\,
\delta(D_2)\,(-2\pi i)\, \delta(D_4) \nn \\
&\times \Bigg\{ X(x,{\bm p}_{\sT}^2)\, \left( 1-\frac{1}{2}\,{\bm
l}_{\sT}^2 + \ds{\frac{{\bm l}_{\sT}\cdot {\bm p}_{\sT}}{{\bm
p}_{\sT}^2}}\, \ds{\frac{{\bm l}_{\sT}^2}{2M_a^2}} - \ds{\frac{{\bm
l}_{\sT}^4}{4M_a^4}}
\right) + \nn \\
&\qquad W(x,{\bm p}_{\sT}^2)\left( \frac{{\bm l}_{\sT}\cdot {\bm
S}_{\sT}}{{\bm S}_{\sT}^2}\right) {\bm S}_{\sT}\cdot \left[ {\bm
l}_{\sT}-2{\bm p}_{\sT}\right] + \nn \\
&\qquad \left(\frac{{\bm l}_{\sT}\cdot {\bm p}_{\sT}}{{\bm
p}_{\sT}^2}\right) \left[ Y_1(x,{\bm p}_{\sT}^2)+ Y_2(x,{\bm
p}_{\sT}^2)\, \frac{{\bm l}_{\sT}\cdot {\bm p}_{\sT}}{{\bm
p}_{\sT}^2} +
Y_3(x,{\bm p}_{\sT}^2)\, {\bm l}_{\sT}^2 \right] \Bigg\} \; , \nn \\
2\,\mathrm{Im}\,J_1^{\prime\,a} &=g_a \intl\,\frac{1}{D_1\,(D_3)^2}\,
(2\pi i)\, \delta(D_2)\,(-2\pi i)\, \delta(D_4) \nn \\
&\times \Bigg\{ -X(x,{\bm p}_{\sT}^2)\, \left( 1+
\frac{{\bm l}_{\sT}^2}{2M_a^2}\right)\, {\bm l}_{\sT}\cdot \left(
{\bm p}_{\sT}-\frac{1}{2}\,{\bm l}_{\sT}\right) + \nn \\
&\qquad \left(\frac{{\bm l}_{\sT}\cdot {\bm p}_{\sT}}{{\bm
p}_{\sT}^2}\right)\Bigg[ Y_1(x,{\bm p}_{\sT}^2)- Y_2(x,{\bm
p}_{\sT}^2)\, {\bm l}_{\sT}\cdot {\bm p}_{\sT} - Y_3(x,{\bm
p}_{\sT}^2)\,
{\bm l}_{\sT}^2 + \nn \\
&\qquad \quad 2M_a^2\,[(m+xM)\,(1-x)-mx] +
M_a^2\,[m\,(\kappa_a-1)-M(1+\kappa_a)\,x]\, {\bm l}_{\sT}^2 \Bigg]
\Bigg\} \; , \label{eq:B-J1a-dip-cov}
\end{align}
where the integrals $X, W, Y_i,\, i=1-3$ are listed in
App.~\ref{sec:C}. Unfortunately, most of the above combinations are
divergent under the $d{\bm l}_{\sT}$ integration, unless a dipolar
form factor is considered with a higher degree, for example
proportional to $[{\bm p}_{\sT}^2+ L_a^2(\Lambda_a^2)]^{-3}$ in
Eq.~(\ref{eq:ffdip}). This would introduce a $1/(D_3)^3$ term inside
Eq.~(\ref{eq:B-J1a-dip-cov}), instead of $1/(D_3)^2$, and grant the
convergence of the various integrals. With this very choice, we
obtain
\begin{align}
2\,\mathrm{Im}\,J_1^a &=g_a \, \frac{(1-x)^2}{2P^+}\,\Bigg\{ X(x,{\bm
p}_{\sT}^2)\, \left[ {\bm p}_{\sT}^2\, \mathcal{I}_1^{\prime\,dip} + \left(
\frac{{\bm p}_{\sT}^2}{2M_a^2}- \frac{1}{2}\right)\,
\mathcal{I}_2^{\prime\,dip} - \frac{1}{4M_a^2}\, \mathcal{I}_5^{\prime\,dip} \right] +
\nn
\\
&\qquad Y_1(x,{\bm p}_{\sT}^2)\, \mathcal{I}_1^{\prime\,dip} + Y_2(x,{\bm
p}_{\sT}^2)
\, \mathcal{I}_3^{\prime\,dip} + Y_3(x,{\bm p}_{\sT}^2)\, \mathcal{I}_2^{\prime\,dip} + \nn \\
&\qquad W(x,{\bm p}_{\sT}^2)\, \left( \mathcal{I}_7^{\prime\,dip} - 2{\bm
p}_{\sT} \cdot {\bm S}_{\sT} \mathcal{I}_6^{\prime\,dip}\right) \Bigg\}
\; , \nn \\
2\,\mathrm{Im}\,J_1^{\prime\,a} &=g_a\, \frac{(1-x)^2}{2P^+}\,\Bigg\{
X(x,{\bm p}_{\sT}^2)\, \left[ -{\bm p}_{\sT}^2\,\mathcal{I}_1^{\prime\,dip} +
\left(\frac{1}{2} - \ds{\frac{{\bm p}_{\sT}^2}{2M_a^2}}\right)
\,\mathcal{I}_2^{\prime\,dip} + \frac{1}{4M_a^2}\,
\mathcal{I}_5^{\prime\,dip} \right] + \nn \\
&\qquad Y_1(x,{\bm p}_{\sT}^2)\, \mathcal{I}_1^{\prime\, dip} - 
Y_2(x,{\bm p}_{\sT}^2)\, \mathcal{I}_3^{\prime\,dip} - Y_3(x,{\bm p}_{\sT}^2)\, 
\mathcal{I}_2^{\prime\,dip} + \nn \\
&\qquad \quad 2M_a^2\,[(m+xM)\,(1-x)-mx]\,\mathcal{I}_1^{\prime\,dip} +
M_a^2\,[m\,(\kappa_a-1)-M(1+\kappa_a)\,x]\,\mathcal{I}_2^{\prime\,dip} \Bigg\}
\; ,
\end{align}
where the integrals $\mathcal{I}_i^{\prime\,dip}, \, i=1-7$, are listed in
App.~\ref{sec:C}.

The final result is, then,
\begin{align}
f_{1T}^{\perp\,q(a)}(x,{\bm p}_{\sT}^2) &=
-\frac{g_a^2}{32}\,\frac{M\,e_c^2}{(2\pi)^3}\, \frac{(1-x)^4}
{(P^+)^2\, [L_a^2(\Lambda_a^2) + {\bm p}_{\sT}^2]^3}
\Bigg\{ X(x,{\bm p}_{\sT}^2)\, \left[ {\bm p}_{\sT}^2\,
\mathcal{I}_1^{\prime\,dip} + \left( \frac{{\bm
p}_{\sT}^2}{2M_a^2}-\frac{1}{2}\right)\, \mathcal{I}_2^{\prime\,dip} -
\frac{1}{4M_a^2}\, \mathcal{I}_5^{\prime\,dip} \right] 
\nn \\ & \quad + Y_1(x,{\bm
p}_{\sT}^2)\, \mathcal{I}_1^{\prime\,dip} +
Y_2(x,{\bm p}_{\sT}^2)\, \mathcal{I}_3^{\prime\,dip} 
+ Y_3(x,{\bm p}_{\sT}^2)\, \mathcal{I}_2^{\prime\,dip} + W(x,{\bm
p}_{\sT}^2)\, \left( \mathcal{I}_7^{\prime\,dip} - 2{\bm p}_{\sT}\cdot {\bm
S}_{\sT}\, \mathcal{I}_6^{\prime\,dip}\right) \Bigg\} \; ,
\nn \\
h_1^{\perp\,q(a)}(x,{\bm p}_{\sT}^2) &=
-\frac{g_a^2}{32}\,\frac{M\,e_c^2}{(2\pi)^3}\, \frac{(1-x)^4}
{(P^+)^2\, [L_a^2(\Lambda_a^2) + {\bm p}_{\sT}^2]^3}
\Bigg\{ X(x,{\bm p}_{\sT}^2)\, \left[ -{\bm
p}_{\sT}^2\,\mathcal{I}_1^{\prime\,dip} + \left(\frac{1}{2} - \ds{\frac{{\bm
p}_{\sT}^2}{2M_a^2}}\right) \,\mathcal{I}_2^{\prime\,dip} +
\frac{1}{4M_a^2}\,\mathcal{I}_5^{\prime\,dip} \right] 
\nn \\ & \quad 
+ Y_1(x,{\bm
p}_{\sT}^2)\, \mathcal{I}_1^{\prime\,dip} -
Y_2(x,{\bm p}_{\sT}^2)\, \mathcal{I}_3^{\prime\,dip} 
- Y_3(x,{\bm p}_{\sT}^2)\, \mathcal{I}_2^{\prime\,dip} +
2M_a^2\,[(m+xM)\,(1-x)-mx]\,\mathcal{I}_1^{\prime\,dip} 
\nn \\ & \quad 
+ M_a^2\,[m\,(\kappa_a-1)-M(1+\kappa_a)\,x]\,\mathcal{I}_2^{\prime\,dip} \Bigg\}
\; . 
\label{eq:SivBM2-a-dip-covar}
\end{align}
}

\item{Exponential form factor
\begin{align}
2\,\mathrm{Im}\,J_1^a &=g_a \intl\,
\frac{e^{[(p-l)^2-m^2]/\Lambda_a^2}}{D_1\,D_3}\,(2\pi i)\,
\delta(D_2)\,(-2\pi i)\, \delta(D_4) \nn \\
&\times \Bigg\{ X(x,{\bm p}_{\sT}^2)\, \left( 1+
\frac{{\bm l}_{\sT}^2}{2M_a^2}\right)\, {\bm l}_{\sT}\cdot \left(
{\bm p}_{\sT}-\frac{1}{2}\,{\bm l}_{\sT}\right)  + \nn \\
&\qquad W(x,{\bm p}_{\sT}^2)\left( \frac{{\bm l}_{\sT}\cdot {\bm
S}_{\sT}}{{\bm S}_{\sT}^2}\right) {\bm S}_{\sT}\cdot \left[ {\bm
l}_{\sT} - 2{\bm p}_{\sT}\right] + \nn \\
&\qquad \left(\frac{{\bm l}_{\sT}\cdot {\bm p}_{\sT}}{{\bm
p}_{\sT}^2}\right)\bigg[ Y_1(x,{\bm p}_{\sT}^2) + {\bm l}_{\sT}\cdot
{\bm p}_{\sT}\, Y_2(x,{\bm p}_{\sT}^2) + {\bm l}_{\sT}^2\,
Y_3(x,{\bm p}_{\sT}^2)
\bigg] \Bigg\}  \nn \\
&=\frac{1}{2P^+}\,\Bigg\{ X(x,{\bm p}_{\sT}^2)\, \left[
{\bm p}_{\sT}^2\, \mathcal{I}_1^{exp} + \left( \frac{{\bm p}_{\sT}^2}{2M_a^2}-
\frac{1}{2}\right)\, \mathcal{I}_2^{exp} - \frac{1}{4M_a^2}\, \mathcal{I}_5^{exp} \right] + \nn\\
&\qquad Y_1(x,{\bm p}_{\sT}^2)\, \mathcal{I}_1^{exp} + Y_2(x,{\bm p}_{\sT}^2)
\, \mathcal{I}_3^{exp} + Y_3(x,{\bm p}_{\sT}^2)\, \mathcal{I}_2^{exp} + \nn \\
&\qquad W(x,{\bm p}_{\sT}^2)\, \left( \mathcal{I}_7^{exp} - 2{\bm p}_{\sT}
\cdot {\bm S}_{\sT}\,\mathcal{I}_6^{exp}\right) \Bigg\}
\; , \nn \\
2\,\mathrm{Im}\,J_1^{\prime\,a} &=g_a \intl\,
\frac{e^{[(p-l)^2-m^2]/\Lambda_a^2}}{D_1\,D_3}\,(2\pi i)\,
\delta(D_2)\,(-2\pi i)\, \delta(D_4) \nn \\
&\times \Bigg\{ -X(x,{\bm p}_{\sT}^2)\,\left( 1+
\frac{{\bm l}_{\sT}^2}{2M_a^2}\right)\,{\bm l}_{\sT}\cdot \left(
{\bm p}_{\sT}-\frac{1}{2}\,{\bm l}_{\sT}\right) + \nn \\
&\qquad \left(\frac{{\bm l}_{\sT}\cdot {\bm p}_{\sT}}{{\bm
p}_{\sT}^2}\right) \Bigg[ Y_1(x,{\bm p}_{\sT}^2) - {\bm
l}_{\sT}\cdot {\bm p}_{\sT}\, Y_2(x,{\bm p}_{\sT}^2) - {\bm
l}_{\sT}^2\, Y_3(x,{\bm p}_{\sT}^2) +
\nn \\
&\qquad \quad + 2M_a^2\,[(m+xM)\,(1-x)-mx] +
M_a^2\,[m(\kappa_a-1)-M(1+\kappa_a)\,x]\,
{\bm l}_{\sT}^2 \Bigg] \Bigg\}\nn \\
&=\frac{1}{2P^+}\,\Bigg\{ X(x,{\bm p}_{\sT}^2)\, \left[ -{\bm
p}_{\sT}^2\,\mathcal{I}_1^{exp} + \left(\frac{1}{2} - \ds{\frac{{\bm
p}_{\sT}^2}{2M_a^2}}\right) \,\mathcal{I}_2^{exp} + \frac{1}{4M_a^2}\,
\mathcal{I}_5^{exp} \right] + \nn \\
&\qquad Y_1(x,{\bm p}_{\sT}^2)\, \mathcal{I}_1^{exp} - Y_2(x,{\bm p}_{\sT}^2)
\, \mathcal{I}_3^{exp} - Y_3(x,{\bm p}_{\sT}^2)\, \mathcal{I}_2^{exp} + \nn \\
&\qquad \quad 2M_a^2\,[(m+xM)\,(1-x)-mx]\,\mathcal{I}_1^{exp} + M_a^2\,
[m(\kappa_a-1)-M(1+\kappa_a)\,x]\,\mathcal{I}_2^{exp} \Bigg\} \; ,
\label{eq:B-J1a-exp-cov}
\end{align}
where the integrals $X, W, Y_i,\, i=1-3$, and $\mathcal{I}_i^{exp}, \, i=1-7$,
are listed in App.~\ref{sec:C}.

The final result is, then,
\begin{align}
f_{1T}^{\perp\,q(a)}(x,{\bm p}_{\sT}^2) &=
-\frac{g_a^2}{32}\,\frac{M\,e_c^2}{(2\pi)^3}\,
\frac{e^{-[{\bm p}_{\sT}^2 + L_a^2(m^2)]/[(1-x)\,\Lambda_a^2]}}
{(P^+)^2\, [L_a^2(m^2) + {\bm p}_{\sT}^2]}
\Bigg\{ X(x,{\bm p}_{\sT}^2)\, \left[ {\bm p}_{\sT}^2\,
\mathcal{I}_1^{exp} + \left( \frac{{\bm
p}_{\sT}^2}{2M_a^2}-\frac{1}{2}\right)\, \mathcal{I}_2^{exp} -
\frac{1}{4M_a^2}\, \mathcal{I}_5^{exp} \right] 
\nn \\ & \quad 
+ Y_1(x,{\bm p}_{\sT}^2)\,
\mathcal{I}_1^{exp} +
Y_2(x,{\bm p}_{\sT}^2)\, \mathcal{I}_3^{exp} 
+ Y_3(x,{\bm p}_{\sT}^2)\, \mathcal{I}_2^{exp} + W(x,{\bm p}_{\sT}^2)\,
\left( \mathcal{I}_7^{exp} - 2{\bm p}_{\sT}\cdot {\bm S}_{\sT}\,
\mathcal{I}_6^{exp}\right) \Bigg\} \; ,
\nn \\
h_1^{\perp\,q(a)}(x,{\bm p}_{\sT}^2) &=-\frac{g_a^2}{32}\,
\frac{M\,e_c^2}{(2\pi)^3}\,
\frac{e^{-[{\bm p}_{\sT}^2 + L_a^2(m^2)]/[(1-x)\,\Lambda_a^2]}}
{(P^+)^2\, [L_a^2(m^2) + {\bm p}_{\sT}^2]}
\Bigg\{ X(x,{\bm p}_{\sT}^2)\, \left[ -{\bm
p}_{\sT}^2\,\mathcal{I}_1^{exp} + \left(\frac{1}{2} - \ds{\frac{{\bm
p}_{\sT}^2}{2M_a^2}}\right) \,\mathcal{I}_2^{exp} +
\frac{1}{4M_a^2}\,\mathcal{I}_5^{exp} \right] 
\nn \\ & \quad 
+ Y_1(x,{\bm p}_{\sT}^2)\,
\mathcal{I}_1^{exp} -
Y_2(x,{\bm p}_{\sT}^2)\, \mathcal{I}_3^{exp} 
- Y_3(x,{\bm p}_{\sT}^2)\, \mathcal{I}_2^{exp} +
2M_a^2\,[(m+xM)\,(1-x)-mx]\,\mathcal{I}_1^{exp} 
\nn \\ & \quad 
+
M_a^2\,[m(\kappa_a-1)-M(1+\kappa_a)\,x] \,\mathcal{I}_2^{exp} \Bigg\} \; .
\label{eq:SivBM2-a-exp-covar}
\end{align}
}
\end{itemize}

\subsection{Axial-vector diquark including also time-like polarization}
\label{sec:toddtimelike}

We have
\begin{align}
f_{1T}^{\perp\,q(a)}(x,{\bm p}_{\sT}^2) &=
\frac{g_a(p^2)}{4}\,\frac{1}{(2\pi)^3}\,
\frac{M\,e_c^2}{4(1-x)P^+}\, \frac{2\,\mathrm{Im}\,J_1^a}{p^2-m^2}
\nn \\
h_1^{\perp\,q(a)}(x,{\bm p}_{\sT}^2) &=\frac{g_a(p^2)}{4}\,
\frac{1}{(2\pi)^3}\, \frac{M\,e_c^2}{4(1-x)P^+}\,
\frac{2\,\mathrm{Im}\,J_1^{\prime\,a}}{p^2-m^2} \; ,
\label{eq:SivBM1-a-pl-feyn}
\end{align}
where the $J_1^a$ and $J_1^{\prime\,a}$ integrals are defined as in
Eqs.~(\ref{eq:J1a-expl}) and (\ref{eq:J1primea-expl}),
respectively, but now the last line in Eq.~(\ref{eq:lc})) 
is employed for the $d_{\mu\nu}(p-l-P)$ and
$d_{\sigma\alpha}(P-p)$ Lorentz structures.

\begin{itemize}
\item{Point-like coupling (to avoid divergences
we assume that the $\bm{p}_{\sT}^2$
integration is extended up to a finite cutoff $\Lambda_a^2$)
\begin{align}
2\,\mathrm{Im}\,J_1^a &=-4P^+x\,\left[
m\,(2\kappa_a+1)+M\,(2\kappa_a\,x + 1)
\right] \, g_a\,\mathcal{I}_1^{p.l.} \nn \\
&=-g_a\, \frac{P^+\,x\,\left[ m\,(2\kappa_a+1)+M\,(2\kappa_a\,x +
1)\right]} {\pi {\bm p}_{\sT}^2}\, \log \left(
\frac{L_a^2(m^2) + {\bm p}_{\sT}^2}{L_a^2(m^2)} \right) \nn \\
2\,\mathrm{Im}\,J_1^{\prime\,a} &=4P^+\,\left[
m\,[(2\kappa_a-1)\,x+2]+xM\,[(\kappa_a-1)\,2x + 3]\right] \,g_a\,
\mathcal{I}_1^{p.l.}
\nn \\
&=g_a\, \frac{P^+\,\left[ m\,[(2\kappa_a-1)\,x+2]+xM\,[(\kappa_a-1)\,2x +
3]\right]} {\pi {\bm p}_{\sT}^2}\, \log \left( \frac{L_a^2(m^2) +
{\bm p}_{\sT}^2}{L_a^2(m^2)} \right) \; , \label{eq:B-J1a-pl-feyn}
\end{align}
where $\mathcal{I}_1^{p.l.}$ is the same integral as in Eq.~(\ref{eq:B-J1s}) but
with the substitution $L_s(m^2) \leftrightarrow L_a(m^2)$.

Using again Eq.~(\ref{eq:offshell}), the final result is
\begin{align}
f_{1T}^{\perp\,q(a)}(x,{\bm p}_{\sT}^2) &= \frac{g_a^2}{8}\,
\frac{M\,e_c^2}{(2\pi)^4}\, \frac{x\,\left[
m\,(2\kappa_a+1)+M\,(2\kappa_a\,x + 1)\right]} {{\bm p}_{\sT}^2\,
[L_a^2(m^2) + {\bm p}_{\sT}^2]}\, \log \left( \frac{L_a^2(m^2) +
{\bm p}_{\sT}^2}{L_a^2(m^2)} \right)
\nn \\
h_1^{\perp\,q(a)}(x,{\bm p}_{\sT}^2) &=-\frac{g_a^2}{8}\,
\frac{M\,e_c^2}{(2\pi)^4}\,
\frac{m\,[(2\kappa_a-1)\,x+2]+xM\,[(\kappa_a-1)\,2x + 3]} {{\bm
p}_{\sT}^2\, [L_a^2(m^2) + {\bm p}_{\sT}^2]}\, \log \left(
\frac{L_a^2(m^2) + {\bm p}_{\sT}^2}{L_a^2(m^2)} \right) \; .
\label{eq:SivBM2-a-pl-feyn}
\end{align}
}
\item{Dipolar form factor
\begin{align}
2\,\mathrm{Im}\,J_1^a &=4P^+x\,(1-x)\,\left[
m\,(2\kappa_a+1)+M\,(2\kappa_a\,x + 1)\right] \,g_a\, \mathcal{I}_1^{dip} \nn \\
&=g_a\, \frac{P^+\,x\,(1-x)\,\left[ m\,(2\kappa_a+1)+M\,(2\kappa_a\,x +
1)\right]} {\pi \, L_a^2(\Lambda_a^2) \, [L_a^2(\Lambda_a^2) + {\bm
p}_{\sT}^2]}
\nn \\
2\,\mathrm{Im}\,J_1^{\prime\,a} &=-4P^+\,(1-x)\,\left[
m\,[(2\kappa_a-1)\,x+2]+xM\,[(\kappa_a-1)\,2x + 3]\right] \,g_a\,
\mathcal{I}_1^{dip}
\nn \\
&=-g_a\, \frac{P^+\,(1-x)\,\left[
m\,[(2\kappa_a-1)\,x+2]+xM\,[(\kappa_a-1)\,2x + 3] \right]} {\pi\,
L_a^2(\Lambda_a^2 )\,[L_a^2(\Lambda_a^2 ) + {\bm p}_{\sT}^2]} \; .
\label{eq:B-J1a-dip-feyn}
\end{align}

The final result is
\begin{align}
f_{1T}^{\perp\,q(a)}(x,{\bm p}_{\sT}^2) &=
\frac{g_a^2}{8}\,\frac{M\,e_c^2}{(2\pi)^4}\, \frac{x\,
(1-x)^2\,\left[ m\,(2\kappa_a+1)+M\,(2\kappa_a\,x + 1)\right]}
{L_a^2(\Lambda_a^2) \, [L_a^2(\Lambda_a^2) + {\bm p}_{\sT}^2]^3} \;
,
\nn \\
h_1^{\perp\,q(a)}(x,{\bm p}_{\sT}^2) &=-\frac{g_a^2}{8}\,
\frac{M\,e_c^2}{(2\pi)^4}\, \frac{(1-x)^2\,\left[
m\,[(2\kappa_a-1)\,x+2]+xM\,[(\kappa_a-1)\,2x + 3]\right]}
{L_a^2(\Lambda_a^2) \, [L_a^2(\Lambda_a^2) + {\bm p}_{\sT}^2]^3} \;
. \label{eq:SivBM2-a-dip-feyn}
\end{align}
We find a discrepancy between these results and those of Eqs.~(18) and (24) in
Ref.~\cite{Bacchetta:2003rz}, probably due to errors in that calculation. 
}
\item{Exponential form factor
\begin{align}
2\,\mathrm{Im}\,J_1^a &=-4P^+x\,\left[
m\,(2\kappa_a+1)+M\,(2\kappa_a\,x + 1)
\right] \,g_a\, \mathcal{I}_1^{exp} \nn \\
&=-g_a\, \frac{P^+\,x\,\left[ m\,(2\kappa_a+1)+M\,(2\kappa_a\,x +
1)\right]} {\pi {\bm p}_{\sT}^2}\, \left[ \Gamma \left( 0,
\frac{L_a^2(m^2)}{(1-x)\,\Lambda_a^2} \right) - \Gamma \left( 0,
\frac{L_a^2(m^2)+{\bm p}_{\sT}^2}{(1-x)\,\Lambda_a^2} \right)
\right]
\; , \nn \\
2\,\mathrm{Im}\,J_1^{\prime\,a} &=4P^+\,\left[
m\,[(2\kappa_a-1)\,x+2]+xM\,[(\kappa_a-1)\,2x + 3]\right] \,g_a\,
\mathcal{I}_1^{exp}
\nn \\
&=g_a\, \frac{P^+\,\left[ m\,[(2\kappa_a-1)\,x+2]+xM\,[(\kappa_a-1)\,2x +
3]\right]} {\pi {\bm p}_{\sT}^2}\, \left[ \Gamma \left( 0,
\frac{L_a^2(m^2)}{(1-x)\,\Lambda_a^2} \right) - \Gamma \left( 0,
\frac{L_a^2(m^2)+{\bm p}_{\sT}^2}{(1-x)\,\Lambda_a^2} \right)
\right] \; . \label{eq:B-J1a-exp-feyn}
\end{align}

The final result is, then,
\begin{align}
f_{1T}^{\perp\,q(a)}(x,{\bm p}_{\sT}^2) &=
\frac{g_a^2}{8}\,\frac{M\,e_c^2}{(2\pi)^4}\, \frac{x\,\left[
m\,(2\kappa_a+1)+M\,(2\kappa_a\,x + 1)\right]} {{\bm p}_{\sT}^2\,
[L_a^2(m^2) + {\bm p}_{\sT}^2]}\,
e^{-[{\bm p}_{\sT}^2 + L_a^2(m^2)]/[(1-x)\,\Lambda_a^2]} \nn \\
&\hspace{3cm} \times
\left[ \Gamma \left( 0,\frac{L_a^2(m^2)}{(1-x)\,\Lambda_a^2} \right) -
\Gamma \left( 0, \frac{L_a^2(m^2)+{\bm p}_{\sT}^2}{(1-x)\,\Lambda_a^2}
 \right) \right] \; , \nn \\
h_1^{\perp\,q(a)}(x,{\bm p}_{\sT}^2) &=-\frac{g_a^2}{8}\,
\frac{M\,e_c^2}{(2\pi)^4}\,
\frac{m\,[(2\kappa_a-1)\,x+2]+xM\,[(\kappa_a-1)\,2x + 3]} {{\bm
p}_{\sT}^2\, [L_a^2(m^2) + {\bm p}_{\sT}^2]}\,
e^{-[{\bm p}_{\sT}^2 + L_a^2(m^2)]/[(1-x)\,\Lambda_a^2]} \nn \\
&\hspace{3cm} \times \left[ \Gamma \left( 0,
\frac{L_a^2(m^2)}{(1-x)\,\Lambda_a^2} \right) - \Gamma \left( 0,
\frac{L_a^2(m^2)+{\bm p}_{\sT}^2}{(1-x)\,\Lambda_a^2} \right) \right]
\; .
\label{eq:SivBM2-a-exp-feyn}
\end{align}
}
\end{itemize}

\section{Useful integrals}
\label{sec:C}

In this appendix, we calculate the relevant integrals that repeatedly
show up in the expressions of T-odd parton densities for all choices
of nucleon-quark-diquark form factors, when vector diquark
propagators are represented in the first and last
forms of Eq.~(\ref{eq:lc}).

We will systematically use the substitutions
${\bm l}_{\sT}' = {\bm l}_{\sT}-{\bm p}_{\sT}$ and
$y = {\bm l}_{\sT}^{\prime\,2}+L_X^2(m^2)$, or
$y={\bm l}_{\sT}^{\prime\,2}+L_X^2(\Lambda_X^2)$ for the dipolar form
factor, where $L_X$ is defined in Eq.~(\ref{eq:offshell}) for $X=s,a$ scalar
and axial-vector diquarks, respectively. We will also encounter the
following angular integrals, where $\theta$ is defined as the angle
between ${\bm l}_{\sT}'$ and ${\bm p}_{\sT}$, and $\phi$, $\phi_S$,
are the azimuthal angles of ${\bm p}_{\sT}$ and ${\bm S}_{\sT}$ with
respect to the scattering plane:
\begin{align}
&\int_0^{2\pi}d\theta\,
\frac{|{\bm l}_{\sT}'||{\bm p}_{\sT}|\cos\theta+{\bm p}_{\sT}^2}
{{\bm l}_{\sT}^{\prime\,2}+{\bm p}_{\sT}^2+2|{\bm l}_{\sT}'|
|{\bm p}_{\sT}|\cos\theta}=  \pi\, \left( 1-\mathrm{sgn}
(|{\bm l}_{\sT}'|-|{\bm p}_{\sT}|) \right) \; , \nn \\[0.2cm]
&\int_0^{2\pi}d\theta\,
\frac{\left[ |{\bm l}_{\sT}'||{\bm p}_{\sT}|\cos\theta+{\bm p}_{\sT}^2
\right]^2}
{{\bm l}_{\sT}^{\prime\,2}+{\bm p}_{\sT}^2+2|{\bm l}_{\sT}'|
|{\bm p}_{\sT}|\cos\theta}= \frac{\pi}{2}\,\left(
-{\bm l}_{\sT}^{\prime\,2} + 3{\bm p}_{\sT}^2 + \big|
{\bm l}_{\sT}^{\prime\,2}-{\bm p}_{\sT}^2\big|\,\right) \; , \nn
\\[0.2cm]
&\int_0^{2\pi}d\theta\,
\frac{|{\bm l}_{\sT}'||{\bm S}_{\sT}| \cos[\theta+(\phi-\phi_S)] +
|{\bm p}_{\sT}||{\bm S}_{\sT}| \cos(\phi-\phi_S)}
{{\bm l}_{\sT}^{\prime\,2}+{\bm p}_{\sT}^2+2|{\bm l}_{\sT}'|
|{\bm p}_{\sT}|\cos\theta}= \pi\,
\frac{|{\bm S}_{\sT}|}{|{\bm p}_{\sT}|}\,\cos(\phi-\phi_S)\,
\left( \mathrm{sgn}(|{\bm p}_{\sT}|-|{\bm l}_{\sT}'|)+1 \right) \; ,
\nn \\[0.2cm]
&\int_0^{2\pi}d\theta\,
\frac{\left( |{\bm l}_{\sT}'||{\bm S}_{\sT}|
\cos[\theta+(\phi-\phi_S)] + |{\bm p}_{\sT}||{\bm S}_{\sT}|
\cos(\phi-\phi_S)\right)^2}
{{\bm l}_{\sT}^{\prime\,2}+{\bm p}_{\sT}^2+2|{\bm l}_{\sT}'|
|{\bm p}_{\sT}|\cos\theta}= \pi\,
\frac{{\bm S}_{\sT}^2}{2{\bm p}_{\sT}^2}\, \Big[\left( {\bm p}_{\sT}^2
-{\bm l}_{\sT}^{\prime\,2} + |{\bm p}_{\sT}^2-
{\bm l}_{\sT}^{\prime\,2}|\right)\, \cos 2(\phi-\phi_S) \nn \\
&\hspace{12cm} + 2{\bm p}_{\sT}^2 \Big] \; .
\label{eq:angles}
\end{align}

\begin{itemize}
\item  Point-like coupling
\begin{align}
\mathcal{I}_1^{p.l.}&=\int \frac{d{\bm l}_{\sT}'}{(2\pi)^2}\,
\frac{({\bm l}_{\sT}'+{\bm p}_{\sT})\cdot {\bm p}_{\sT}}
{{\bm p}_{\sT}^2}\,
\frac{1}{({\bm l}_{\sT}'+{\bm p}_{\sT})^2\,[{\bm l}_{\sT}^{\prime\,2}
+L_X^2(m^2)]} \nn \\
&=\frac{1}{(2\pi)^2{\bm p}_{\sT}^2} \int_0^\infty d|{\bm l}_{\sT}'|
|{\bm l}_{\sT}'|\,\frac{1}{{\bm l}_{\sT}^{\prime\,2}+L_X^2(m^2)}
\int_0^{2\pi}d\theta\,
\frac{|{\bm l}_{\sT}'||{\bm p}_{\sT}|\cos\theta+{\bm p}_{\sT}^2}
{{\bm l}_{\sT}^{\prime\,2}+{\bm p}_{\sT}^2+2|{\bm l}_{\sT}'|
|{\bm p}_{\sT}|\cos\theta} \nn \\
&=\frac{1}{(2\pi)\,{\bm p}_{\sT}^2} \int_0^{|{\bm p}_{\sT}|}
d|{\bm l}_{\sT}'||{\bm l}_{\sT}'|\,
\frac{1}{{\bm l}_{\sT}^{\prime\,2}+L_X^2(m^2)} \nn \\
&=\frac{1}{4\pi{\bm p}_{\sT}^2}
\int_{L_X^2(m^2)}^{L_X^2(m^2)+{\bm p}_{\sT}^2}\frac{dy}{y} =
\frac{1}{4\pi{\bm p}_{\sT}^2} \,\log \left(
\frac{L_X^2(m^2)+{\bm p}_{\sT}^2}{L_X^2(m^2)} \right) \; ;
\label{eq:I1pl}
\end{align}

\item  Dipolar form factor
\begin{align}
\mathcal{I}_1^{dip}&=-\int \frac{d{\bm l}_{\sT}'}{(2\pi)^2}\,
\frac{({\bm l}_{\sT}'+{\bm p}_{\sT})\cdot {\bm p}_{\sT}}
{{\bm p}_{\sT}^2}\,
\frac{(1-x)}{({\bm l}_{\sT}'+{\bm p}_{\sT})^2\,[{\bm l}_{\sT}^{\prime\,2}+
L_X^2(\Lambda_X^2)]^2} \nn \\
&=-\frac{(1-x)}{4\pi{\bm p}_{\sT}^2}
\int_{L_X^2(\Lambda_X^2)}^{L_X^2(\Lambda_X^2)+{\bm p}_{\sT}^2}
\frac{dy}{y^2} = -
\frac{(1-x)}{4\pi L_X^2(\Lambda_X^2)\,[L_X^2(\Lambda_X^2)+{\bm p}_{\sT}^2]}
\; ;
\label{eq:I1dip}
\end{align}

\item  Exponential form factor
\begin{align}
\mathcal{I}_1^{exp}&=\int \frac{d{\bm l}_{\sT}'}{(2\pi)^2}\,
\frac{({\bm l}_{\sT}'+{\bm p}_{\sT})\cdot {\bm p}_{\sT}}
{{\bm p}_{\sT}^2}\,
\frac{e^{-[{\bm l}_{\sT}^{\prime\,2}+L_X^2(m^2)]/[(1-x)\Lambda_X^2]}}
{({\bm l}_{\sT}'+{\bm p}_{\sT})^2\,[{\bm l}_{\sT}^{\prime\,2}+
L_X^2(m^2)]} \nn\\
&=\frac{1}{(2\pi)\,{\bm p}_{\sT}^2} \int_0^{|{\bm p}_{\sT}|}
d|{\bm l}_{\sT}'||{\bm l}_{\sT}'|\,
\frac{e^{-[{\bm l}_{\sT}^{\prime\,2}+L_X^2(m^2)]/[(1-x)\Lambda_X^2]}}
{{\bm l}_{\sT}^{\prime\,2}+L_X^2(m^2)} \nn \\
&=\frac{1}{(4\pi)\,{\bm p}_{\sT}^2}
\int_{L_X^2(m^2)}^{L_X^2(m^2)+{\bm p}_{\sT}^2} \frac{dy}{y}\,
e^{-y/[(1-x)\Lambda_X^2]} \nn \\
&=\frac{1}{4\pi {\bm p}_{\sT}^2}\, \left[ \Gamma \left( 0,
\frac{L_X^2(m^2)}{(1-x)\Lambda_X^2} \right) -
\Gamma \left( 0, \frac{L_X^2(m^2)+{\bm p}_{\sT}^2}{(1-x)\Lambda_X^2}
\right) \right] \; .
\label{eq:I1exp}
\end{align}

\end{itemize}

Next, we list the coefficients and calculate the relevant
integrals that are needed to construct T-odd parton densities for all
choices of nucleon-quark-diquark form factors, when vector
diquarks are represented in the second form in
Eq.~(\ref{eq:lc}).

\begin{align}
X &=-\frac{16 (m + M)\, (P^+)^2 \, \kappa_a}{M_a^2} \; , \nn \\
W &=-\frac{8 (m + M)\, (P^+)^2 \,(\kappa_a-1) \, (1-x)}{M_a^2} \; , \nn \\
Y_1 &=-\frac{8 (P^+)^2}{M_a^2} \Bigg[ \kappa_a \,
       \Big( 2m^3 - 3xm^3 +Mx^2m^2 + 2Mm^2 -3 Mxm^2 -M^2xm +2M_a^2xm
              + M^3x^2  -M^3x\nn \\
    &\hspace{2cm} - [{\bm p}_{\sT}^2 + L_a^2(m^2) ]\, (M+m)
       \Big) + (1+\kappa_a)M_a^2x\, [M(1+x)+m] + [{\bm p}_{\sT}^2 + L_a^2(m^2) ]
        \, m\,(\kappa_a-1)\,(x-1) \nn \\
    &\hspace{2cm} + m\,(\kappa_a-1)\,x^2\,(M_a^2+m^2)+m^3\,(2x-1)+M^2m\,
           (1-2x+(1+\kappa_a)x^2) \Bigg] \; , \nn \\
Y_2 &=-\frac{8 m \,(P^+)^2\, (\kappa_a - 1) \,(1-x)}{M_a^2} \; , \nn \\
Y_3 &=\frac{4 (P^+)^2}{M_a^4} \, \Bigg[  \kappa_a \,
          \Big( -m^3 + x m^3 - M x^2 m^2 - 2 M m^2 + 3 M x m^2 -
                M^2 m - 2 M^2 x^2 m
                + 3 M^2 x m - 2 M_a^2 x m \nn \\
     &\hspace{2cm} - M^3 x^2 + [{\bm p}_{\sT}^2 + L_a^2(m^2)] \,
          (m + M) + M^3 x \Big) + M_a^2 m (\kappa_a-1) - M M_a^2 (1+\kappa_a) x \Bigg]
\; , \nn \\
\label{eq:Dcoeffs}
\end{align}

\begin{align}
\mathcal{I}_1^{\prime\,dip}&=\int \frac{d {\bm l}_{\sT}'}{(2\pi)^2}
\frac{({\bm l}_{\sT}'+{\bm p}_{\sT})\cdot {\bm p}_{\sT}}{{\bm p}_{\sT}^2}
\frac{1}{({\bm l}_{\sT}'+{\bm p}_{\sT})^2\,[{\bm l}_{\sT}^{\prime\,2} +L_a^2(\Lambda_a^2)]^3}
\nn\\
&=\int_0^\infty \frac{d|{\bm l}_{\sT}'||{\bm l}_{\sT}'|}{(2\pi)^2{\bm p}_{\sT}^2}
\frac{1}{[{\bm l}_{\sT}^{\prime\,2}+L_a^2(\Lambda_a^2)]^3}
\int_0^{2\pi}d\theta\,
\frac{|{\bm l}_{\sT}'||{\bm p}_{\sT}|\cos\theta+{\bm p}_{\sT}^2}{{\bm l}_{\sT}^{\prime\,2}+{\bm p}_{\sT}^2+2|{\bm l}_{\sT}'||{\bm p}_{\sT}|\cos\theta}
\nn\\
&=\int_0^{|{\bm p}_{\sT}|} \frac{d|{\bm l}_{\sT}'||{\bm l}_{\sT}'|}{(2\pi){\bm p}_{\sT}^2}
\frac{1}{[{\bm l}_{\sT}^{\prime\,2} +L_a^2(\Lambda_a^2)]^3}\;\;=\;\;
\frac{1}{4\pi{\bm p}_{\sT}^2}\int_{L_a^2(\Lambda_a^2)}^{L_a^2(\Lambda_a^2)+{\bm p}_{\sT}^2}\;\frac{dy}{y^3}
\nn\\
&=\frac{1}{8\pi{\bm p}_{\sT}^2}\left[\frac{1}{L_a^4(\Lambda_a^2)}-
\frac{1}{[L_a^2(\Lambda_a^2)+{\bm p}_{\sT}^2]^2}\right]  \nn\\
&= \frac{2L_a^2(\Lambda_a^2)+{\bm p}_{\sT}^2}{8\pi
L_a^4(\Lambda_a^2)\,[L_a^2(\Lambda_a^2)+{\bm p}_{\sT}^2]^2}\;, \nn
\end{align}

\begin{align}\mathcal{I}_2^{\prime\,dip}&=\int \frac{d {\bm l}_{\sT}'}{(2\pi)^2}
\frac{({\bm l}_{\sT}'+{\bm p}_{\sT})^2}{({\bm l}_{\sT}'+{\bm p}_{\sT})
^2\,[{\bm l}_{\sT}^{\prime\,2} +L_a^2(\Lambda_a^2)]^3}\nn\\
&=\int_0^\infty \frac{d|{\bm l}_{\sT}'||{\bm
l}_{\sT}'|}{(2\pi)^2}\frac{1}{[{\bm
l}_{\sT}^{\prime\,2}+L_a^2(\Lambda_a^2)]^3}\int_0^{2\pi}d\theta\;\;=\;\;
\frac{1}{4\pi}\int_{L_a^2(\Lambda_a^2)}^\infty\;\frac{dy}{y^3}\hspace{2cm}\;\nn\\
&=\frac{1}{8\pi L_a^4(\Lambda_a^2)}\;, 
\end{align}

\begin{align}\mathcal{I}_3^{\prime\,dip}&=\int \frac{d^2 {\bm l}_{\sT}'}{(2\pi)^2}
\frac{\left[({\bm l}_{\sT}'+{\bm p}_{\sT})\cdot {\bm p}_{\sT}\right]^2}{{\bm p}_{\sT}^2}
\frac{1}{({\bm l}_{\sT}'+{\bm p}_{\sT})^2\,[{\bm l}_{\sT}^{\prime\,2} +L_a^2(\Lambda_a^2)]^3}
\nn\\
&=\int_0^\infty \frac{d|{\bm l}_{\sT}'||{\bm l}_{\sT}'|}{(2\pi)^2{\bm p}_{\sT}^2}
\frac{1}{[{\bm l}_{\sT}^{\prime\,2}+L_a^2(\Lambda_a^2)]^3}
\int_0^{2\pi}d\theta\,
\frac{\Big[|{\bm l}_{\sT}'||{\bm p}_{\sT}|\cos\theta+{\bm p}_{\sT}^2\Big]^2}{{\bm l}_{\sT}^{\prime\,2}+{\bm p}_{\sT}^2+2|{\bm l}_{\sT}'||{\bm p}_{\sT}|\cos\theta}
\nn\\
&=\frac{1}{4\pi{\bm p}_{\sT}^2}\,\Bigg\{\int_0^{|{\bm
p}_{\sT}|}\frac{d|{\bm l}_{\sT}'||{\bm l}_{\sT}'|}{[{\bm
l}_{\sT}^{\prime\,2}+L_a^2(\Lambda_a^2)]^3}\,(2{\bm p}_{\sT}^2-{\bm
l}_{\sT}^{\prime\,2})+\,{\bm p}_{\sT}^2
\int_{|{\bm p}_{\sT}|}^\infty \frac{d|{\bm l}_{\sT}'||{\bm l}_{\sT}'|}{[{\bm l}_{\sT}^{\prime\,2}+L_a^2(\Lambda_a^2)]^3}\Bigg\}
\nn\\
&=\frac{1}{8\pi{\bm
p}_{\sT}^2}\,\Bigg\{\int_{L_a^2(\Lambda_a^2)}^{L_a^2(\Lambda_a^2)+{\bm p}_{\sT}^2}\,dy\,\frac{2{\bm p}_{\sT}^2+L_a^2(\Lambda_a^2)-y}{y^3}
\,+\,{\bm p}_{\sT}^2\int_{L_a^2(\Lambda_a^2)+{\bm p}_{\sT}^2}^\infty
\;\frac{dy}{y^3}\Bigg\}\nn\\
&=\frac{1}{8\pi{\bm p}_{\sT}^2}\,\Bigg\{\left(
\frac{2{\bm p}_{\sT}^{\,6}+3L_a^2(\Lambda_a^2)\,{\bm
p}_{\sT}^{\,4}}{2L_a^2(\Lambda_a^2)\,[L_a^2(\Lambda_a^2)+{\bm
p}_{\sT}^2]^2} \right)\,+\,{\bm p}_{\sT}^2
\left(\frac{1}{2\,[L_a^2(\Lambda_a^2)+{\bm p}_{\sT}^2]^2}\right)\Bigg\}\nn\\
&=\frac{L_a^2(\Lambda_a^2)+2{\bm p}_{\sT}^2}{16\pi
L_a^4(\Lambda_a^2)\,[L_a^2(\Lambda_a^2)+{\bm p}_{\sT}^2]}\;,\nn
\end{align}

\begin{align}
\mathcal{I}_4^{\prime\,dip}&=\int \frac{d {\bm l}_{\sT}'}{(2\pi)^2}\frac{({\bm
l}_{\sT}'+{\bm p}_{\sT})\cdot
{\bm p}_{\sT}}{{\bm p}_{\sT}^2}\frac{{({\bm l}_{\sT}'+{\bm p}_{\sT})}^2}
{({\bm l}_{\sT}'+{\bm p}_{\sT})^2\,[{\bm l}_{\sT}^{\prime\,2} +L_a^2(\Lambda_a^2)]^3}\nn\\
&=\int_0^\infty \frac{d|{\bm l}_{\sT}'||{\bm
l}_{\sT}'|}{(2\pi)^2{\bm p}_{\sT}^2}\frac{1}{[{\bm
l}_{\sT}^{\prime\,2}+L_a^2(\Lambda_a^2)]^3}\int_0^{2\pi}d\theta\,
\Big(|{\bm l}_{\sT}'||{\bm p}_{\sT}|\cos\theta+{\bm p}_{\sT}^2\Big)\qquad\quad\nn\\
&=\int_0^\infty \frac{d|{\bm l}_{\sT}'||{\bm
l}_{\sT}'|}{(2\pi)}\frac{1}{[{\bm
l}_{\sT}^{\prime\,2}+L_a^2(\Lambda_a^2)]^3}\;\;=\;\;
\frac{1}{4\pi}\int_{L_a^2(\Lambda_a^2)}^\infty\;\frac{dy}{y^3}\nn\\
&=\frac{1}{8\pi L_a^4(\Lambda_a^2)}\equiv \mathcal{I}_2^{\prime\,dip}\;,\nn
\end{align}

\begin{align}
\mathcal{I}_5^{\prime\,dip}&=\int \frac{d {\bm l}_{\sT}'}{(2\pi)^2}\frac{{({\bm l}_{\sT}'+{\bm p}_{\sT})}^4}
{({\bm l}_{\sT}'+{\bm p}_{\sT})^2\,[{\bm l}_{\sT}^{\prime\,2} +L_a^2(\Lambda_a^2)]^3}\nn\\
&=\int_0^\infty \frac{d|{\bm l}_{\sT}'||{\bm
l}_{\sT}'|}{(2\pi)^2}\frac{1}{[{\bm
l}_{\sT}^{\prime\,2}+L_a^2(\Lambda_a^2)]^3}\int_0^{2\pi}d\theta\,
\Big({\bm l}_{\sT}^{\prime\,2}+{\bm p}_{\sT}^2+2|{\bm l}_{\sT}'||{\bm p}_{\sT}|\cos\theta\Big)\nn\\
&=\int_0^\infty \frac{d|{\bm l}_{\sT}'||{\bm l}_{\sT}'|}{(2\pi)}\frac{({\bm l}_{\sT}^{\prime\,2}+{\bm p}_{\sT}^2)}
{[{\bm l}_{\sT}^{\prime\,2}+L_a^2(\Lambda_a^2)]^3}\;\;=\;\frac{1}{4\pi}\int_{L_a^2(\Lambda_a^2)}^\infty dy\,\frac{y-L_a^2(\Lambda_a^2)+{\bm p}_{\sT}^2}{y^3}\nn\\
&=\frac{L_a^2(\Lambda_a^2)+{\bm p}_{\sT}^2}{8\pi
L_a^4(\Lambda_a^2)}\;,\nn
\end{align}

\begin{align}
\mathcal{I}_6^{\prime\,dip}&=\int \frac{d {\bm l}_{\sT}'}{(2\pi)^2}\frac{({\bm l}_{\sT}'+{\bm p}_{\sT})\cdot {\bm S}_{\sT}}
{{\bm S}_{\sT}^2}\frac{1}{({\bm l}_{\sT}'+{\bm p}_{\sT})^2\,[{\bm l}_{\sT}^{\prime\,2} +L_a^2(\Lambda_a^2)]^3}\nn\\
&=\int_0^\infty \frac{d|{\bm l}_{\sT}'||{\bm
l}_{\sT}'|}{(2\pi)^2{\bm p}_{\sT}^2}\,\frac{1}{[{\bm
l}_{\sT}^{\prime\,2}+L_a^2(\Lambda_a^2)]^3} \int_0^{2\pi}d\theta\,
\frac{|{\bm l}_{\sT}'||{\bm S}_{\sT}|\cos[\theta+(\phi-\phi_S)]+|{\bm
p}_{\sT}||{\bm S}_{\sT}|\cos(\phi-\phi_S)}{{\bm l}_{\sT}^{\prime\,2}+{\bm p}_{\sT}^2+2|{\bm l}_{\sT}'||{\bm p}_{\sT}|\cos\theta}
\nn\\
&=\frac{1}{4\pi{\bm S}_{\sT}^2}\,\frac{|{\bm S}_{\sT}|}{|{\bm
p}_{\sT}|}\,\cos(\phi-\phi_S)\int_{L_a^2(\Lambda_a^2)}^{L_a^2(\Lambda_a^2)+{\bm
p}_{\sT}^2}\frac{dy}{y^3}
=\frac{1}{8\pi{\bm S}_{\sT}^2}\,\frac{|{\bm S}_{\sT}|}{|{\bm p}_{\sT}|}\left[\frac{1}{L_a^4(\Lambda_a^2)}-\frac{1}
{[L_a^2(\Lambda_a^2)+{\bm p}_{\sT}^2]^2}\right]\,\cos(\phi-\phi_S)\nn\\
&=\frac{1}{8\pi{\bm S}_{\sT}^2}\,({\bm p}_{\sT}\cdot{\bm
S}_{\sT})\,\frac{2L_a^2(\Lambda_a^2)+{\bm
p}_{\sT}^2}{L_a^4(\Lambda_a^2)\,[L_a^2(\Lambda_a^2)+{\bm
p}_{\sT}^2]^2}\;,\nn
\end{align}

\begin{align}
\mathcal{I}_7^{\prime\,dip}&=\int \frac{d {\bm l}_{\sT}'}{(2\pi)^2}\frac{{\left[({\bm l}_{\sT}'+{\bm p}_{\sT})\cdot {\bm S}_{\sT}\right]}^2}{{\bm S}_{\sT}^2}
\frac{1}{({\bm l}_{\sT}'+{\bm p}_{\sT})^2\,[{\bm l}_{\sT}^{\prime\,2} +L_a^2(\Lambda_a^2)]^3}\nn\\
&=\int_0^\infty \frac{d|{\bm l}_{\sT}'||{\bm
l}_{\sT}'|}{(2\pi)^2{\bm S}_{\sT}^2}\,\frac{1}{[{\bm
l}_{\sT}^{\prime\, 2}+L_a^2(\Lambda_a^2)]^3} \int_0^{2\pi}d\theta\,
\frac{\Big[|{\bm l}_{\sT}'||{\bm
S}_{\sT}|\cos[\theta+(\phi-\phi_S)]+|{\bm p}_{\sT}||{\bm
S}_{\sT}|\cos(\phi-\phi_S)\Big]^2}
{{\bm l}_{\sT}^{\prime\,2}+{\bm p}_{\sT}^2+2|{\bm l}_{\sT}'||{\bm p}_{\sT}|\cos\theta}
\nn\\
&=\frac{1}{{(2\pi)}^2{\bm
S}_{\sT}^2}\Bigg\{\int_{L_a^2(\Lambda_a^2)}^{L_a^2(\Lambda_a^2)+{\bm
p}_{\sT}^2}\frac{dy}{2\,y^3}\;\pi\,\frac{{\bm S}_{\sT}^2}{{\bm
p}_{\sT}^2}\, \left[{\bm p}_{\sT}^2+\left({\bm
p}_{\sT}^2+L_a^2(\Lambda_a^2)-y\right)\cos 2(\phi-\phi_S)\right]\;+\;\pi{\bm
S}_{\sT}^2\int_{L_a^2(\Lambda_a^2)+{\bm
p}_{\sT}^2}^\infty\frac{dy}{2\,y^3}
\Bigg\}\nn\\
&=\frac{1}{{(2\pi)}^2{\bm S}_{\sT}^2}\Bigg\{\Bigg[\pi\frac{{\bm
S}_{\sT}^2\,{\bm
p}_{\sT}^2}{4L_a^4(\Lambda_a^2)\,[L_a^2(\Lambda_a^2)+{\bm
p}_{\sT}^2]^2}\Big(
(\cos 2(\phi-\phi_S)+1)\,{\bm p}_{\sT}^2+(\cos 2(\phi-\phi_S)+2)\,L_a^2(\Lambda_a^2)\Big)\Bigg]\nn\\
&\qquad\qquad\quad\,+\;\Bigg[\pi\,\frac{{\bm S}_{\sT}^2}
{4\,[L_a^2(\Lambda_a^2)+{\bm p}_{\sT}^2]^2}\Bigg]\Bigg\}\nn\\
&=\frac{1}{16\pi L_a^4(\Lambda_a^2)\,[L_a^2(\Lambda_a^2)+{\bm
p}_{\sT}^2]}\Big[ L_a^2(\Lambda_a^2)+{\bm
p}_{\sT}^2\Big(1+\cos2(\phi-\phi_S)\Big)\Big]\nn\\
&=\frac{1}{16\pi L_a^4(\Lambda_a^2)\,[L_a^2(\Lambda_a^2)+{\bm
p}_{\sT}^2]}\Big[L_a^2(\Lambda_a^2)+2\frac{({\bm p}_{\sT}\cdot{\bm S}_{\sT})^2}{{\bm S}_{\sT}^2}\Big]\;,\nn\\
\end{align}

\begin{align}
\mathcal{I}_2^{exp}&=\int \frac{d{\bm l}_{\sT}'}{(2\pi)^2}\,
\frac{e^{-[{\bm l}_{\sT}^{\prime\, 2} +L_a^2(m^2)]/[(1-x)\Lambda_a^2]}}
{({\bm l}_{\sT}'+{\bm p}_{\sT})^2\,[{\bm l}_{\sT}^{\prime\, 2} +
L_a^2(m^2)]}\, ({\bm l}_{\sT}+{\bm p}_{\sT})^2  \nn \\
&=\int_0^\infty \frac{d|{\bm l}_{\sT}'||{\bm l}_{\sT}'|}{(2\pi)}\,
\frac{e^{-[{\bm l}_{\sT}^{\prime\, 2} +L_a^2(m^2)]/[(1-x)\Lambda_a^2]}}
{[{\bm l}_{\sT}^{\prime\, 2} +L_a^2(m^2)]} \nn \\
&=\frac{1}{4\pi} \int_{L_a^2(m^2)}^\infty dy\,
\frac{e^{-y/[(1-x)\Lambda_a^2]}}{y} = \frac{1}{4\pi}\;\Gamma\left(0,
\frac{L_a^2(m^2)}{(1-x)\Lambda_a^2}\right) \; , \nn
\end{align}

\begin{align}
\mathcal{I}_3^{exp}&=\int \frac{d{\bm l}_{\sT}'}{(2\pi)^2}\,
\frac{e^{-[{\bm l}_{\sT}^{\prime\, 2} +L_a^2(m^2)]/[(1-x)\Lambda_a^2]}}
{({\bm l}_{\sT}'+{\bm p}_{\sT})^2\,[{\bm l}_{\sT}^{\prime\, 2} +
L_a^2(m^2)]}\,
\frac{\left[({\bm l}_{\sT}'+{\bm p}_{\sT})\cdot {\bm p}_{\sT}
\right]^2}{{\bm p}_{\sT}^2} \nn \\
&=\int_0^\infty \frac{d|{\bm l}_{\sT}'||{\bm l}_{\sT}'|}
{(2\pi)^2\,{\bm p}_{\sT}^2} \,
\frac{e^{-[{\bm l}_{\sT}^{\prime\, 2} +L_a^2(m^2)]/[(1-x)\Lambda_a^2]}}
{[{\bm l}_{\sT}^{\prime\, 2} +L_a^2(m^2)]} \, \int_0^{2\pi}d\theta\,
\frac{\left[ |{\bm l}_{\sT}'||{\bm p}_{\sT}|\cos\theta+{\bm p}_{\sT}^2
\right]^2}
{{\bm l}_{\sT}^{\prime\, 2}+{\bm p}_{\sT}^2+2|{\bm l}_{\sT}'|
|{\bm p}_{\sT}|\cos\theta} \nn\\
&=\frac{1}{4\pi{\bm p}_{\sT}^2}\,\left\{ \int_0^{|{\bm p}_{\sT}|}
\frac{d|{\bm l}_{\sT}'||{\bm l}_{\sT}'|}{[{\bm l}_{\sT}^{\prime\, 2} +
L_a^2(m^2)]}\, (2{\bm p}_{\sT}^2-{\bm l}_{\sT}^{\prime\, 2})\,
e^{-[{\bm l}_{\sT}^{\prime\, 2} +L_a^2(m^2)]/[(1-x)\Lambda_a^2]} \right.
\nn\\
&\qquad\qquad \left. +\; {\bm p}_{\sT}^2 \int_{|{\bm p}_{\sT}|}^\infty
\frac{d|{\bm l}_{\sT}'||{\bm l}_{\sT}'|}{[{\bm l}_{\sT}^{\prime\, 2} +
L_a^2(m^2)]}\;
e^{-[{\bm l}_{\sT}^{\prime\, 2} +L_a^2(m^2)]/[(1-x)\Lambda_a^2]} \right\}
\nn\\
&=\frac{1}{8\pi{\bm p}_{\sT}^2}\,\left\{
\int_{L_a^2(m^2)}^{L_a^2(m^2)+{\bm p}_{\sT}^2} dy\,
\frac{2{\bm p}_{\sT}^2+L_a^2(m^2)-y}{y} \; e^{-y/[(1-x)\Lambda_a^2]}
+\;{\bm p}_{\sT}^2 \int_{L_a^2(m^2)+{\bm p}_{\sT}^2}^\infty dy\,
\frac{e^{-y/[(1-x)\Lambda_a^2]}}{y}  \right\}   \nn\\
&=\frac{1}{8\pi{\bm p}_{\sT}^2} \left\{ [L_a^2(m^2)+2{\bm p}_{\sT}^2]
\, \Gamma\left( 0,\frac{L_a^2(m^2)}{(1-x)\Lambda_a^2}\right) -
[L_a^2(m^2)+{\bm p}_{\sT}^2]\,\Gamma\left( 0,
\frac{L_a^2(m^2)+{\bm p}_{\sT}^2}{(1-x)\Lambda_a^2} \right) \right.
\nn\\
&\qquad\quad\quad \left. +\;(1-x)\,\Lambda_a^2\,\left(
e^{-[L_a^2(m^2)+{\bm p}_{\sT}^2]/[(1-x)\Lambda_a^2]}-
e^{-L_a^2(m^2)/[(1-x)\Lambda_a^2]}\right) \right\} \; , \nn
\end{align}

\begin{align}
\mathcal{I}_4^{exp}&=\int \frac{d{\bm l}_{\sT}'}{(2\pi)^2}\,
\frac{e^{-[{\bm l}_{\sT}^{\prime\, 2} +L_a^2(m^2)]/[(1-x)\Lambda_a^2]}}
{({\bm l}_{\sT}'+{\bm p}_{\sT})^2\,[{\bm l}_{\sT}^{\prime\, 2} +
L_a^2(m^2)]}\,
\frac{\left[({\bm l}_{\sT}'+{\bm p}_{\sT})\cdot {\bm p}_{\sT}
\right]^2}{{\bm p}_{\sT}^2} \, ({\bm l}_{\sT}'+{\bm p}_{\sT})^2  \nn
\\
&=\int_0^\infty \frac{d|{\bm l}_{\sT}'||{\bm l}_{\sT}'|}
{(2\pi)^2\,{\bm p}_{\sT}^2} \,
\frac{e^{-[{\bm l}_{\sT}^{\prime\, 2} +L_a^2(m^2)]/[(1-x)\Lambda_a^2]}}
{[{\bm l}_{\sT}^{\prime\, 2} +L_a^2(m^2)]} \, \int_0^{2\pi} d\theta\,
\left( |{\bm l}_{\sT}'||{\bm p}_{\sT}|\cos\theta + {\bm p}_{\sT}^2
\right) \nn\\
&=\int_0^\infty \frac{d|{\bm l}_{\sT}'||{\bm l}_{\sT}'|}{2\pi} \,
\frac{e^{-[{\bm l}_{\sT}^{\prime\, 2} +L_a^2(m^2)]/[(1-x)\Lambda_a^2]}}
{[{\bm l}_{\sT}^{\prime\, 2} +L_a^2(m^2)]}  \nn \\
&=\frac{1}{4\pi}\int_{L_a^2(m^2)}^\infty dy\,
\frac{e^{-y/[(1-x)\Lambda_a^2]}}{y} = \frac{1}{4\pi}\;\Gamma\left( 0,
\frac{L_a^2(m^2)}{(1-x)\Lambda_a^2} \right) \equiv \mathcal{I}_2^{exp} \; ,
\nn
\end{align}

\begin{align}
\mathcal{I}_5^{exp}&=\int \frac{d{\bm l}_{\sT}'}{(2\pi)^2}\,
\frac{e^{-[{\bm l}_{\sT}^{\prime\, 2} +L_a^2(m^2)]/[(1-x)\Lambda_a^2]}}
{({\bm l}_{\sT}'+{\bm p}_{\sT})^2\,[{\bm l}_{\sT}^{\prime\, 2} +
L_a^2(m^2)]}\, ({\bm l}_{\sT}'+{\bm p}_{\sT})^4  \nn \\
&=\int_0^\infty \frac{d|{\bm l}_{\sT}'||{\bm l}_{\sT}'|}
{(2\pi)^2} \,
\frac{e^{-[{\bm l}_{\sT}^{\prime\, 2} +L_a^2(m^2)]/[(1-x)\Lambda_a^2]}}
{[{\bm l}_{\sT}^{\prime\, 2} +L_a^2(m^2)]} \, \int_0^{2\pi}d\theta\,
\left( {\bm l}_{\sT}^{\prime\, 2}+{\bm p}_{\sT}^2+2|{\bm l}_{\sT}'|
|{\bm p}_{\sT}| \cos\theta \right)  \nn  \\
&=\int_0^\infty \frac{d|{\bm l}_{\sT}'||{\bm l}_{\sT}'|}{2\pi} \,
\frac{e^{-[{\bm l}_{\sT}^{\prime\, 2} +L_a^2(m^2)]/[(1-x)\Lambda_a^2]}}
{[{\bm l}_{\sT}^{\prime\, 2} +L_a^2(m^2)]} \,
({\bm l}_{\sT}^{\prime\, 2} + {\bm p}_{\sT}^2) \nn  \\
&=\frac{1}{4\pi}\int_{L_a^2(m^2)}^\infty dy\,
\frac{y-L_a^2(m^2)+{\bm p}_{\sT}^2}{y^3}\,e^{-y/[(1-x)\Lambda_a^2]}
\nn \\
&=\frac{1}{4\pi}\,\left\{ [{\bm p}_{\sT}^2-L_a^2(m^2)]\,
\Gamma\left( 0, \frac{L_a^2(m^2)}{(1-x)\Lambda_a^2} \right) +(1-x)\,
\Lambda_a^2\, e^{-L_a^2(m^2)/[(1-x)\Lambda_a^2]} \right\}  \; , \nn
\end{align}

\begin{align}
\mathcal{I}_6^{exp}&=\int \frac{d{\bm l}_{\sT}'}{(2\pi)^2}\,
\frac{e^{-[{\bm l}_{\sT}^{\prime\, 2} +L_a^2(m^2)]/[(1-x)\Lambda_a^2]}}
{({\bm l}_{\sT}'+{\bm p}_{\sT})^2\,[{\bm l}_{\sT}^{\prime\, 2} +
L_a^2(m^2)]}\,
\frac{({\bm l}_{\sT}'+{\bm p}_{\sT})\cdot {\bm S}_{\sT}}
{{\bm S}_{\sT}^2}  \nn  \\
&=\int_0^\infty \frac{d|{\bm l}_{\sT}'||{\bm l}_{\sT}'|}
{(2\pi)^2{\bm S}_{\sT}^2}\,
\frac{e^{-[{\bm l}_{\sT}^{\prime\, 2} +L_a^2(m^2)]/[(1-x)\Lambda_a^2]}}
{[{\bm l}_{\sT}^{\prime\, 2} +L_a^2(m^2)]} \, \int_0^{2\pi}d\theta\,
\frac{|{\bm l}_{\sT}'||{\bm S}_{\sT}|\cos(\theta+\phi-\phi_S)+
|{\bm p}_{\sT}||{\bm S}_{\sT}|\cos(\phi-\phi_S)}
{{\bm l}_{\sT}^{\prime\, 2}+{\bm p}_{\sT}^2+2|{\bm l}_{\sT}'|
|{\bm p}_{\sT}|\cos\theta}  \nn  \\
&=\frac{|{\bm p}_{\sT}||{\bm S}_{\sT}|\, \cos(\phi - \phi_S)}
{4\pi\,{\bm S}_{\sT}^2 {\bm p}_{\sT}^2}\,
\int_{L_a^2(m^2)}^{L_a^2(m^2)+{\bm p}_{\sT}^2} dy\,
\frac{e^{-y/[(1-x)\Lambda_a^2]}}{y}  \nn \\
&=\frac{{\bm p}_{\sT} \cdot {\bm S}_{\sT}}
{4\pi{\bm S}_{\sT}^2 {\bm p}_{\sT}^2}\,
\left\{ \Gamma\left( 0,\frac{L_a^2(m^2)}{(1-x)\Lambda_a^2}\right) -
\Gamma\left( 0,\frac{L_a^2(m^2)+{\bm p}_{\sT}^2}{(1-x)\Lambda_a^2}
\right) \right\} \; , \nn
\end{align}

\begin{align}
\mathcal{I}_7^{exp}&=\int \frac{d{\bm l}_{\sT}'}{(2\pi)^2}\,
\frac{e^{-[{\bm l}_{\sT}^{\prime\, 2} +L_a^2(m^2)]/[(1-x)\Lambda_a^2]}}
{({\bm l}_{\sT}'+{\bm p}_{\sT})^2\,[{\bm l}_{\sT}^{\prime\, 2} +
L_a^2(m^2)]}\,
\frac{\left[ ({\bm l}_{\sT}'+{\bm p}_{\sT})\cdot {\bm S}_{\sT}
\right]^2}{{\bm S}_{\sT}^2}  \nn  \\
&=\int_0^\infty \frac{d|{\bm l}_{\sT}'||{\bm l}_{\sT}'|}
{(2\pi)^2{\bm S}_{\sT}^2}\,
\frac{e^{-[{\bm l}_{\sT}^{\prime\, 2} +L_a^2(m^2)]/[(1-x)\Lambda_a^2]}}
{[{\bm l}_{\sT}^{\prime\, 2} +L_a^2(m^2)]}\,  \int_0^{2\pi}d\theta\,
\frac{|{\bm l}_{\sT}'||{\bm S}_{\sT}|\cos(\theta+\phi-\phi_S)+
|{\bm p}_{\sT}||{\bm S}_{\sT}|\cos(\phi-\phi_S)}
{{\bm l}_{\sT}^{\prime\, 2}+{\bm p}_{\sT}^2+2|{\bm l}_{\sT}'|
|{\bm p}_{\sT}|\cos\theta} \nn \\
&=\frac{1}{(2\pi)^2{\bm S}_{\sT}^2} \Bigg\{
\int_{L_a^2(m^2)}^{L_a^2(m^2)+{\bm p}_{\sT}^2} dy\,
\frac{e^{-y/[(1-x)\Lambda_a^2]}}{y}\,\pi\,
\frac{{\bm S}_{\sT}^2}{{\bm p}_{\sT}^2}\,
\left[ {\bm p}_{\sT}^2 +( {\bm p}_{\sT}^2+L_a^2(m^2)-y)\,
\cos 2(\phi-\phi_S)\right]  \nn \\
&\qquad\qquad\quad+\pi {\bm S}_{\sT}^2
\int_{L_a^2(m^2)+{\bm p}_{\sT}^2}^\infty dy\,
\frac{e^{-y/[(1-x)\Lambda_a^2]}}{y}  \Bigg\}  \nn  \\
&=\frac{1}{8\pi{\bm p}_{\sT}^2} \Bigg\{ (1-x)\, \Lambda_a^2 \,
\left( e^{-[L_a^2(m^2)+{\bm p}_{\sT}^2]/[(1-x)\Lambda_a^2]}-
e^{-L_a^2(m^2)/[(1-x)\Lambda_a^2]} \right)\, \cos 2(\phi-\phi_S) \nn
\\
&\qquad\qquad+\Gamma\left( 0,\frac{L_a^2(m^2)}{(1-x)\,\Lambda_a^2}
\right) \left[ [{\bm p}_{\sT}^2+L_a^2(m^2)]\, \cos2(\phi-\phi_S)+
{\bm p}_{\sT}^2 \right] \nn  \\
&\qquad\qquad-\Gamma\left( 0,
\frac{L_a^2(m^2)+{\bm p}_{\sT}^2}{(1-x)\Lambda_a^2} \right)\,
[{\bm p}_{\sT}^2+L_a^2(m^2)]\, \cos 2(\phi-\phi_S) \Bigg\} \; .
\label{eq:Dcoeffs2}
\end{align}


\bibliographystyle{h-physrev}
\bibliography{alebiblio}


\end{document}